\documentclass[final,5p,times,twocolumn]{elsarticle}

\usepackage{graphicx}
\usepackage{dcolumn}
\usepackage{microtype}
\usepackage[hyperindex=true,colorlinks=true]{hyperref}
\hypersetup{citecolor=[rgb]{0.125,0.679,0.196}} 
\usepackage{url}

\usepackage{amsmath}
\usepackage{amssymb}
\usepackage{braket}
\usepackage{mathrsfs} 
\usepackage{bbm} 
\usepackage{stmaryrd} 
\usepackage{slashed}
\usepackage{bm} 
\usepackage{color} 
\usepackage{physics}
\usepackage{empheq}
\usepackage{simplewick} 

\DeclareMathOperator{\GL}{GL}
\DeclareMathOperator{\U}{U}
\newcommand*{\transpose}{^{\mkern-1.5mu\mathsf{T}}} 
\def\mean#1{\left< #1 \right>}
\makeatletter
\newsavebox\myboxA
\newsavebox\myboxB
\newlength\mylenA
\newcommand*\xoverline[2][0.75]{%
    \sbox{\myboxA}{$\m@th#2$}%
    \setbox\myboxB\null
    \ht\myboxB=\ht\myboxA%
    \dp\myboxB=\dp\myboxA%
    \wd\myboxB=#1\wd\myboxA
    \sbox\myboxB{$\m@th\overline{\copy\myboxB}$}
    \setlength\mylenA{\the\wd\myboxA}
    \addtolength\mylenA{-\the\wd\myboxB}%
    \ifdim\wd\myboxB<\wd\myboxA%
       \rlap{\hskip 0.5\mylenA\usebox\myboxB}{\usebox\myboxA}%
    \else
        \hskip -0.5\mylenA\rlap{\usebox\myboxA}{\hskip 0.5\mylenA\usebox\myboxB}%
    \fi}
\makeatother
\usepackage{xparse}
\NewDocumentCommand{\evalat}{sO{\big}mm}{%
  \IfBooleanTF{#1}
   {\mleft. #3 \mright|_{#4}}
   {#3#2|_{#4}}%
}

\usepackage{feynmp-auto}

\def\orcid#1{\href{https://orcid.org/#1}{\!\includegraphics[keepaspectratio,width=0.7em]{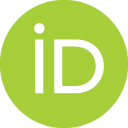}}}

\journal{Annals of Physics}

\begin{document}

\begin{frontmatter}


\title{Nambu-Covariant Many-Body Theory II: Self-Consistent Approximations}

\author[aff1,aff2]{M.~Drissi \orcid{0000-0001-9472-6280}\corref{cor}}
\ead{mdrissi@triumf.ca}

\author[aff3,aff2]{A.~Rios \orcid{0000-0002-8759-3202} }

\author[aff4,aff5,aff2]{C.~Barbieri \orcid{0000-0001-8658-6927} }

\affiliation[aff1]{organization={TRIUMF},
        addressline={4004 Wesbrook Mall},
        city={Vancouver},
        postcode={V6T 2A3},
        state={British Columbia},
        country={Canada}}

\affiliation[aff2]{organization={Department of Physics, University of Surrey},
        city={Guildford},
        postcode={GU2 7XH},
        country={United Kingdom}}

\affiliation[aff3]{
        organization={Departament de Fsíca Quàntica i Astrofísica,
            Institut de Ciències del Cosmos (ICCUB), Universitat de Barcelona}, 
        addressline=\hbox{Martí Franquès 1},
        postcode={E08028},
        city={Barcelona},
        country={Spain}}

\affiliation[aff4]{organization={Dipartimento di Fisica "Aldo Pontremoli", Università degli studi di Milano},
            addressline={Via Celoria 16}, 
            city={Milano},
            postcode={20133}, 
            country={Italy}}

\affiliation[aff5]{organization={INFN, Sezione di Milano},
            addressline={Via Celoria 16}, 
            city={Milano},
            postcode={20133}, 
            country={Italy}}

\cortext[cor]{Corresponding author}


\date{\today}

\begin{abstract}
The theory of Self-Consistent Green's Function (SCGF) is reformulated in an explicit Nambu-covariant fashion
for applications to many-body systems at non-zero temperature in symmetry-broken phases.
This is achieved by extending the Nambu-covariant formulation of perturbation theory, presented in the first part of this work,
to non-perturbative schemes based on self-consistently dressed propagators and vertices.
We work out in detail 
the self-consistent ladder approximation,
motivated by a trade-off between numerical complexity
and many-body phenomenology.
Taking a complex general Hartree-Fock-Bogoliubov (HFB) propagator as a
starting point, we also formulate and prove a sufficient condition
on the stability of the HFB self-energy to ensure the convergence
of the initial series of ladders at any energy.
The self-consistent ladder approximation
is written purely in terms of spectral functions and
the resulting set of equations, when expressed in terms of Nambu tensors,
are remarkably similar to those in the symmetry-conserving case. 
This puts the application of the self-consistent ladder approximation
to symmetry-broken phases of infinite nuclear matter within reach.
\end{abstract}

\begin{keyword}
Quantum many-body theory \sep Symmetry-breaking \sep Self-consistent Green's functions theory \sep Superfluidity



\end{keyword}

\end{frontmatter}

\tableofcontents


\section{Introduction}

In the first part of this work, henceforth referred to as Part~I~\cite{part1},
we have formulated the perturbative expansion of the 
Green's functions of a given many-body system in terms of Nambu tensors.
This formulation, referred to as
Nambu-Covariant Perturbation Theory (NCPT), allows for a straightforward
design of Bogoliubov-invariant perturbative approximations of many-body observables.
We have also shown that the un-oriented Feynman diagrams, 
indexing perturbative contributions in NCPT, factorise a multitude
of Feynman diagrams that occur in other diagrammatic formalisms
accounting for particle-number symmetry-breaking~\cite{Soma2011,Duguet2016}.
We have argued that such factorisation can be
used to develop new algorithms with improved scalability
for massively parallel computational architectures.
Nevertheless, as powerful as they are, approximations that remain perturbative
in the bare interaction are not well suited to tackle strongly
correlated many-body systems, including atomic nuclei and neutron-star matter. 
To overcome this difficulty, a multitude of approaches have been
developed in \emph{ab initio} nuclear physics.

A first strategy consists in improving the starting point of the perturbative
expansion. This leads to the development of algorithms that tackle directly
more and more complex model Hamiltonians. From a given Hamiltonian $H$,
the aim is to design another Hamiltonian, $H_{\text{ref}}$, together with the means
of calculating associated observables. $H_{\text{ref}}$ is engineered such that 
the observables associated to it are as close as possible to those associated to $H$.
This is the case, for example, of the No-Core Shell Model~\cite{Navratil2009,Roth2011,Barrett2013}
which aims at calculating eigenenergies and eigenstates of atomic nuclei
via a partial diagonalisation of a given $H_{\text{ref}}$, which equals $H$ on a carefully
chosen model space.
Another approach along these lines is mean-field theory,
and its many refinements including spontaneous symmetry-breaking, projections and
the Generator Coordinate Method~\cite{Griffin1957}. 
These approaches ultimately amount to design an $H_{\text{ref}}$ not equal to $H$,
but optimised to best reproduce the ground-state energy of
$H$ associated to a system of interest.
For a recent review combining projections and spontaneous symmetry-breaking
in the theory of mean fields see Ref.~\cite{Bally2021}.
Similarity Renormalisation Group
techniques~\cite{Glazek1993,Wegner1994,Bogner2010}, and their in-medium
counterparts like the In-Medium Similarity Renormalisation Group (IMSRG)~\cite{Tsukiyama2011,Tsukiyama2012,Hergert2013,Bogner2014},
can also be included in this line of research. 
A complementary strategy consists in building a series
of corrections, depending on the difference $H-H_{\text{ref}}$, with the best possible
rate of convergence to the exact value of the observables of interest.
Standard Many-Body Perturbation Theory (MBPT), in its many flavours~\cite{Shavitt2009},
is a prime example of this strategy.
One can also treat complex many-body systems using 
summations of infinite 
subsets of Feynman diagrams. 
Among these refinements, let us mention
the Coupled Cluster (CC)~\cite{Coester1958,Coester1960,Kowalski2004,Hagen2014}
and Self-Consistent Green's Function (SCGF)
approaches~\cite{Luttinger1960,Dickhoff1992,Dickhoff2004,Barbieri2017lnp}.
Additionally, the techniques of resummation of a series can also be
included in this line of research as they aim at improving the rate
of convergence of a given series. For example, let us mention the resummations based on
analytic continuation or Padé approximants which are compared in Ref.~\cite{Demol2020}.

Modern \emph{ab initio} approaches have grown more sophisticated
by hybridising several of the techniques mentioned above.
To name a few of those recent hybridisations, we mention
the combination of particle-number symmetry-breaking with MBPT, CC, SCGF and IMSRG
which lead, respectively, to the development of Bogoliubov MBPT~\cite{Tichai2018a};
Bogoliubov CC~\cite{Signoracci2015,Duguet2016};
Gorkov SCGF~\cite{Soma2011,Soma2013,Soma2014,Barbieri2022} and
Bogoliubov IMSRG~\cite{Tichai2021}.
In general, the aim of such hybridisation techniques is to increase the range of applicability
while reducing the global numerical cost.
The typical price to pay is the development of an increasingly complex formalism
and a more intricate numerical implementation.
For example, a reliable and precise calculation of thermodynamical
properties of infinite superfluid nuclear matter is expected to require
such a hybrid method. Neutron-star physics requires the application 
of
an approach that can tackle anisotropic pairing gaps, which can be dealt with
rotational and particle-number symmetry-breaking~\cite{Dean2003,Sigrist1991}.
Moreover, the strong repulsion (or hard core) of several
nuclear interactions can be dealt with by 
performing an infinite, ladder diagrammatic
summation~\cite{Brueckner1954,Brueckner1955,Bethe1956,Day1985,Rios2014}.
While softer interactions with increasing precision have been derived based on $\chi$EFT
with a low cutoff~\cite{Epelbaum2009,Entem2020,Epelbaum2020},
their applicability to the regime of high nuclear density
(several times higher than the saturation density) necessary to describe the outer core of
neutron stars is questionable.
This is especially true concerning small quantities such as
the ${}^{3}PF_2$ gap in pure neutron matter~\cite{Drischler2017}.
To reduce the sensitivity to regulator artefacts, high cutoff interactions
should be considered and 
the original challenge of the strong repulsion of the nuclear interaction is re-encountered.
Finally, to ensure the thermodynamical consistency of macroscopic observables
computed \emph{ab initio}, a SCGF approximation is required~\cite{Baym1962}.
While this
is not the only possible \emph{ab initio} route
to deal with the thermodynamics of superfluid nuclear matter,
the combination of spontaneous symmetry-breaking, ladder diagrammatic summation,
and self-consistent propagators at non-zero temperature 
represents a substantial step forward compared to existing treatments. 
Unsurprisingly, these requirements increase non-negligibly the formal complexity.
To mitigate this complexity, one could use 
software to automatically expand more and more complex many-body approximations.
For example, a software toolchain is under development which so far automatises
diagrammatic generation at zero temperature~\cite{Arthuis2019} and angular momentum
reduction~\cite{Tichai2020}.
In our case, where we want to sum an infinite number of diagrams at non-zero temperature,
automatised approaches in existence fall short
and further developments are required.

Instead, we address directly the additional complexity at the formal level. 
In Part~I, we reformulated the many-body problem
in terms of Nambu tensors, which allowed us to derive approximations of many-body observables
without specifying the field basis we are working with.
This extra mathematical abstraction not only allows us to derive equations
valid up to any Bogoliubov transformation, but also removes an unnecessary
surplus of formal complexity. The field basis should be specified
only when it brings necessary extra properties\footnote{For example, we will see that
the Galitskii-Migdal-Koltun (GMK) sum rule as given in Eq.~\eqref{GMKsumRule}
is only valid for a certain subset of field bases.}.
In the case of the self-consistent ladder approximation at non-zero temperature
with symmetry-breaking, this will lead to a self-consistent set of equations
of similar complexity compared to the symmetry-conserving case.

The aim of this paper is therefore to formulate the theory of SCGF
in a Nambu-covariant fashion. We refer to this particular approach of SCGF
as Nambu-Covariant Self-Consistent Green's Function (NC-SCGF).
The present paper is organised as follows.
In Sec.~\ref{Sec:NambuProp}, we derive exact properties of the propagator
and emphasise their associated covariance group.
Self-consistent approximations of the propagator are then expressed within
the Nambu-covariant formalism in Sec.~\ref{Sec:SCprop}.
Finally, in Sec.~\ref{Sec:SCvert}, we study a self-consistent dressing of the two-body
interaction via a Bethe-Salpeter equation~\cite{Salpeter1951}. 
We also derive the ladder approximation as a particular case.
We defer the discussion of specific applications to infinite nuclear matter
into a subsequent paper.

\section{Nambu tensor propagator}\label{Sec:NambuProp}
In this section,
we define the propagator for a given
Hamiltonian, $H$, as a Nambu tensor.
For simplicity, we assume that $H$ is Hermitian and time-independent.
We introduce the spectral representation of the propagator and
detail the generalisation of standard exact properties
to the symmetry-breaking case.

\subsection{Definitions}
\subsubsection{Bases}
Let us consider a many-body system of fermions in a statistical ensemble
described by the Hermitian Hamiltonian $H$ and the inverse 
temperature $\beta$.
Let $\mathcal{B}^f$ be a basis of the field vector space 
$\mathscr{H}^f$ as discussed in Part~I.
For convenience, we choose to work with a field basis $\mathcal{B}^f$
based on an orthonormal single-particle basis $\mathcal{B} = \Set{\ket{b}}$, i.e.\
\begin{equation}\label{WorkingFieldBasis}
    \mathcal{B}^f \equiv \set{a^\dagger_b} \cup \set{a_c} \ ,
\end{equation}
where ${}^\dagger$ denotes the usual Hermitian conjugation,
and where $a^\dagger_b (a_b)$ denote creation (annihilation) operators
associated to the orthonormal single-particle basis, $\mathcal{B}$.
We also index the field basis over global indices
\begin{equation}
    \mu \equiv (b,l_b)
\end{equation}
where $b$ indexes single-particle states, and $l_b$ are Nambu indices.
Despite working with a specific field basis $\mathcal{B}^f$, most of the equations
that follow will be equalities between two Nambu tensors. As such, 
the equations will remain valid
if one works in a different field basis, ${\mathcal{B}^f}'$.
In some particular cases, however, equations will be valid only for
a subset of field bases.
In such cases, we will specify under the action of which
sub-group of $\GL(\mathscr{H}^f)$ the equations remain valid.

Let $\mathrm{A}^{\mu}$ and $\mathrm{A}_{\mu}$
be the Nambu fields associated to $\mathcal{B}^f$.
These fields verify\footnote{We stress that, in a general field basis ${\mathcal{B}^f}'$,
Nambu fields are linear combinations of creation and annihilation operators.
In addition, without the assumed orthonormality of $\mathcal{B}$,
creation and annihilation operators would not be Hermitian
conjugated to each other. See Part~I for more details.}
\begin{subequations}\label{NambuFieldsDef}
\begin{align}
    \mathrm{A}_{(b, 1)} &\equiv a^\dagger_b \ , \\
    \mathrm{A}_{(b, 2)} &\equiv a_{b} \ , \\
    \mathrm{A}^{(b, 1)} &\equiv a_b \ , \\
    \mathrm{A}^{(b, 2)} &\equiv a^\dagger_{b}\ .
\end{align}
\end{subequations}
Let also $g^{\mu\nu}$ be the metric tensor verifying 
\begin{subequations}
\begin{align}
 \Set{ \mathrm{A}^\mu , \mathrm{A}^\nu } 
    &= g^{\mu\nu} \label{CAR_UU} \ , \\
 \Set{ \mathrm{A}^\mu , \mathrm{A}_\nu } 
    &= {g^{\mu}}_{\nu} \label{CAR_UD} \ , \\
 \Set{ \mathrm{A}_\mu , \mathrm{A}_\nu } 
    &= g_{\mu\nu} \label{CAR_DD} \ .
\end{align}
\end{subequations}
We recall that the metric tensor can be used to raise or lower indices,
\begin{subequations}\label{raiseIndex}
\begin{align}
    \mathrm{A}^{\mu} = \sum_{\nu} g^{\mu\nu} \ \mathrm{A}_{\nu} \ , \\
    \mathrm{A}_{\mu} = \sum_{\nu} g_{\mu\nu} \ \mathrm{A}^{\nu} \ ,
\end{align}
\end{subequations}
and that it fulfils the relation
\begin{equation}
    \sum_{\lambda} g^{\mu\lambda} g_{\lambda\nu}
        = {g^{\mu}}_{\nu} = \delta_{\mu\nu} \ .
\end{equation}
For more details on the foundations of the Nambu-covariant formalism,
we refer the reader to Part~I.

\subsubsection{Nambu tensor propagator}
The Hamiltonian $H$ is expressed as a polynomial of contravariant 
Nambu fields according to
\begin{equation}
\label{Hamiltonian}
    H 
    \equiv 
        \sum^{k_{\text{max}}}_{k=0} \ \frac{1}{(2k)!}
                        \sum_{\mu_1 \dots \mu_{2k}}
                              v^{(k)}_{\mu_1 \dots \mu_{2k}} \
                              \mathrm{A}^{\mu_1} \dots \mathrm{A}^{\mu_{2k}} \ ,
\end{equation}
where $v^{(k)}_{\mu_1 \dots \mu_{2k}}$ are covariant Nambu tensors.
We note that $k$ denotes the $k$-body nature of the tensor, in the sense
that it involves $2k$ Nambu fields. In the symmetry-conserving case, 
$k=2$ would be associated to a two-body interaction; $k=3$, to a three-body interaction, and so on. 

The exact contravariant, mixed and covariant versions of the Nambu tensor
propagator are defined by\footnote{As for Part~I,
we assume natural units where $\hbar = c = k_\text{B} = 1$.}
\begin{subequations}\label{DefTensorProp}
\begin{align}
    - \mathcal{G}^{\mu\nu}(\tau, \tau')
        &\equiv
        \mean{\mathrm{T}\left[ 
                \mathrm{A}^{\mu}(\tau) \mathrm{A}^{\nu}(\tau') 
            \right]} \ , \label{DefContravProp} \\
    - {\mathcal{G}^{\mu}}_{\nu}(\tau, \tau')
        &\equiv
        \mean{\mathrm{T}\left[ 
                \mathrm{A}^{\mu}(\tau) \mathrm{A}_{\nu}(\tau') 
            \right]} \ , \label{DefMixedProp} \\  
    - {\mathcal{G}_{\mu}}^{\nu}(\tau, \tau')
        &\equiv
        \mean{\mathrm{T}\left[ 
                \mathrm{A}_{\mu}(\tau) \mathrm{A}^{\nu}(\tau') 
            \right]} \ , \label{DefMixedPropBIS} \\
    - \mathcal{G}_{\mu\nu}(\tau, \tau')
        &\equiv
        \mean{\mathrm{T}\left[ 
                \mathrm{A}_{\mu}(\tau) \mathrm{A}_{\nu}(\tau') 
            \right]} \ , \label{DefCovProp}
\end{align}
\end{subequations}
where the imaginary-time dependence is with respect to $H$ and
where $\mathrm{T}\left[ \dots \right]$ denotes
the time-ordering from right to left when increasing imaginary-time $\tau$.
The ensemble average $\mean{\dots}$ is defined with respect to $H$.
The exact density matrix, $\rho$, and the partition function, $Z$, are defined by
\begin{subequations}
\begin{align}
    \rho &\equiv \frac{1}{Z} e^{-\beta H} \ , \\
    Z &\equiv \Tr\left( e^{-\beta H} \right) \ .
\end{align}
\end{subequations}

We stress that the raising and lowering of indices via metric contractions
is compatible with the definitions given in Eqs.~\eqref{DefTensorProp}.
For instance, the fully covariant and fully contravariant propagators are related by
the expression
\begin{equation}
    \mathcal{G}^{\mu\nu}(\tau,\tau')
    = \sum_{\lambda_1\lambda_2}
        g^{\mu\lambda_1} g^{\nu\lambda_2} \ 
        \mathcal{G}_{\lambda_1\lambda_2}(\tau,\tau') \ .
\end{equation}
In the following, most equations will remain valid after any raising/lowering
of indices. Whenever there is no ambiguity, we choose to drop
tensor indices and write equalities between tensors
in an intrinsic fashion.

Since we assume $H$ to be time-independent, the propagator depends
only on the difference of its two times and we can use, without ambiguity,
the one-time notation
\begin{equation}
    \mathcal{G}(\tau) 
        \equiv \mathcal{G}(\tau + \tau', \tau')
        = \mathcal{G}(\tau, 0) \ .
\end{equation}
The exact propagator verifies an antisymmetry property,
and is extended into a $\beta$-quasiperiodic function
so that
\begin{subequations}
\begin{align}
    \label{AntisymProp}
    \mathcal{G}^{\mu\nu}(\tau) &= - \mathcal{G}^{\nu\mu}(-\tau) \ , \\
    \label{QuasiPeriodicProp}
    \mathcal{G}^{\mu\nu}(\tau + \beta) &= - \mathcal{G}^{\nu\mu}(\tau) \ .
\end{align}
These equations remain valid under the action of $\GL(\mathscr{H}^f)$.
The propagator also fulfils the Hermitian property,
\begin{align}
    \label{HermitianProp_Ortho}
    \mathcal{G}^{\mu\nu}(\tau)^* &= \mathcal{G}_{\nu\mu}(\tau) \ , 
\end{align}
which, however, remains valid only under
the action of the unitary group $\U(\mathscr{H}^f)$.
As discussed in~\ref{App:HermitianConjugation},
we replace Eq.~\eqref{HermitianProp_Ortho} with 
\begin{equation}\label{HermitianProp}
    (\mathcal{G}^\dagger)^{\mu\nu}(\tau) = \mathcal{G}^{\mu\nu}(\tau)
\end{equation}
\end{subequations}
where ${}^\dagger$ denotes the Hermitian conjugation of a tensor as
defined in~\ref{App:HermitianConjugation}.
The advantage of Eq.~\eqref{HermitianProp} over Eq.~\eqref{HermitianProp_Ortho}
is that it remains valid under the action of $\GL(\mathscr{H}^f)$.
With this, we can write the antisymmetry, $\beta$-quasiperiodic and Hermitian properties of the propagator  
in an intrinsic fashion as
\begin{subequations}\label{intrinsicSymProps}
\begin{align}
    \mathcal{G}\transpose(\tau) &= -\mathcal{G}(-\tau) \ , \\
    \mathcal{G}(\tau + \beta) &= - \mathcal{G}\transpose(\tau) \ , \\
        \mathcal{G}^\dagger(\tau) &= \mathcal{G}(\tau) \ ,
\end{align}
\end{subequations}
where $\transpose$ denotes the transposition defined in Part~I.

As a consequence of Eqs.~\eqref{intrinsicSymProps},
we can define the energy representation of the exact propagator 
as the following Fourier transform
\begin{subequations}
\begin{align}
    \mathcal{G}(\omega_p)
        &\equiv \int_{0}^{\beta} \mathrm{d}\tau \ e^{i\omega_p \tau} \
                            \mathcal{G}(\tau) \ , \\
    \mathcal{G}(\tau)
        &= \frac{1}{\beta} \sum_{\omega_p}
                              \ e^{-i\omega_p \tau} \  
                              \mathcal{G}(\omega_p) \ ,
\end{align}
\end{subequations}
where $\omega_p \equiv (2p+1) \frac{\pi}{\beta}$ are fermionic Matsubara frequencies.
In the energy representation, the Hermitian and antisymmetry properties read
\begin{subequations}
\begin{align}
    \mathcal{G}(\omega_p)
        &= \mathcal{G}^\dagger(-\omega_p) \ , 
        \label{HermitianPropEnergy} \\
    \mathcal{G}(\omega_p)
        &= -\mathcal{G}\transpose(-\omega_p) \ . \label{AntisymPropEnergy}
\end{align}
\end{subequations}

\subsection{Spectral representation}
Let $\ket{\Psi_n}$ and $E_n$ denote exact 
orthonormal eigenstates and real eigenvalues of $H$, i.e.\ 
\begin{equation}
   \forall n , \ H \ket{\Psi_n} = E_n \ket{\Psi_n} \ .
\end{equation}
The set of $\ket{\Psi_n}$ forms a complete basis of the Fock space.
Using this orthonormal basis to express traces over the Fock space, 
the Fourier transform of the exact propagator reads
\begin{multline}
\label{LehmRepContravProp}
  \mathcal{G}^{\mu\nu}(\omega_p)
  =\frac{1}{Z}
    \sum_{m,n} \Braket{
                \Psi_m|
                  \mathrm{A}^{\mu}
                |\Psi_n}
                \Braket{\Psi_n|
                  \mathrm{A}^{\nu}
                |\Psi_m} \, \\ \times
                e^{-\beta E_m}
                \frac{ e^{-\beta(E_n - E_m)} + 1}
                {i\omega_p - (E_n - E_m)} \ ,
\end{multline}
which is the so-called Lehman's representation of the propagator.
Defining the spectral function by
\begin{align}\label{DefSpectralFunction}
  S^{\mu\nu} (\omega)
  &\equiv
    \frac{1}{Z} \sum_{m,n}
        \Braket{\Psi_m|\mathrm{A}^{\mu}|\Psi_n}
        \Braket{\Psi_n|\mathrm{A}^{\nu}|\Psi_m} \nonumber \\
    &\phantom{= \sum}
        \times
        e^{-\beta E_m}\left(1 + e^{-\beta\omega} \right)\
        2\pi \ \delta\left(E_n - E_m - \omega \right) \ ,
\end{align}
the \emph{spectral representation} of the propagator reads 
\begin{equation}\label{SpectralRepProp}
    \mathcal{G}(\omega_p)
      =
      \int_{-\infty}^{+\infty} \frac{\mathrm{d}\omega'}{2\pi} \
        \frac{S(\omega')}{i\omega_p - \omega'} \ .
\end{equation}
The spectral function verifies
the Hermitian and antisymmetry properties
\begin{subequations}\label{SymmSp}
\begin{align}
        S(\omega) &= S^\dagger(\omega) \ , \label{HermitianSp} \\
        S(\omega) &= S\transpose(-\omega) \ . \label{AntisymSp}
\end{align}
\end{subequations}

From the spectral function, we define the analytic propagator $\mathcal{G}(z)$
as the Nambu tensor verifying
\begin{equation}\label{AnalyticProp}
    \mathcal{G}(z)
      \equiv
      \int_{-\infty}^{+\infty} \frac{\mathrm{d}\omega'}{2\pi} \
        \frac{S(\omega')}{z - \omega'} \ ,
\end{equation}
where the energy $z$ is now generically complex.
This analytic continuation into the complex plane is unique~\cite{Baym1961a},
so long as the tensor coordinates
are functions of energy that are
analytic off the real axis, vanish at infinity and verify
\begin{equation}
    \mathcal{G}^{\mu\nu}(z = i\omega_p) = \mathcal{G}^{\mu\nu}(\omega_p) \ .
\end{equation}
The spectral function is recovered from the analytic propagator 
as the discontinuity across the real axis,
\begin{equation}\label{PropToSpectralFunction}
    S(\omega) =
      i\left[ \mathcal{G}(z=\omega+i\eta)
        - \mathcal{G}(z=\omega-i\eta)
      \right] \ ,
\end{equation}
where $\omega$ is a real frequency 
and $\eta$ is to be understood in
the limit $\lim_{\eta\to0^+}$.
The Hermitian and antisymmetry properties
of the analytic propagator read
\begin{subequations}
\begin{align}
    \mathcal{G}(z) = \mathcal{G}^\dagger(z^*) \ , \label{HermitianAnalytic} \\
    \mathcal{G}(z) = -\mathcal{G}\transpose(-z) \ . \label{AntisymAnalytic}
\end{align}
\end{subequations}

From the analytic propagator, we define the retarded and advanced
propagators as the Nambu tensors verifying
  \begin{equation}\label{AnalyticToGFs}
    {G^{R/A}}(\omega) = \mathcal{G}(z=\omega \pm i \eta) \ .
  \end{equation}
In other words, retarded (advanced) components of the propagator are obtained
as the limits toward the real axis of the analytic propagator in the complex 
upper (lower) halfplane. 
The retarded and advanced propagators verify the Hermitian and antisymmetry 
properties
\begin{subequations}
\begin{align}
    {G^{R}}(\omega) &= {G^{A}}^\dagger(\omega)
        \ , \label{RetardedAdvancedHermitianCjg} \\
    {G^{R}}(\omega) &= -{G^{A}}\transpose(-\omega)
        \ . \label{RetardedAdvancedAntisymProp}
\end{align}
\end{subequations}

The retarded and advanced propagators can be recovered directly from
the spectral function by plugging Eq.~\eqref{AnalyticProp} into
Eq.~\eqref{AnalyticToGFs}, so that
  \begin{equation}\label{SpFToGFs}
    G^{R/A}(\omega) =
        \int_{-\infty}^{+\infty}
                    \frac{\mathrm{d}\omega'}{2\pi}
                    \frac{S(\omega')}{\omega-\omega' \pm i\eta} \ .
  \end{equation}
%
Combining Eq.~\eqref{SpFToGFs} and the Sokhotski-Plemelj identity
\begin{equation}\label{SokhotskiPlemelj}
    \frac{1}{x \pm i \eta} = \mathcal{P}\frac{1}{x} \mp i \pi \delta(x) \ ,
\end{equation}
we obtain the following dispersion relations
\begin{subequations}\label{DispersionProp}
\begin{align}
 \xoverline{\Re} \ G^{R/A}(\omega)
    &= \mathcal{P}\int_{-\infty}^{+\infty}
              \frac{\mathrm{d}\omega'}{2\pi}
              \frac{S(\omega')}{\omega-\omega'} \ , \\
 \xoverline{\Im} \ G^{R/A}(\omega)
    &= \mp \frac{1}{2} S(\omega) \ ,
\end{align}
\end{subequations}
where $\mathcal{P}$ denotes the principal value and where
\begin{subequations}
\begin{align}
    \xoverline{\Re} \ t 
        \equiv \frac{t + t^\dagger}{2} \label{HermitianPart} \ , \\
    \xoverline{\Im} \ t 
        \equiv \frac{t - t^\dagger}{2i}
        \label{AntiHermitianPart} \ ,
\end{align}
\end{subequations}
define respectively the Hermitian and anti-Hermitian parts of a tensor $t$,
which are proper Nambu tensors.
Conversely, we can find $t$ and its Hermitian conjugate, $t^\dagger$,
from its Hermitian and anti-Hermitian parts
\begin{subequations}
\begin{align}
    t = 
        \xoverline{\Re} \ t 
        + i \ \xoverline{\Im} \ t \ , \\
    t^\dagger = 
        \xoverline{\Re} \ t 
        - i \ \xoverline{\Im} \ t \ .
\end{align}
\end{subequations}
We note that, compared to the standard symmetry-conserving case~\cite{economou_2006,stefanucci_van_leeuwen_2013}, the real and imaginary parts
appearing in the dispersion relations have been replaced by Hermitian
and anti-Hermitian parts.

\subsection{Sum rules and positivity bounds}
In addition to the symmetries described by Eqs.~\eqref{SymmSp},
the spectral function of the propagator verifies additional exact properties
which take the form of sum rules and positivity bounds.
The energy weighted sum rules relate the different moments of the 
spectral function, $S(\omega)$, to the Hamiltonian, $H$.
The $0^{\text{th}}$ moment is deduced from the anticommutation rule in Eq.~\eqref{CAR_UU}
and reads
\begin{equation}
\label{sumrule0}
    \int_{-\infty}^{+\infty} \frac{\mathrm{d}\omega}{2\pi} \ S(\omega)
        = g \ .
\end{equation}
More generally, the $n^{\text{th}}$ moment of the spectral function satisfies the
relation
\begin{align}
m^{\mu\nu}_{n}&\equiv \int_{-\infty}^{+\infty} \frac{\mathrm{d}\omega}{2\pi} \
    \omega^n \ S^{\mu\nu}(\omega) \nonumber \\ 
    &=
 \big\langle 
    \big\{
       \underbrace{ \left[ \dots [\mathrm{A}^{\mu} , H ], \dots , H \right] }_{n\text{ commutators}},
        \mathrm{A}^{\nu}
    \big\}
 \big\rangle \ ,
\end{align}
where the right-hand side defines the thermal average of a series of $n$ nested 
commutators involving $\mathrm{A}^{\mu}$ and $H$, and an additional anti-commutator 
with $\mathrm{A}^{\nu}$. 

An interesting property of the spectral representation is the relation between
the moments of the spectral function and the asymptotic expansion around infinity
of the analytic propagator. 
For any $n \in \mathbb{N}$, we find 
\begin{equation}\label{AsymptoticExpProp}
  \mathcal{G}(z)
    = \sum^{n}_{k=0} \frac{m_{k}}{z^{k+1}}
      + O\left(\frac{1}{z^{n+2}}\right) \ .
\end{equation}
For $n=0$, we recover
the well known asymptotic behaviour of the diagonal components
\begin{equation}\label{DiagAsymptotProp}
  {\mathcal{G}^{\mu}}_{\mu}(z)
    = \frac{1}{z}
      + O\left(\frac{1}{z^{2}}\right) \, ,
\end{equation}
while the asymptotic behaviour of the off-diagonal components verify
\begin{equation}\label{OffDiagAsymptotProp}
  {\mathcal{G}^{\mu}}_{\nu}(z)
    = \frac{{(m_{1})^{\mu}}_{\nu}}{z^2}
      + O\left(\frac{1}{z^{3}}\right) \ ,
\end{equation}
with $\mu \neq \nu$\footnote{Eq.~\eqref{OffDiagAsymptotProp}
is an equality between tensor coordinates, but not all of them. The equality does not hold  when
$\mu=\nu$. Thus Eq.~\eqref{OffDiagAsymptotProp} should not be understood as
an equality of tensors. Any associated raising or lowering of indices requires extra care.}.
We stress that
the diagonal and off-diagonal
components of the $(1,1)$-type propagator have different asymptotic behaviours
when $\abs{z} \to \infty$. 

Another important sum rule is the so-called
Galitskii-Migdal-Koltun (GMK) sum rule~\cite{Galitskii1958,Koltun1974}, which connects 
the one-body spectral function
to the expectation value of the Hamiltonian. 
We emphasise that the validity of this sum rule, as given here,
depends on the choice of the field basis. In other words, the sum rule is not an equality
between two Nambu tensors.

The GMK sum rule stems from the following relation between a sum of commutators of Nambu tensors
and the Hamiltonian,
\begin{multline}
    - \sum_{\mu}
        \mean{[\mathrm{A}_{\mu}, H] \mathrm{A}^{\mu}}
        = \\
        \sum^{k_{\text{max}}}_{k=1} \frac{1}{(2k-1)!} 
            \sum_{\mu_1 \dots \mu_{2k}}
            v^{(k)}_{[\dot{\mu}_1 \dots \dot{\mu}_{2k-1} \mu_{2k}]}
            \mean{\mathrm{A}^{\mu_1} \dots \mathrm{A}^{\mu_{2k}}} \ .
\end{multline}
On the right-hand side, the partially antisymmetric part of
a $k$-body interaction,  
$v^{(k)}_{[\dot{\mu}_1 \dots \dot{\mu}_{2k-1} \mu_{2k}]}$, 
appears. This partially antisymmetric part is defined by
\begin{equation}\label{PartialAntisymPartTensor}
  v^{(k)}_{[\dots \dot{\mu}_1 \dots \dot{\mu}_2 \dots \dot{\mu}_p \dots]}
  \equiv
  \frac{p!}{(2k)!}
  \sum_{\sigma \in S_{2k}/S_{p}}
  \epsilon(\sigma) \ 
  v^{(k)}_{\mu_{\sigma(1)} \dots \mu_{\sigma(2k)}} \ ,
\end{equation}
where $S_{2k}/S_{p}$ is the set of permutations of the $2k$ indices
keeping the order of the $p$ dotted indices fixed.
More details about totally and partially antisymmetrisations are given in Part~I of our work. 
The problem in establishing a GMK-type sum rule is that, in general, the partially antisymmetric parts
are different from the original interaction terms, i.e.\
\begin{equation}
    v^{(k)}_{[\dot{\mu}_1 \dots \dot{\mu}_{2k-1} \mu_{2k}]} 
    \neq v^{(k)}_{\mu_1 \dots \mu_{2k-1} \mu_{2k}} \ ,
\end{equation}
which would normally appear in the expectation value of the energy.

To go further, we rely on the fact that, in our particular choice
of field basis, defined in Eq.~\eqref{WorkingFieldBasis},
Nambu fields are either pure creation
or pure annihilation operators of single-particle states.
We therefore use Eqs.~\eqref{NambuFieldsDef}, combined with
the assumption that $H$ conserves the number of particles,
to show that
\begin{equation}\label{SimplificationNbConserving}
    - \sum_{c} \mean{[\mathrm{A}_{(c,1)}, H] \mathrm{A}^{(c,1)}}
        = \sum^{k_{\text{max}}}_{k=1} k \mean{H^{[k]}} \ ,
\end{equation}
where $H^{[k]}$ represents an individual $k$-body term in the Hamiltonian, 
Eq.~\eqref{Hamiltonian}, i.e.\ 
\begin{equation}
    H^{[k]}
    \equiv
    \frac{1}{(2k)!}
        \sum_{\mu_1 \dots \mu_{2k}}
          v^{(k)}_{\mu_1 \dots \mu_{2k}} \
          \mathrm{A}^{\mu_1} \dots \mathrm{A}^{\mu_{2k}} \ .
\end{equation}
Note that in Eq.~\eqref{SimplificationNbConserving} the contraction
is only made on the single-particle index $c$ so that the resulting term
is \emph{not} a Nambu scalar tensor. We emphasise this aspect using
the explicit $(b,l_b)$ notation, rather than the global indices, $\mu$.
Assuming a two-body interaction only, $k_{\text{max}} = 2$, we have
\begin{align}\label{2B_Decompo_Expect}
    \mean{H}
        &= \mean{H^{[1]}} + \mean{H^{[2]}} \nonumber \\
        &= \frac{1}{2} \mean{H^{[1]}} 
            + \frac{1}{2} \sum^{k_{\text{max}}}_{k=1} k \mean{H^{[k]}} \ .
\end{align}
The $k=1$ term can be computed directly as an expectation value
over the propagator. 
Using the spectral representation of the propagator,
one can show that
\begin{equation}\label{Expec_1B}
    \mean{H^{[1]}} = \frac{1}{2}
        \sum_{bc} \int^{+\infty}_{-\infty} \frac{\mathrm{d}\omega}{2\pi} \ 
        {v^{(1)(b,1)}}_{(c,1)} \ f(\omega) \  {S^{(c,1)}}_{(b,1)}(\omega)
\end{equation}
where $f(\omega) \equiv \frac{1}{1+ e^{\beta \omega}}$
is the Fermi-Dirac distribution. This distribution
arises when performing the Matsubara sum in
\begin{equation}
    \mean{\mathrm{A}_{(b,1)} \mathrm{A}^{(c,1)}}
    = -\frac{1}{\beta} \sum_{\omega_p}
        {\mathcal{G}_{(b,1)}}^{(c,1)}(\omega_p) e^{-i \omega_p \eta} \ ,
\end{equation}
after the propagator has been replaced by its spectral representation.
Since $v^{(1)}$ encodes the one-body part of the Hamiltonian,
it typically includes kinetic terms and mean-field-like potentials. 
Similarly, the expectation value in Eq.~\eqref{SimplificationNbConserving}
reads
\begin{multline}
\label{Expec_Remainder}
    - \sum_{c}
        \mean{[\mathrm{A}_{(c,1)}, H] \mathrm{A}^{(c,1)}}
        = \\
        \sum_{c} \int^{+\infty}_{-\infty} \frac{\mathrm{d}\omega}{2\pi} \ 
        \omega \ f(\omega) \ {S^{(c,1)}}_{(c,1)}(\omega) \ .
\end{multline}
Combining Eqs.~\eqref{2B_Decompo_Expect},~\eqref{Expec_1B}
and~\eqref{Expec_Remainder}, we find the GMK sum rule at
non-zero temperature with symmetry-breaking, namely
\begin{multline}
\label{GMKsumRule}
    \mean{H}
    =
    \frac{1}{2} \sum_{bc} \int^{+\infty}_{-\infty} 
        \frac{\mathrm{d}\omega}{2\pi} \ 
        f(\omega) \ {S^{(c,1)}}_{(b,1)}(\omega) \, \\ \times
        \left( \frac{1}{2}{v^{(1)(b,1)}}_{(c,1)} + \omega \ \delta_{bc} \right) \ .
\end{multline}
In the zero-temperature limit, we recover the same sum rule given
by the Gorkov-Green's function formalism in Ref.~\cite{Soma2011}.
This is also formally equivalent to the finite-temperature GMK sum rule
obtained in the symmetry-conserving case~\cite{Soma2006,Rios2006a,Rios2007}.

The GMK sum rule in Eq.~\eqref{GMKsumRule}
is only valid for a Hamiltonian conserving the particle-number symmetry
and containing only two-body interactions.
To derive this equation, we also have relied
on the fact that we work in a field basis, $\mathcal{B}^f$, made
of pure single-particle creation and annihilation operators
as given in Eq.~\eqref{WorkingFieldBasis}.
The latter implies that the GMK sum rule, as given in Eq.~\eqref{GMKsumRule},
remains invariant under the action of the sub-group $\GL(\mathscr{H}_1)$
but not under the action of the whole group $\GL(\mathscr{H}^f)$.
An extension to include three-body interactions may be derived
following the steps drawn for non-superfluid systems in Ref.~\cite{Carbone2013b}. 

In addition to symmetry properties and sum rules, the exact spectral function
of the propagator fulfils a series of relevant positivity inequalities. 
From the spectral function in Eq.~\eqref{DefSpectralFunction}, we can show that,
for any $(1,0)$-tensor
$X$ and any orthogonal field basis, 
\begin{equation}\label{DefPositiveSp}
    \sum_{\mu\nu} (X^\mu)^* \ {S^{\mu}}_{\nu}(\omega) X^\nu > 0 \ .
\end{equation}
Eq.~\eqref{HermitianSp} together with the previous equality
amount to state that \emph{the spectral function is Hermitian definite positive}.
As a shorthand notation to denote Hermitian definite positiveness, we write
\begin{equation}
    S(\omega) \succ 0 \ .
\end{equation}
In practice, the positive definiteness of $S(\omega)$ is equivalent to
stating that all the principal minors of the matrix obtained from
the tensor coordinates ${S^{\mu}}_{\nu}(\omega)$ in any orthogonal field basis
${\mathcal{B}^f}'$ are strictly positive. 

In our case, we work in a field basis $\mathcal{B}^f$ which is orthogonal,
as discussed in~\ref{App:HermitianConjugation}.
For the first principal minors, Hermitian definite positiveness means that
\begin{equation}\label{1stPrincipalMinorPos}
    \forall \mu, \ {S^{\mu}}_{\mu}(\omega) > 0 \ .
\end{equation}
The inequality~\eqref{1stPrincipalMinorPos} is equivalent to the standard positivity property
obtained in the symmetry-conserving case which, together
with the sum rule in Eq.~\eqref{sumrule0}, endows the
(diagonal elements) of the spectral function
with a probabilistic interpretation~\cite{stefanucci_van_leeuwen_2013}. 
In the symmetry-conserving case, the spectral function is diagonal and
only the first principal minors are setting non-trivial constraints
on the spectral function.
In the symmetry-breaking case, in contrast, the positivity condition 
of higher principal minors yields a set of non-trivial inequalities that must be fulfilled.
As an example, from the positivity of the second principal minors,
we find that the spectral function must satisfy
\begin{equation}\label{2ndPrincipalMinorPos}
    \forall \mu \neq \nu, \;
      \abs{{S^{\mu}}_{\nu}(\omega)}
        < \sqrt{{S^{\mu}}_{\mu}(\omega){S^{\nu}}_{\nu}(\omega)} \ ,
\end{equation}
where the Hermitian property of the spectral function, Eq.~\eqref{HermitianSp},
has been used together with the orthogonality of $\mathcal{B}^f$.

We stress that inequalities~\eqref{1stPrincipalMinorPos}
and~\eqref{2ndPrincipalMinorPos}
remain valid only up to the action of the sub-group $\U(\mathscr{H}^f)$.
For more details on the orthogonality of a field basis and
the unitary group $\U(\mathscr{H}^f)$ see~\ref{App:HermitianConjugation}.
In our case, where $\mathcal{B}^f$ is made of pure single-particle creation/annihilation
operators,
inequality~\eqref{2ndPrincipalMinorPos} can be recast as
\begin{subequations}\label{Anomalous2ndSpectralBounds}
  \begin{align}
    \abs{{S^{(b,1)}}_{(c,2)}(\omega)}
      &< \sqrt{{S^{(b,1)}}_{(b,1)}(\omega){S^{(c,2)}}_{(c,2)}(\omega)}  \ , \\
    \abs{{S^{(b,1)}}_{(b,2)}(\omega)}
      &< \sqrt{{S^{(b,1)}}_{(b,1)}(\omega){S^{(b,2)}}_{(b,2)}(\omega)}  \ , \\
      \abs{{S^{(b,1)}}_{(c,1)}(\omega)}
        &< \sqrt{{S^{(b,1)}}_{(b,1)}(\omega){S^{(c,1)}}_{(c,1)}(\omega)}  \ .
  \end{align}
\end{subequations}
We note that the positivity of higher minors yields additional bounds,
which are not displayed here for conciseness.

\section{Self-consistent propagator}\label{Sec:SCprop}
In the previous section, we have introduced
a series of properties for the exact propagator
in a manifestly Nambu-covariant fashion.
In this section, we study self-consistent approximations to the propagator, 
much as one would do in the symmetry-conserving case. 
First, we introduce the self-energy via a Dyson-Schwinger equation. 
We also detail its analytical properties, which will be useful for future applications.
Second, we introduce self-consistent approximations on the self-energy 
and the propagator.
Last, we use the Hartree-Fock-Bogoliubov~(HFB) approximation
as an example of such self-consistent approximations.

\subsection{Dyson-Schwinger equation}\label{subsec:DysonSchwinger}

Let us first consider the partitioning of the Hamiltonian
\begin{subequations}\label{GeneralPartitionPT}
  \begin{align}
    H &\equiv H_0 + H_1 \ , \\
    H_0 &\equiv
      \frac{1}{2} \sum_{\mu\nu} U_{\mu\nu} \mathrm{A}^\mu \mathrm{A}^\nu \ , 
      \label{unperturbedHamiltonian}\\
    H_1 &\equiv
      \sum^{k_{\text{max}}}_{k=0} \ \frac{1}{(2k)!}
                        \sum_{\mu_1 \dots \mu_{2k}}
                              v^{(k)}_{\mu_1 \dots \mu_{2k}} \
                              \mathrm{A}^{\mu_1} \dots \mathrm{A}^{\mu_{2k}} \ , 
		\label{perturbedHamiltonian}
  \end{align}
\end{subequations}
with the assumption that $U_{\mu\nu}$ is traceless and antisymmetric\footnote{This
assumption can be made without loss of generality to the price of shifting
$H$ by a global constant. For more details, see Part~I.}.

The Dyson-Schwinger equation relates the exact propagator
to the unperturbed one. The unperturbed analytic propagator
$\mathcal{G}^{(0)}(z)$ is defined as the analytic propagator
associated to $H_0$. The analytic self-energy
$\Sigma(z)$ associated to the partitioning of Eq.~\eqref{GeneralPartitionPT}
is defined as the Nambu tensor verifying\footnote{The analytic self-energy
$\Sigma(z)$ denotes a tensor of type $p+q=2$. To simplify notations,
we use whenever possible the intrinsic notation, $\Sigma(z)$,
as we did for the propagator and its spectral function.}
\begin{equation}\label{SelfEnergyDef}
    \Sigma(z) \equiv 
        {\mathcal{G}^{(0)}}^{-1}(z)
        -
        \mathcal{G}^{-1}(z) \ .
\end{equation}
Consequently, the analytic self-energy is related to the exact and
unperturbed propagators via the \emph{Dyson-Schwinger} equations
\begin{subequations}\label{EnergyDysonEqs}
\begin{align}
  \mathcal{G}(z)
    &= \mathcal{G}^{(0)}(z)
    +
      \mathcal{G}^{(0)}(z) \
      \Sigma(z) \
      \mathcal{G}(z) \\
  \mathcal{G}(z)
    &= \mathcal{G}^{(0)}(z)
    +
      \mathcal{G}(z) \
      \Sigma(z) \
      \mathcal{G}^{(0)}(z) \ .
\end{align}
\end{subequations}
In the previous equation, the inverse of a tensor $\mathcal{G}^{-1}(z)$ and the
products of tensors $\mathcal{G}^{(0)}(z) \Sigma(z) \mathcal{G}(z)$ are to be understood as
functions of tensors as defined in~\ref{App:FunctionalCalculus}. 
As such, the products of tensors involve implicit sums over global indices
that we do not make explicit for the sake of conciseness. 

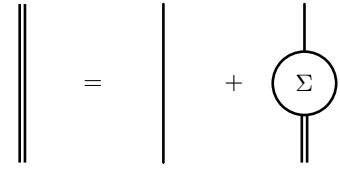
\begin{figure}[t]
  \centering
  \parbox{40pt}{\begin{fmffile}{DressedProp}
    \begin{fmfgraph*}(40,60)
        \fmfkeep{DressedProp}
     \fmfset{arrow_len}{2.8mm}
     \fmfset{arrow_ang}{20}
     \fmfbottom{i}
     \fmftop{o}
     \fmf{dbl_plain}{i,o}
    \end{fmfgraph*}
  \end{fmffile}}
  =
\parbox{40pt}{\begin{fmffile}{BareProp}
    \begin{fmfgraph*}(40,60)
        \fmfkeep{BareProp}
        \fmfbottom{i}
        \fmftop{o}
        \fmf{plain}{i,o}
    \end{fmfgraph*}
\end{fmffile}}
+
\parbox{40pt}{\begin{fmffile}{SelfContrib}
  \begin{fmfgraph*}(40,60)
    \fmfkeep{SelfContrib}
  \fmfbottom{i}
  \fmftop{o}
    \fmf{dbl_plain}{i,v1}
    \fmf{plain}{v1,o}
    \fmfv{d.sh=circle,d.f=empty, d.si=0.6w, b=(1,,1,,1), label=$\Sigma$, label.dist=0}{v1}
\end{fmfgraph*}
\end{fmffile}}
\caption{Diagrammatic representation of the first Dyson-Schwinger 
equation given in Eqs.~\eqref{EnergyDysonEqs}. The unperturbed propagator
$\mathcal{G}^{(0)}$ and the exact propagator $\mathcal{G}$ are
represented by simple and double plain lines, respectively.}
\label{Fig:DysonEqs}
\end{figure}

We show the first equation in Eqs.~\eqref{EnergyDysonEqs} in
Fig.~\ref{Fig:DysonEqs} using un-oriented Feynman diagrams.
From now on, whenever there is no ambiguity,
un-oriented Feynman diagrams obtained from the diagrammatics detailed in
Part~I will be simply referred to as Feynman diagrams.

If the Hamiltonian is a quadratic polynomial of Nambu fields,
the propagator can be explicitly computed.
For example, the unperturbed propagator, associated to $H_0$ defined in
Eq.~\eqref{unperturbedHamiltonian} reads
\begin{equation}
    \mathcal{G}^{(0)}(z)
    = \left( z - U \right)^{-1} \ .
\end{equation}
In the case of the exact propagator associated to $H$, the explicit expression
of the propagator depends on the self-energy and reads
\begin{equation}\label{DirectPropFromSelf}
    \mathcal{G}(z)
    = \left( z - (U + \Sigma(z)) \right)^{-1} \ .
\end{equation}

In analogy to the propagator, the retarded, advanced, and imaginary-frequency
components of the self-energy are obtained from the analytic self-energy
$\Sigma(z)$ by 
\begin{subequations}
\begin{align}
    \Sigma^{R/A}(\omega) &\equiv \Sigma(z = \omega \pm i\eta) \ , \\
    \Sigma(\omega_p) &\equiv  \Sigma(z = i\omega_p) \ ,
\end{align}
\end{subequations}
where $\omega_p$ are fermionic Matsubara frequencies.

Finally, although we mostly focus on the energy representation of the
self-energy, it will be sometimes convenient to use the imaginary-time representation,
$\Sigma(\tau,\tau')$. Like the propagator,
the self-energy only depends on the time difference, $\tau-\tau'$. 
The energy representation is thus related to the time components via
the Fourier transform\label{TimeSelfEnergyDef}
\begin{subequations}
\begin{align}
  \Sigma(\omega_p) \ \beta \delta_{\omega_p,-\omega_p'} &\equiv
      \int_0^\beta \mathrm{d}\tau \int_0^\beta \mathrm{d}\tau' \
      e^{i\omega_p \tau} e^{i\omega_p' \tau'}
      \Sigma(\tau,\tau') \ , \\
  \Sigma(\omega_p) &=
      \int_0^\beta \mathrm{d}\tau \
      e^{i\omega_p \tau} \Sigma(\tau) \ ,
\end{align}
\end{subequations}
where $\Sigma(\tau) \equiv \Sigma(\tau,0)$.
For completeness, we provide the Dyson-Schwinger equations in the time representation
\begin{subequations}\label{TimeDysonEqs}
\begin{align}
  \mathcal{G}(\tau,\tau')
    &= \mathcal{G}^{(0)}(\tau,\tau') \nonumber \\
    &\phantom{=} + \int_0^\beta \mathrm{d}\tau_1 \mathrm{d}\tau_2 \
      \mathcal{G}^{(0)}(\tau,\tau_1) \
      \Sigma(\tau_1,\tau_2) \
      {\mathcal{G}}(\tau_2,\tau') \ , \\
  \mathcal{G}(\tau,\tau')
    &= \mathcal{G}^{(0)}(\tau,\tau') \nonumber \\
    &\phantom{=} + \int_0^\beta \mathrm{d}\tau_1 \mathrm{d}\tau_2 \
      \mathcal{G}(\tau,\tau_1) \
      \Sigma(\tau_1,\tau_2) \
      \mathcal{G}^{(0)}(\tau_2,\tau') \ .
\end{align}
\end{subequations}

\subsection{Properties of the exact self-energy}\label{subsec:SelfProperties}
The tensor coordinates of the self-energy are
analytic in the upper and lower complex energy half-planes.
Similarly as for the propagator, a spectral representation
for the self-energy can be derived.
First, we decompose the self-energy into an instantaneous (energy-independent) part
and a continuous (energy-dependent, and vanishing at infinity) part,
\begin{align}\label{instantContinuousDecompo}
  \Sigma_{\mu\nu}(z)
    &\equiv \Sigma^{\infty}_{\mu\nu} + \Sigma^{C}_{\mu\nu}(z) \ , \\
  \lim_{\abs{z}\to\infty} \Sigma^{C}_{\mu\nu}(z)
    &\equiv 0 \ .
\end{align}
The instantaneous and continuous part of the self-energy are proper Nambu tensors.
The continuous part has a spectral representation, namely\footnote{Here we
are assuming for simplicity that
$\forall \mu,\nu, \; \Sigma^{C}_{\mu\nu}(z) = o\left( \frac{1}{z}\right)$.
A counterexample of the spectral representation of Eq.~\eqref{continuousSelfSpectralDecopmpo}
is given in chapter 14 of Ref.~\cite{Blaizot1986}. Note, however,
that a generalised integral representation always holds.
We refer the reader to Ref.~\cite{Kac1974} for more details on necessary and sufficient conditions to have a
spectral representation.}
\begin{equation}\label{continuousSelfSpectralDecopmpo}
  \Sigma^{C}(z) \equiv
    \int^{+\infty}_{-\infty} \frac{\mathrm{d}\omega}{2\pi}\
      \frac{\Gamma(\omega)}{z-\omega} \ ,
\end{equation}
where $\Gamma(\omega)$ is a Nambu tensor commonly referred to as
the \emph{width} of the self-energy.
Plugging Eq.~\eqref{continuousSelfSpectralDecopmpo} into
Eq.~\eqref{instantContinuousDecompo},
the \emph{spectral representation of the self-energy} reads
\begin{equation}\label{spectralRepSelf}
    \Sigma(z)
    =
    \Sigma^{\infty}
    +
    \int^{+\infty}_{-\infty} \frac{\mathrm{d}\omega}{2\pi}\
      \frac{\Gamma(\omega)}{z-\omega} \ .
\end{equation}
Similarly to the asymptotic expansion of the propagator, Eq.~\eqref{AsymptoticExpProp}, 
an asymptotic expansion
can be worked out for the self-energy, namely
\begin{equation}\label{AsymptoticExpSelf}
  \Sigma(z)
    = 
    \Sigma^{\infty}
    +
    \sum^{n}_{k=0} \frac{s_{k}}{z^{k+1}}
      + O\left(\frac{1}{z^{n+2}}\right) \ ,
\end{equation}
as a function of the moments of the width, $s_k$, defined by
\begin{equation}
    s_{k} \equiv
        \int_{-\infty}^{+\infty} \frac{\mathrm{d}\omega}{2\pi} \
            \omega^k \ \Gamma(\omega) \ ,
\end{equation}
for any $k \in \mathbbm{N}$.
The width can be recovered from the discontinuity of the analytic self-energy across the real axis,
\begin{equation}\label{WidthFromSelfEnergy}
  \Gamma(\omega) =
    i \left[ \Sigma(z = \omega+i\eta) - \Sigma(z = \omega-i\eta) \right] \ .
\end{equation}

From the Hermitian and antisymmetry properties of the propagators,
and from the Dyson-Schwinger equations~\eqref{EnergyDysonEqs}
relating the self-energy to the propagators,
we can obtain useful symmetry properties of the self-energy. 
For the analytic self-energy, these symmetry properties read
\begin{subequations}
\begin{align}
    \Sigma(z) &= \Sigma^\dagger(z^*) \ , \\
    \Sigma(z) &= - \Sigma\transpose(-z) \ .
\end{align}
\end{subequations}
As a consequence, we have the following relations between the retarded and advanced
components of the self-energy,
\begin{subequations}
\begin{align}
    \Sigma^{R}(\omega) &= {\Sigma^{A}}^\dagger(\omega) \ , \\
    \Sigma^{R}(\omega) &= -{\Sigma^{A}}\transpose(-\omega) \ .
\end{align}
\end{subequations}
From these, one finds that 
\begin{equation}
        \xoverline{\Im} \ \Sigma^{\infty}
            =  \xoverline{\Im} \ \Gamma(\omega)
            = 0 \ ,
\end{equation}
or, in other words, the instantaneous self-energy and the width are necessarily Hermitian.
Since the retarded and advanced self-energies are Hermitian conjugates of each other, 
it is
convenient to define their common Hermitian part $R(\omega)$
\begin{equation}
\label{commonHermitian}
  R(\omega)
    \equiv \xoverline{\Re} \ \Sigma^{R}(\omega) 
    = \xoverline{\Re} \ \Sigma^{A}(\omega)  \ ,
\end{equation}
so that
\begin{equation}
  \Sigma^{R/A}(\omega)  
    = R(\omega) \mp i \frac{1}{2} \Gamma(\omega) \ .
\end{equation}    

Finally, we note that the width also fulfils a series of interesting symmetry 
and positivity properties\footnote{The Hermitian
definite positiveness property of the width is the least trivial one to derive.
It stems from the Hermitian definite positiveness property of the spectral function
$S(\omega)$ combined with Eq.~\eqref{SPFunctionFromSelfEnergy} reformulated into 
\begin{equation}
    S(\omega) 
        = G^R(\omega) \ \Gamma(\omega) \ {G^R}^\dagger(\omega) \ .
\end{equation}}:
\begin{subequations}
\begin{align}
    \Gamma(\omega) &= \Gamma^\dagger(\omega) \ , \label{HermitianWidth}\\
    \Gamma(\omega) &= \Gamma\transpose(-\omega) \ , \label{AntisymWidth}\\
    \Gamma(\omega) &\succ 0 \label{HermitianPositiveDefWidth}\ .
\end{align}
\end{subequations}
We stress, in particular, that Eqs.~\eqref{HermitianWidth}
and~\eqref{HermitianPositiveDefWidth} means that the width
is Hermitian positive definite. 

From the previous properties, 
the relation between the width and the
retarded and advanced self-energies takes the form of the dispersion
relations
\begin{subequations}\label{DispSelfEnergy}
\begin{align}
    \xoverline{\Re} \ \Sigma^{R/A}(\omega)
        &=
        \Sigma^{\infty}
        + \mathcal{P}\int^{+\infty}_{-\infty}
           \frac{\mathrm{d}\omega'}{2\pi} \
            \frac{\Gamma(\omega')}{\omega-\omega'} \ , \\
    \xoverline{\Im} \ \Sigma^{R/A}(\omega)
        &= \mp \frac{1}{2} \Gamma(\omega) \ .
\end{align}
\end{subequations}
As usual, these dispersion relations can be used to build the full retarded and
advanced components of the self-energy from $\Sigma^{\infty}$ and $\Gamma(\omega)$.

\subsection{From the self-energy to the spectral function}\label{subsec:SpectralFromSelf}
In this subsection, we look at the relation between
the spectral function of the propagator, $S(\omega)$,
and the width $\Gamma(\omega)$, combined with the instantaneous
self-energy $\Sigma^{\infty}$. We derive two different sets of relations. 
First, we provide the general relations between the energy moments
of these quantities.
Then, we work out a direct relationship between these two tensors,
leading us to refine the traditional physical interpretation
of the spectral function.

\subsubsection{Relations between energy moments}
As discussed in Ref.~\cite{Polls1994} for the symmetry-conserving case,
the energy moments of the spectral function, $m_k$, and of the width, $s_k$, 
are related to each other. A link can be established
by matching the asymptotic expansion of the propagator, Eq.~\eqref{AsymptoticExpProp},
to the one obtained by plugging the asymptotic expansion of the
self-energy, Eq.~\eqref{AsymptoticExpSelf}, into Eq.~\eqref{DirectPropFromSelf}.
In the symmetry-breaking case, for any $n \in \mathbbm{N}^*$,
the $n^{\text{th}}$ moment of the spectral function is related
to the moments of the width according to
\begin{equation}\label{EnergySumRule}
    m_{n}
    =
    \sum^{n}_{p=1}
    \sum_{\substack{k_1 + \dots + k_p = n \\ k_1, \dots, k_p \in \mathbbm{N}^*}}
    s_{k_1 - 2} \dots s_{k_p - 2} \ ,
\end{equation}
where the inner sum runs over ordered partitions of $n$,
and the tensor $s_{-1}$ is defined, for convenience, as
\begin{equation}
    s_{-1} \equiv U + \Sigma^{\infty} \ .
\end{equation}
Let us stress that tensors $s_k$ do not commute in general.

We provide two examples of these relations. The first moment of the
spectral function equals $s_{-1}$, 
\begin{align}
    m_{1} &= s_{-1}=U + \Sigma^{\infty}  \label{1stMomentSp} \ .
\end{align}
The second moment, in contrast, involves an energy integral over $\Gamma$ and reads
\begin{align}
    m_{2} &= 
        \left(U + \Sigma^{\infty}\right)^2 
        + 
        \int_{-\infty}^{+\infty} 
            \frac{\mathrm{d}\omega}{2\pi} \ \Gamma(\omega) \ . \label{2ndMomentSp}
\end{align}
We note that, when no symmetries are broken, we recover the relations
given in Ref.~\cite{Polls1994}.

The sum rules given in Eq.~\eqref{EnergySumRule}
are of importance both for the physical insight they provide and 
for their usefulness in numerical implementations of SCGF calculations.
The $0^{\text{th}}$ moment of $S(\omega)$ is, essentially, a 
normalisation condition. In practice, this normalisation condition can be used
to perform a quasiparticle-background separation of the spectral function,
so that quasiparticle resonances are separately and carefully handled~\cite{Bozek2002}.
The $1^{\text{st}}$ moment, $m_1$ in Eq.~\eqref{1stMomentSp},
defines, through its eigenvalues, the effective single-particle energies (ESPEs)
introduced by French and Baranger~\cite{French1966,Baranger1970}.
ESPEs have been shown to be scale-dependent, thus hampering their traditional
interpretation in terms of nuclear shells~\cite{Duguet2012,Duguet2015}.
Still, they provide an insightful approximate static picture of nuclei
at a fixed resolution scale.
The $2^{\text{nd}}$ moment, $m_2$ in Eq.~\eqref{2ndMomentSp},
characterises, after subtraction of the static part, the integrated
fragmentation around the quasiparticle resonances~\cite{Rios2017b}.
The verification of the sum rules~\eqref{EnergySumRule} in SCGF numerical implementations
provide a good test of the numerical accuracy and consistency
regarding both the static and the dynamical part of
the self-energy~\cite{Rios2006b}.
For example,
in zero-temperature calculations, one finds that the dominant
features of the spectral function converge quickly with respect to
reproducing its lowest moments, $m_n$~\cite{Soma2014}.
In practical applications, this fact enables devising optimised
simplifications of dressed propagators~\cite{Raimondi2019rpa}
and exploiting Krylov subspace projection methods~\cite{Schirmer1989,Soma2014},
both being crucial to converge large scale computations of medium-mass isotopes.
More generally, the connection with the high-energy asymptotic expansion
of the propagator, Eq.~\eqref{AsymptoticExpProp}, indicates that a good convergence
of the first moments is essential to ensure the reproduction of the high-energy behaviour
of the propagator.

\subsubsection{Direct relation}\label{subsubsec:DirectSpectralFromSelf}
Using the Nambu-covariant formalism, we can also derive a direct,
formal relation between the $S(\omega)$ and $\Gamma(\omega)$ tensors.
Using Eqs.~\eqref{PropToSpectralFunction},~\eqref{DirectPropFromSelf}
and~\eqref{spectralRepSelf}, we have
\begin{multline}
\label{DirectSpFromSelfEnergy}
  S(\omega) = \\
    i\left( (\omega+i\eta)
            - U - \Sigma^{\infty}
            - \int^{+\infty}_{-\infty}
                \frac{\mathrm{d}\omega'}{2\pi}\
                \frac{\Gamma(\omega')}{\omega+i\eta-\omega'}
    \right)^{-1} 
    \\ 
    -i\left( (\omega-i\eta)
            - U - \Sigma^{\infty}
          - \int^{+\infty}_{-\infty}
              \frac{\mathrm{d}\omega'}{2\pi}\
              \frac{\Gamma(\omega')}{\omega-i\eta-\omega'}
    \right)^{-1} \, .
\end{multline}
Using the common Hermitian part of the advanced and retarded self-energy components, 
$R(\omega)$ in Eq.~\eqref{commonHermitian}, as well as the dispersion relations, 
we can conveniently rewrite the previous
expression as
\begin{align}\label{SPFunctionFromSelfEnergy}
  S(\omega) &=
    i\left(
      \omega - U - R(\omega)
      + i \frac{\Gamma(\omega)}{2}
    \right)^{-1}   \nonumber  \\ &\phantom{=}
    -i\left(
      \omega - U - R(\omega)
      - i \frac{\Gamma(\omega)}{2}
    \right)^{-1}
    \ .
\end{align}

At this stage, one typically assumes that the tensors
$U+R(\omega)$ and $\Gamma(\omega)$ are simultaneously
diagonalisable (see, for example, Chap.~14 of~\cite{Blaizot1986})
so that their commutator vanishes:
\begin{equation}\label{PropertyForSimplSpFunction}
  \left[ U + R(\omega) , \Gamma(\omega) \right] = 0 \ ,
\end{equation}
where the bracket $[\ . \ ,\ .\ ]$ denotes the standard commutator.
Eq.~\eqref{PropertyForSimplSpFunction} is equivalent to assuming
that $U+\Sigma^{R/A}(\omega)$ is normal, which, itself, is equivalent to assuming
that its eigenbasis is orthogonal.
In this special case, we recover the well-known formula
\begin{equation}\label{SimpleSpFunctionFormula}
  S(\omega) =
      \frac{\Gamma(\omega)}
      { \left(\omega - U - R(\omega) \right)^2
              + \left(\frac{\Gamma(\omega)}{2}
        \right)^2
      }  \ .
\end{equation}

The commuting hypothesis in Eq.~\eqref{PropertyForSimplSpFunction} is, however, not
necessarily fulfilled in the symmetry-broken case.
To generalise it, we introduce a new
\emph{line-shape tensor}
\begin{align}\label{FanoTensorDef}
  \Theta(\omega)
    &\equiv
    \left(
    \left(\omega - U - R(\omega)\right)^2
    +
    \left(\frac{\Gamma(\omega)}{2}\right)^2
    \right)^{-1} \nonumber \\
    &\phantom{=}\times
    \left[
      U + R(\omega)
    ,
      \frac{\Gamma(\omega)}{2}
    \right] \ ,
\end{align}
which involves the commutator in Eq.~\eqref{PropertyForSimplSpFunction}.
This tensor is in general non-zero and 
allows us to easily generalise
Eq.~\eqref{SimpleSpFunctionFormula} to
the non-commutative case: 
\begin{multline}
\label{NCSpFunctionFormula}
  S(\omega) =
    \left[\vphantom{\left(\left(\frac{\Gamma(\omega)}{2}\right)^2\right)}
      \Gamma(\omega)
      +
      2 \left( \omega - U - R(\omega) \right)
      \Theta(\omega)
    \right]
    \\ \times
    \left[
      \left(
      \left(\omega - U - R(\omega)\right)^2
      +
      \left(\frac{\Gamma(\omega)}{2}\right)^2
      \right)
      \left(\vphantom{\left(\frac{\Gamma(\omega)}{2}\right)^2}
        1 + \Theta^2(\omega)
      \right)
    \right]^{-1} \ .
\end{multline}

Eq.~\eqref{SimpleSpFunctionFormula} is often interpreted in terms of sharp
quasiparticle resonances embedded in a smooth background~\cite{Blaizot1986,Ramos1989,Bozek1999}. 
In the symmetry-conserving case, it is common to analyse the spectral function
in different approximations, which support this interpretation.
We concentrate here on two possibilities. The first one,
the so-called \emph{peak approximation}, assumes that the spectral function
has a single quasiparticle peak at a real energy $\omega=\omega_\text{qp}$, 
and discards the dispersive energy dependence of $R$ and $\Gamma$ around
$\omega_\text{qp}$. The second approximation, the \emph{quasiparticle
approximation}, 
assumes the propagator has a simple isolated pole in the complex energy plane, $z_\text{qp}$,
and derives an approximated spectral function taking into account the soft energy dependence
of $R$ and $\Gamma$ around the peak. 
These two standard analyses lead to \emph{Lorentzian} shapes, and 
motivate the interpretation of $U + R(\omega_\text{qp})$
and $\Gamma(\omega_\text{qp})$ as tensors characterising, respectively,
the position and the width of Lorentzian resonances associated to quasiparticle states.
These resonances are embedded in a smooth background associated to a residual medium, 
which accounts for the strength that is not concentrated on the quasiparticle peaks.
Physically, the position and width of these Lorentzian-like resonances are related,
respectively, to the energy and life-time of the \emph{damped propagation} 
of quasiparticle states in the residual medium~\cite{Rios2012}.
Together, the quasiparticle resonances and background make up the spectral function, $S(\omega)$.

We reproduce in~\ref{App:ThetaInterpretation} an equivalent analysis
for the generalisation of the spectral function, Eq.~\eqref{NCSpFunctionFormula}.
We work out both the peak and the quasiparticle approximations.
Instead of the Lorentzian resonance line-shape, 
characteristic of the symmetry-conserving case, 
we find that, in general, the resonant part of the spectral function
is best described by a \emph{Fano line-shape}~\cite{Fano1961}. 
In the peak approximation, there is a clear analogy between the 
Fano line-shape parameter, $q$, and the inverse of the tensor $\Theta(\omega_\text{qp})$. 
We use this analogy to provide a physical
interpretation for the line-shape tensor, $\Theta(\omega)$, which
we regard
as describing the additional effect of \emph{interferences} between
the damped propagation of quasiparticle states in the residual medium
and the excitation of a continuum of non-resonant modes displayed by 
the residual medium.
The Fano resonances have their line-shape
controlled by $\Theta^{-1}(\omega_\text{qp})$, whereas their positions and widths
are still dictated by $U+R(\omega_\text{qp})$ and $\Gamma(\omega_\text{qp})$, respectively.
In the case where the line-shape tensor is vanishingly small, we recover the standard Lorentzian picture.
A similar conclusion is drawn in the quasiparticle approximation,
providing further support for our interpretation. 

Let us stress that the line-shape tensor $\Theta(\omega)$ vanishes in any
mean-field approximation, where $\Gamma(\omega) = 0$, 
and in any symmetry-conserving approximation, where $U + R(\omega)$
and $\Gamma(\omega)$ are simultaneously diagonal and, hence, commute. 
As a consequence, one can take the line-shape tensor 
$\Theta(\omega)$ as a theoretical indicator
of the \emph{combined} importance of correlations and symmetry-breaking.
In analogy to three-point mass differences, which can be used to probe
the importance of pairing gaps, relating $\Theta(\omega)$
to a physical observable could help us find quantitative measures
to detect whether a physical system is in a phase where both symmetry-breaking
and correlation effects are important.
This effort lies beyond the remit of our initial work.

\subsection{Self-consistent schemes}
\begin{figure}[t]
  \centering
  \includegraphics[scale=0.318, trim= 170 100 170 100]{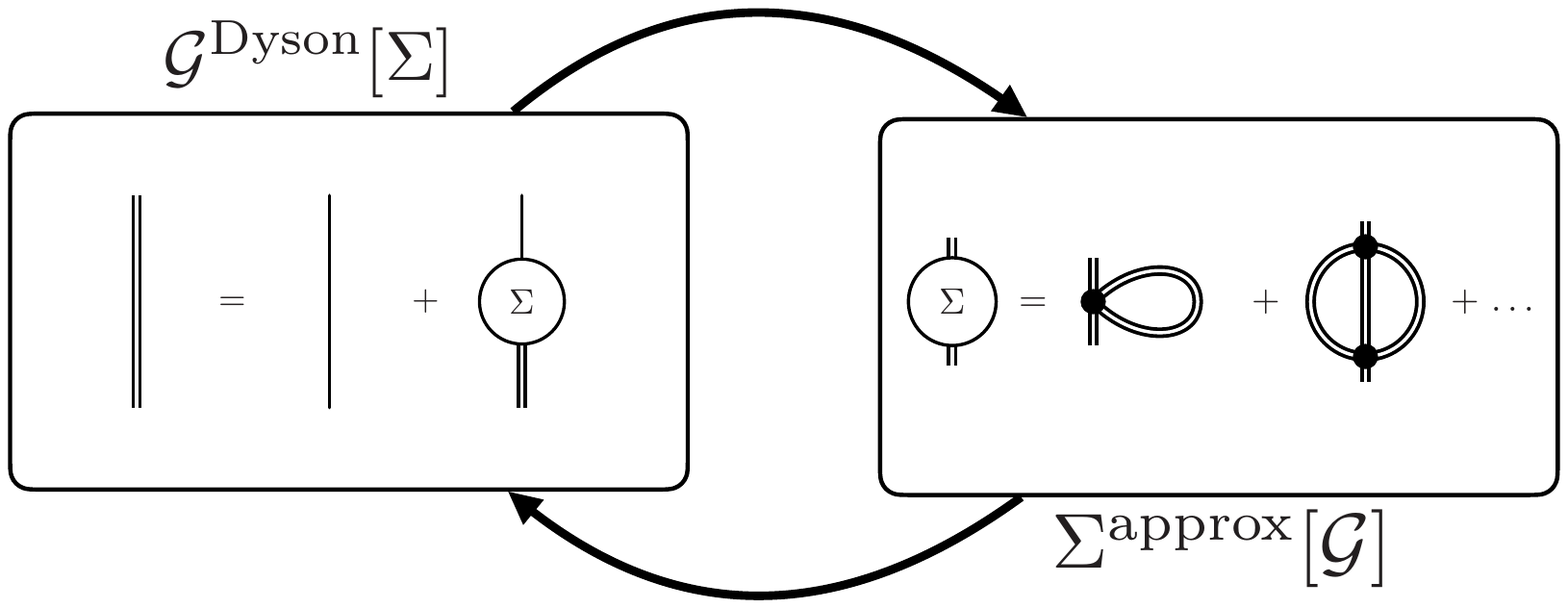}
\caption{Flowchart representing the self-consistent cycle solved by iterating
$\mathcal{G}^{\text{Dyson}}[\Sigma]$
and $\Sigma^{\text{approx}}[\mathcal{G}]$.}
\label{Fig:SCGF_cycle_general}
\end{figure}

Let us define $\mathcal{G}^{\text{Dyson}}[\Sigma]$
as the solution of the Dyson-Schwinger equation, Eq.~\eqref{EnergyDysonEqs},
for a given self-energy $\Sigma$.
A SCGF approximation relies on approximating the exact self-energy by another functional,
$\Sigma^{\text{approx}}[\mathcal{G}]$.
Combined with the Dyson-Schwinger equation, the self-consistent propagator
and self-energy are thus defined as the solutions
$(\mathcal{G}^{\text{SC}},\Sigma^{\text{SC}})$ of 
\begin{equation}\label{SelfConsistentMasterEq}
    \begin{pmatrix}
        \mathcal{G}^{\text{Dyson}}[\Sigma^{\text{SC}}] \\ ~ \\
        \Sigma^{\text{approx}}[\mathcal{G}^{SC}]
    \end{pmatrix}
    =
    \begin{pmatrix}
        \mathcal{G}^{\text{SC}} \\ ~ \\
        \Sigma^{\text{SC}}
    \end{pmatrix} \ .
\end{equation}
A self-consistent scheme aims at solving Eq.~\eqref{SelfConsistentMasterEq}
by iteration from a certain initial guess,
until convergence to a fixed point is reached.
Secs.~\ref{subsec:DysonSchwinger},~\ref{subsec:SelfProperties}
and~\ref{subsec:SpectralFromSelf} were focused on studying $\mathcal{G}^{\text{Dyson}}[\Sigma]$
and its consequences.
We now discuss a class of approximations defined by a functional 
$\Sigma^{\text{approx}}[\mathcal{G}]$ such that
the Nambu tensor character of the self-energy is preserved.
From now on, we refer to such self-consistent
approximations as NC-SCGF approximations.
A general self-consistent cycle is given by the diagram in
Fig.~\ref{Fig:SCGF_cycle_general}. 

Let us discuss first the exact case. The complete self-energy
can be expressed as an infinite sum of Feynman diagrams
derived from NCPT, as detailed in Part~I.
The exact (contravariant) propagator then reads 
\begin{equation}\label{PertExpPropagatorFromBare}
    -\mathcal{G}^{\mu\nu}(\omega_{p})
        =
        \sum_{\mathscr{G} \in \mathcal{S}}
        \mathcal{A}^{\mu\nu}[\mathcal{G}^{(0)}](\omega_p) \ ,
\end{equation}
where $\mathcal{A}^{\mu\nu}[\mathcal{G}^{(0)}](\omega_p)$
denotes the amplitude associated to a diagram, $\mathscr{G}$, where each 
fermion line
corresponds to an unperturbed propagator, $\mathcal{G}^{(0)}$;
$\mu,\nu$ are the external global indices; 
and $\omega_p$ is the external
Matsubara frequency.
For the exact case, the sum runs over the complete set of Feynman diagrams with two external lines,
which we denote by $\mathcal{S}$.
Combining Eq.~\eqref{PertExpPropagatorFromBare} with the Dyson-Schwinger equation,
the exact (covariant) self-energy is expressed also in terms of Feynman diagrams, 
\begin{equation}\label{SelfFromBareAmplitudes}
    -\Sigma_{\mu\nu}(\omega_{p})
        =
        \sum_{\mathscr{G}  \in \mathcal{S}'_{1\text{PI}}}
        \mathcal{A}_{\mu\nu}[\mathcal{G}^{(0)}](\omega_p) \ ,
\end{equation}
with $\mathcal{S}'_{1\text{PI}}$ the complete set of one-particle irreducible
(1PI) Feynman diagrams with two amputated external lines. In other words,
just as in the symmetry-conserving case, the Dyson-Schwinger equation allows us to reduce 
the number of diagrams. With the Dyson-Schwinger equation, we go
from the complete set $\mathcal{S}$ of diagrams for the one-body Green's functions,
to the 1PI subset $\mathcal{S}'_{1\text{PI}}$ for the self-energy.

Equation~\eqref{SelfFromBareAmplitudes} allows us to define approximations
of the self-energy by specifying a finite subset of Feynman
diagrams, $\mathcal{S}'_{\text{approx}} \subseteq \mathcal{S}'_{1\text{PI}}$.
Such approximations depend on the unperturbed propagator, $\mathcal{G}^{(0)}$,
through the diagrammatic expressions in Eq.~\eqref{SelfFromBareAmplitudes}.
To build a NC-SCGF approximation, the self-energy is instead
expressed directly in terms of diagrams where lines correspond
to exact propagators, $\mathcal{G}$.
This so-called \emph{dressing} of propagator lines should help to
account for correlations by incorporating medium effects into the propagator. 
Similarly to the symmetry-conserving case~\cite{Luttinger1960,Carbone2013b},
we avoid double-counting of NCPT diagrams by
restricting the sum to the subset $\mathcal{S}'_{SK}$
of skeleton (SK) diagrams with two amputated external lines.
By definition, a Feynman diagram is said to be of skeleton type if it does not contain
any self-energy insertion.
The self-energy can then be expressed as the sum
\begin{equation}\label{SelfFromDressedAmplitudes}
    -\Sigma_{\mu\nu}(\omega_{p})
        =
        \sum_{\mathscr{G}  \in \mathcal{S}'_{SK}}
        \mathcal{A}_{\mu\nu}[\mathcal{G}](\omega_p) \ .
\end{equation}
We note that the amplitudes $\mathcal{A}_{\mu\nu}$ are now functionals
of the fully dressed propagator, $\mathcal{G}$, as opposed to the unperturbed
propagator, $\mathcal{G}^{(0)}$.
Eq.~\eqref{SelfFromDressedAmplitudes} allows us to specify
NC-SCGF approximations to the self-energy defined
by a finite subset of Feynman diagrams,
$\mathcal{S}'_{\text{approx}} \subseteq \mathcal{S}'_{SK}$.
Such a self-consistent approximation amounts to summing
an infinite subset of Feynman diagrams in terms of the unperturbed propagator,
thus going beyond standard perturbation theory. For the case of symmetry-conserving theories,  Refs.~\cite{Baym1961b,Baym1962} discussed how the self-consistency requirement implies thermodynamic consistency and the satisfaction of the conservation laws associated with symmetries of the Hamiltonian.
Note that the number of skeleton diagrams can be further reduced to those containing effective interactions whenever many-body forces are present~\cite{Carbone2013b}.

\subsection{Hartree-Fock-Bogoliubov approximation}
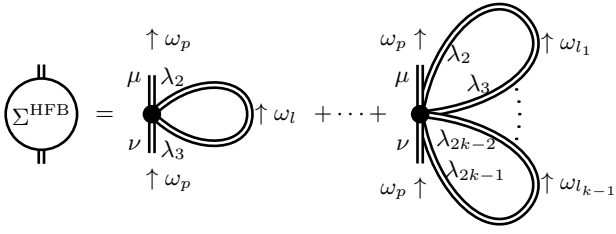
\begin{figure}[t]
  \centering
  \parbox{50pt}{\begin{fmffile}{SelfEnergyHFB}
     \begin{fmfgraph*}(30,80)
        \fmfkeep{SelfEnergyHFB}
     \fmfset{arrow_len}{2.8mm}
     \fmfset{arrow_ang}{20}
     \fmfbottom{b} \fmftop{t}
     \fmf{phantom, tag=2}{b,v1}
     \fmf{phantom, tag=1}{v1,t}
     \fmfv{d.sh=circle,d.f=empty, d.si=0.9w, b=(1,,1,,1), label=$\Sigma^{\text{HFB}}$, label.dist=0}{v1}
    \fmfposition
    \fmfipath{p[]}
    \fmfiset{p1}{vpath1(__v1,__t)}
    \fmfiset{p2}{vpath2(__b,__v1)}
    \fmfi{dbl_plain}{subpath (0,length(p1)/5) of p1}
    \fmfi{dbl_plain}{subpath (4length(p2)/5,length(p2)) of p2}
    \end{fmfgraph*}
  \end{fmffile}}
  \hspace{-0.35cm}
  =
  \hspace{-0.65cm}\parbox{60pt}{\begin{fmffile}{HFBSelfEnergy_2body}
  \begin{fmfgraph*}(50,100)
    \fmfkeep{HFBSelfEnergy_2body}
  \fmfbottom{i}
  \fmftop{o}
  \fmf{phantom, tension=1, tag=1}{v1,o}
  \fmf{dbl_plain, tension=0.9, tag=2}{v1,v1}
  \fmf{phantom, tension=1, tag=3}{i,v1}
  \fmfv{d.shape=circle,d.filled=full,d.size=3thick,l.angle=180,l.dist=8thick}{v1}
  \fmfposition
  \fmfipath{p[]}
  \fmfiset{p1}{vpath1(__v1,__o)}
  \fmfiset{p2}{vpath2(__v1,__v1)}
  \fmfiset{p3}{vpath3(__i,__v1)}
  \fmfi{dbl_plain}{subpath (0,length(p1)/4) of p1}
  \fmfi{dbl_plain}{subpath (3length(p3)/4,length(p3)) of p3}
  \fmfiv{label=$\mu$,l.dist=2thick,l.angle=170}{point length(p1)/5 of p1}
  \fmfiv{label=$\lambda_2$,l.dist=2thick,l.angle=100}{point 14length(p2)/15 of p2}
  \fmfiv{label=$\lambda_3$,l.dist=1.6thick,l.angle=-100}{point 1length(p2)/15 of p2}
  \fmfiv{label=$\nu$,l.dist=2thick,l.angle=170}{point 4length(p3)/5 of p3}

  \fmfiv{label=$\uparrow \omega_l$,l.dist=1thick,l.angle=0}{point length(p2)/2 of p2}
  \fmfiv{label=$\uparrow \omega_p$,l.dist=-0.75mm,l.angle=0}{point 8length(p1)/15 of p1}
  \fmfiv{label=$\uparrow \omega_p$,l.dist=-0.75mm,l.angle=0}{point 8length(p3)/15 of p3}
\end{fmfgraph*}
\end{fmffile}}
  \hspace{1cm}$+ \dots +$\hspace{-2.3cm}
  \parbox{150pt}{\begin{fmffile}{HFBSelfEnergy_kbody}
  \begin{fmfgraph*}(150,100)
    \fmfkeep{HFBSelfEnergy_kbody}
  \fmfbottom{i}
  \fmftop{o}
  \fmfright{r1,r2,r3}
  \fmf{phantom, tension=1, tag=1}{v1,o}
  \fmf{phantom, tension=1, tag=4}{i,v1}
  \fmfv{d.shape=circle,d.filled=full,d.size=3thick,l.angle=180,l.dist=8thick}{v1}
  \fmffreeze
  \fmfipath{p[]}
  \fmfiset{p1}{vpath1(__v1,__o)}
  \fmfiset{p2}{vloc(__v1){dir 5} .. tension 1
                .. vloc(__r3) shifted (-7mm,-5mm) .. tension 1
                .. vloc(__v1){dir -100}}
  \fmfiset{p3}{vloc(__v1){dir -80} .. tension 1
                .. vloc(__r1) shifted (-7mm,5mm)  .. tension 1
                .. vloc(__v1){dir 175}}
  \fmfiset{p4}{vpath4(__i,__v1)}
  \fmfiset{p5}{fullcircle scaled .5w shifted (.5w,.5h)}
  \fmfi{dbl_plain, tag=2}{p2}
  \fmfi{dbl_plain, tag=3}{p3}
  \fmfi{dbl_plain}{subpath (0,length(p1)/3) of p1}
  \fmfi{dbl_plain}{subpath (2length(p4)/3,length(p3)) of p4}
  \fmfi{dots, tag=5}{subpath (-length(p5)/25,length(p5)/25) of p5}
  \fmfiv{label=$\mu$,l.dist=2thick,l.angle=170}{point length(p1)/5 of p1}
  \fmfiv{label=$\lambda_2$,l.dist=1.4thick,l.angle=20}{point 13length(p2)/15 of p2}
  \fmfiv{label=$\lambda_3$,l.dist=1.4thick,l.angle=90}{point 2length(p2)/15 of p2}
  \fmfiv{label=$\lambda_{2k-2}$,
         l.dist=1thick,l.angle=-170}{point 12length(p3)/15 of p3}
  \fmfiv{label=$\lambda_{2k-1}$,
         l.dist=1thick,l.angle=10}{point 2length(p3)/15 of p3}
  \fmfiv{label=$\nu$,l.dist=2thick,l.angle=170}{point 4length(p4)/5 of p4}

  \fmfiv{label=$\uparrow \omega_{l_1}$,l.dist=1thick,l.angle=0}{point 4length(p2)/10 of p2}
  \fmfiv{label=$\uparrow \omega_{l_{k-1}}$,l.dist=1thick,l.angle=0}{point 6length(p2)/10 of p3}
  \fmfiv{label=$\omega_p \uparrow$,l.dist=-5.25mm,l.angle=0}{point 8length(p1)/15 of p1}
  \fmfiv{label=$\omega_p \uparrow$,l.dist=-5.25mm,l.angle=0}{point 7length(p4)/15 of p4}
\end{fmfgraph*}
\end{fmffile}}
\vspace{0.5cm}
\caption{Labelled diagrams contributing to the HFB self-energy with $2$- up to
$k$-body interactions. The orientation convention for the energy flow is made
explicit. Double lines denote self-consistent propagators. Amputated external lines
are shown, for clarity, as shortened double lines.}
\label{fig:HFB_SelfEnergy}
\end{figure}

Let us now discuss a simple example of NC-SCGF approximation, based on
a first-order expansion of the self-energy. We shall see that this gives rise to
the traditional Hartree-Fock-Bogoliubov (HFB) approximation.
To be specific, we approximate the self-energy by Feynman diagrams containing at most
one vertex from $2$- up to $k$-body interactions. 
The associated diagrams are shown in Fig.~\ref{fig:HFB_SelfEnergy}. 
In this approximation, the self-energy is energy-independent and reads, as a function of
the dressed propagator, $\mathcal{G}$, 
\begin{align}
    \Sigma_{\mu\nu}^{\text{approx}}[\mathcal{G}] 
    =
    \sum^{k_{\text{max}}}_{k=2}
    &\left[
        \frac{1}{2^{k-1}(k-1)!}
        \sum_{\lambda_2 \dots \lambda_{2k-1}}
        v^{(k)}_{[\mu \dot{\lambda}_2 \dot{\lambda}_3
                  \ddot{\lambda}_4 \ddot{\lambda}_5 \dots \nu]}
    \right. \nonumber \\
    &\left.
        \times \prod^{k-1}_{j=1}
            \frac{1}{\beta}
            \sum_{\omega_{l_j}} -\mathcal{G}^{\lambda_{2j}\lambda_{2j+1}}(\omega_{l_j})
                                                e^{-i\omega_{l_j} \eta_j}
    \right] \ .
\end{align}
We refer the reader to Part~I for the Feynman rules and for more details on the notation convention,
including the partial antisymmetrisation of vertices denoted by dotted indices. 
We obtain an equivalent equation in terms of the spectral function
of the propagator by explicitly summing over Matsubara frequencies. 
The approximated self-energy reads, for a general spectral function, $S$,
\begin{align}
    \Sigma_{\mu\nu}^{\text{approx}}[S] 
    =
    &\sum^{k_{\text{max}}}_{k=2}
    \left[
        \frac{1}{2^{k-1}(k-1)!}
        \sum_{\lambda_2 \dots \lambda_{2k-1}}
        v^{(k)}_{[\mu \dot{\lambda}_2 \dot{\lambda}_3
                  \ddot{\lambda}_4 \ddot{\lambda}_5 \dots \nu]}
    \right. \nonumber \\
    &\left.
        \times \prod^{k-1}_{j=1}
            \int^{+\infty}_{-\infty} \frac{\mathrm{d}\omega_j}{2\pi} \
                f(-\omega_j) \ S^{\lambda_{2j}\lambda_{2j+1}}(\omega_j)
    \right] \ . \label{DressedSelfHFBkBody_from_SpProp}
\end{align}

Since the approximation of the HFB self-energy is particularly simple,
we can also derive a convenient implicit equation, 
where the propagator is altogether removed
from the equation. This is achieved by using the Dyson-Schwinger equation,
Eq.~\eqref{DirectPropFromSelf}.
Since the self-energy is energy independent, Matsubara sums can be 
straightforwardly performed and we obtain
\begin{align}\label{ImplicitHFBSelfenergy}
    \Sigma_{\mu\nu}^{\text{HFB}}
    =
    \sum^{k_{\text{max}}}_{k=2}
    &\left[
        \frac{1}{2^{k-1}(k-1)!}
        \sum_{\lambda_2 \dots \lambda_{2k-1}}
        v^{(k)}_{[\mu \dot{\lambda}_2 \dot{\lambda}_3
                  \ddot{\lambda}_4 \ddot{\lambda}_5 \dots \nu]}
    \right. \nonumber \\
    &\left.
        \times \prod^{k-1}_{j = 1}
            f\left(
                - (U + \Sigma^{\text{HFB}})
            \right)^{\lambda_{2j}\lambda_{2j+1}}
    \right] \ ,
\end{align}
where the Fermi-Dirac distribution, $f(\omega)$, is extended into
a tensor function as defined in~\ref{App:FunctionalCalculus}.

A key advantage of the Nambu-covariant formalism arises
from the simplicity of the associated expressions.
For the $k_{\text{max}}=2$ case, for instance, the implicit equation Eq.~\eqref{ImplicitHFBSelfenergy}
involves the contraction of a two-body matrix element with a Fermi-Dirac distribution
evaluated at the quasiparticle (tensor) energy $\omega=- (U + \Sigma^{\text{HFB}})$. 
Eq.~\eqref{ImplicitHFBSelfenergy} is deceptively close to the expression
of the symmetry-conserving (self-consistent) Hartree-Fock approximation~\cite{Rios2007}.
We stress, however, that this expression incorporates all the complexity of the symmetry-broken case.

We also want to stress that Eq.~\eqref{ImplicitHFBSelfenergy} generalises straightforwardly
the standard HFB equations to the case of interactions with $k>2$.
For nuclear physics applications, three-body interactions ($k=3$) are essential.
The HFB contribution of three-body interactions arises, as expected,
as a double fermion contraction over a partially antisymmetrised
three-body interaction.
From a numerical point of view, we stress that the partial antisymmetrisation of two- or three-body 
interactions does not contribute to the self-consistent cycle and can thus be factorised 
as a one-off pre-computing step.
One can then iterate the self-consistent cycle until convergence is reached.
Having determined
$\Sigma^{\text{HFB}}$, one can immediately obtain the propagator via Eq.~\eqref{DirectPropFromSelf}
and, in turn, the ground-state energy of the system through the GMK sum rule. 

At this point, we have discussed the lowest-order self-consistent approximation to the self-energy. 
NC-SCGF approximations of the self-energy can be refined by adding more and more
Feynman diagrams, combined with self-consistently dressed propagators.
One could do this, for instance, by considering the second- and third-order
NCPT skeleton diagrams given in Part~I and dressing the corresponding lines. 
To go beyond any finite order SCGF approximation, several approaches have been proposed.
For example, the algebraic diagrammatic construction has been applied
to devise approximations summing ladder and rings diagram at zero-temperature
in the symmetry-conserving case~\cite{Barbieri2017lnp,Cipollone2015}.
Instead, we follow here a more versatile approach.
We go beyond any finite order NC-SCGF approximation 
by dressing self-consistently not only the fermion lines,
but also the interaction vertices.

\section{Self-consistent interactions}\label{Sec:SCvert}
In this section, we discuss the self-consistent procedure associated to the dressing of vertices
in a Nambu-covariant formalism.
First, we derive the full equations of motion
for the one-body Green's function. This allows us to formulate
NC-SCGF schemes as approximations of exact many-body vertices.
Second, we discuss the self-consistent dressing of a two-body interaction via a
Bethe-Salpeter equation~\cite{Salpeter1951} in Sec.~\ref{subsec:BSeq}.
Last, we work out explicitly the example of the ladder dressing of the two-body interaction
in Sec.~\ref{subsec:ladderApprox}.
We stress that we only discuss specific examples of self-consistent
interactions.
For a more general discussion on self-consistent vertices,
we refer the reader to Refs.~\cite{DeDominicis1964a,DeDominicis1964b},
where they are introduced by means of a Legendre transformations
and related to a diagrammatic expansion.

\subsection{Equations of motion}\label{subsec:EoMs}
Solving the $\mathrm{A}$-body problem for a given physical system
is equivalent to computing all $k$-body
Green's functions associated to the Hamiltonian $H$ describing the physical system.
The different $k$-body Green's functions are, however,
not independent from one another. The relations between
different $k$-body Green's functions take the form of a hierarchy
of equations of motion derived originally by Kadanoff, Martin and  
Schwinger (KMS)~\cite{Martin1959,Kadanoff1961,stefanucci_van_leeuwen_2013}.
For simplicity we focus in this section on the equation of motion coupling
the one-body to higher $k$-body Green's functions through generic $v^{(k)}$ interaction terms.

\subsubsection{Equation of motion of the propagator}
We now proceed to derive the equations of motion for the one-body Green's function, i.e.\ the propagator. 
We work with the Hamiltonian partitioning described in Eqs.~\eqref{GeneralPartitionPT}.
We start by considering the equation of motion of a simple Nambu field,
\begin{equation}\label{EoM_NambuField}
  \partial_\tau \mathrm{A}^{\mu}(\tau)
    = \left[H, \mathrm{A}^{\mu}(\tau) \right] \ ,
\end{equation}
with $\tau$ an imaginary time. 
Using Eq.~\eqref{DefContravProp}, 
the equation of motion for the propagator reads
\begin{align}
-\partial_\tau
  \mathcal{G}^{\mu\nu}(\tau, \tau')
  =
  \delta\left( \tau - \tau' \right) g^{\mu\nu} 
  + \mean{\mathrm{T}
  \left[
    \left[ H, \mathrm{A}^{\mu}(\tau) \right]
    \mathrm{A}^{\nu}(\tau')
  \right]} \ .
\end{align}
To compute the commutator in the right hand side of the previous
expression, we use its derivation property, namely
\begin{equation}\label{MasterCommutator}
  \left[ \prod^{2k}_{j=1} \mathrm{A}^{\mu_j}, \mathrm{A}^{\mu} \right]
  =
  \sum^{2k}_{i=1}
    (-1)^{i}
    g^{\mu_i\mu}
    \prod^{2k}_{\substack{j=1 \\ j \neq i}}
    \mathrm{A}^{\mu_j}
   \ ,
\end{equation}
where the products are to be written
from left to right with increasing $j$.
With this, the equation of motion is seen to couple the one-body Green's function
to  higher $k$-body Green's functions,
\begin{multline}
\label{GeneralEoM}
  \sum_{\mu_1} \left(- {g^{\mu}}_{\mu_1} \partial_\tau - U{^{\mu}}_{\mu_1} \right)
     \mathcal{G}^{\mu_1\nu}(\tau, \tau') \\
    =
    \delta\left( \tau - \tau' \right) g^{\mu\nu} 
    - \sum^{k_{\text{max}}}_{k=1} \frac{(-1)^k}{(2k-1)!}
      \sum_{\alpha \mu_1 \dots \mu_{2k-1}}
      g^{\alpha\mu} \
      v^{(k)}_{[\alpha \dot{\mu}_1 \dots \dot{\mu}_{2k-1}]} 
      \\ \times
     \mathcal{G}^{\mu_1 \dots \mu_{2k-1} \nu}(\tau, \tau') \ ,
\end{multline}
where $U$ is the traceless antisymmetric tensor associated to
the unperturbed Hamiltonian, Eq.~\eqref{unperturbedHamiltonian}.
Further, we use the notation
$\mathcal{G}^{\mu_1 \dots \mu_{2k}}(\tau, \tau')$
to denote the time limit
\begin{align}
\label{NbodyGF_2Times}
  (-1)^{k}&\mathcal{G}^{\mu_1 \dots \mu_{2k}}(\tau, \tau') \nonumber \\
  \equiv&
  \mean{
        \mathrm{T}
        \left[
          \mathrm{A}^{\mu_1}(\tau^{+\dots+})
          \dots
          \mathrm{A}^{\mu_{2k-1}}(\tau^{+})
          \mathrm{A}^{\mu_{2k}}(\tau')
        \right]
      } \\  \nonumber
  \equiv&
  \lim_{\tau_{1} > \dots > \tau_{2k-1} \to \tau^+}
  (-1)^{k}\mathcal{G}^{\mu_1 \dots \mu_{2k}}(\tau_1, \dots, \tau_{2k-1}, \tau') \ .
\end{align}
We also use the notation defined in Eq.~\eqref{PartialAntisymPartTensor} to denote
a partially antisymmetric part of $v^{(k)}$.

To express Eq.~\eqref{GeneralEoM} in terms of $k$-body Green's functions
and of the self-energy, we contract it with the unperturbed propagator 
$\mathcal{G}^{(0)}$ and integrate over time to find
\begin{multline}
 \mathcal{G}^{\mu\nu}(\tau, \tau')
 = \mathcal{G}^{(0) \mu\nu}(\tau, \tau') \  \\
   - \sum^{k_{\text{max}}}_{k=1} \frac{(-1)^k}{(2k-1)!}
     \sum_{\mu_1 \dots \mu_{2k}}
     \int^{\beta}_{0} \mathrm{d}s \
     \mathcal{G}^{(0) \mu\mu_1}(\tau, s) \
     v^{(k)}_{[\mu_1\dot{\mu}_2 \dots \dot{\mu}_{2k}]}  \  
     \\ \times 
     \mathcal{G}^{\mu_2 \dots \mu_{2k} \nu}(s, \tau') \ .
\end{multline}
Note that the right-hand-side couples the propagator 
to all $k$-body Green's functions via the $k$-body interaction.
In the case where $k_{\text{max}}=2$, the equation
couples the one- and the two-body Green's functions. 
This is equivalent to the first equation of the KMS hierarchy.

The general equation can be further simplified by 
using the Dyson-Schwinger equation, Eq.~\eqref{TimeDysonEqs}, to find
\begin{multline}
\label{Sigma_kBGF}
 \sum_{\mu_1}\int^{\beta}_{0} \mathrm{d}s \
  \Sigma_{\mu\mu_1}(\tau,s)
  \mathcal{G}^{\mu_1\nu}(s, \tau') \\
  =
  -\sum^{k_{\text{max}}}_{k=1} \frac{(-1)^k}{(2k-1)!}
        \sum_{\mu_2 \dots \mu_{2k}}
        v^{(k)}_{[\mu \dot{\mu}_2 \dots \dot{\mu}_{2k}]} \
        \mathcal{G}^{\mu_2 \dots \mu_{2k} \nu}(\tau, \tau') \ .
\end{multline}
With Eq.~\eqref{Sigma_kBGF}, we have a relation between 
the one-body Green’s function, the self-energy and higher-order Green's functions.
Instead of defining a NC-SCGF approximation at the level of the self-energy,
we can design NC-SCGF approximations from diagrammatic truncations of (higher) 
$k$-body Green's functions. These truncations are then
brought back to approximations on the self-energy using Eq.~\eqref{Sigma_kBGF}.
Shifting our focus to higher $k$-body Green's functions
will allow us to easily design richer many-body approximations
that incorporate self-consistent interactions.
To do this, however, we first need to introduce
the exact many-body interaction vertices on which these approximations are based.
Ultimately, we will be able to rewrite Eq.~\eqref{Sigma_kBGF}
in terms of such self-consistent interaction vertices, 
rather than the associated $k$-body Green's functions. 

\subsubsection{Exact many-body vertices}

\begin{figure}[t]
  \centering
$\mathcal{G}^{\mu_1\mu_2\mu_3\mu_4}(\tau_1,\tau_2,\tau_3,\tau_4) = \ \ $
\parbox{25pt}{\begin{fmffile}{Direct2B}
  \begin{fmfgraph*}(18,50)
    \fmfkeep{Direct2B}
  \fmfbottom{i1,i2}
  \fmftop{o1,o2}
  \fmf{dbl_plain}{i1,o1}
  \fmf{dbl_plain}{i2,o2}
\end{fmfgraph*}
\end{fmffile}}
+
\parbox{25pt}{\begin{fmffile}{Exchange2B}
  \begin{fmfgraph*}(20,50)
    \fmfkeep{Exchange2B}
  \fmfbottom{i1,i2}
  \fmftop{o1,o2}
  \fmf{dbl_plain}{i1,o2}
  \fmf{dbl_plain}{i2,o1}
\end{fmfgraph*}
\end{fmffile}}
+
\parbox{20pt}{\begin{fmffile}{Anomalous2B}
  \begin{fmfgraph*}(22,50)
    \fmfkeep{Anomalous2B}
  \fmfbottom{i1,i2}
  \fmftop{o1,o2}
  \fmf{dbl_plain, left=0.75}{i1,i2}
  \fmf{dbl_plain, right=0.75}{o1,o2}
\end{fmfgraph*}
\end{fmffile}}
+
\hspace{1mm}
\parbox{30pt}{\begin{fmffile}{Exct2BVertex}
  \begin{fmfgraph*}(23,50)
    \fmfkeep{Exct2BVertex}
  \fmfbottom{i1,i2}
  \fmftop{o1,o2}
  \fmf{dbl_plain}{i1,v1}
  \fmf{dbl_plain}{i2,v1}
  \fmf{dbl_plain}{v1,o1}
  \fmf{dbl_plain}{v1,o2}
  \fmfv{d.sh=circle,d.f=empty, d.si=1w, b=(1,,1,,1), label=$\Gamma^{(2)}$, label.dist=0}{v1}
\end{fmfgraph*}
\end{fmffile}}

\caption{Diagrammatic representation of Eq.~\eqref{Def2BExactVertex}.}
\label{Fig:Def2BExactVertex}
\end{figure}

We define exact many-body vertices as the 
amputated connected part of the corresponding many-body Green's function.
For example, the exact two-body vertex
$\Gamma^{(2)}_{\mu_1\mu_2\mu_3\mu_4}(\tau_1,\tau_2,\tau_3,\tau_4)$
is defined implicitly by the following equation,
\begin{multline}\label{Def2BExactVertex}
  \mathcal{G}^{\mu_1\mu_2\mu_3\mu_4}(\tau_1,\tau_2,\tau_3,\tau_4)
  \equiv
  \mathcal{G}^{\mu_1\mu_4}(\tau_1,\tau_4) \mathcal{G}^{\mu_2\mu_3}(\tau_2,\tau_3) \ \\
  - \ \mathcal{G}^{\mu_1\mu_3}(\tau_1,\tau_3) \mathcal{G}^{\mu_2\mu_4}(\tau_2,\tau_4)
  + \ \mathcal{G}^{\mu_1\mu_2}(\tau_1,\tau_2) \mathcal{G}^{\mu_3\mu_4}(\tau_3,\tau_4)
  \\
    - \sum_{\lambda_1\lambda_2\lambda_3\lambda_4}
      \int_{0}^{\beta} \mathrm{d}\tau'_1 \mathrm{d}\tau'_2 \mathrm{d}\tau'_3 \mathrm{d}\tau'_4 \
      \mathcal{G}^{\mu_1\lambda_1}(\tau_1,\tau'_1) \mathcal{G}^{\mu_2\lambda_2}(\tau_2,\tau'_2) \\
      \times \Gamma^{(2)}_{\lambda_1\lambda_2\lambda_3\lambda_4}(\tau'_1,\tau'_2,\tau'_3,\tau'_4) \times \
      \mathcal{G}^{\lambda_4\mu_4}(\tau'_4,\tau_4) \mathcal{G}^{\lambda_3\mu_3}(\tau'_3,\tau_3) \ ,
\end{multline}
involving exact one- and two-body propagators, 
$\mathcal{G}^{\mu_1\mu_2}$ and $\mathcal{G}^{\mu_1\mu_2\mu_3\mu_4}$, respectively.
For a diagrammatic representation of Eq.~\eqref{Def2BExactVertex},
see Fig.~\ref{Fig:Def2BExactVertex}. The exact two-body vertex is clearly the remaining 
diagrammatic component after all disconnected contributions to $\mathcal{G}^{\mu_1\mu_2\mu_3\mu_4}$
have been eliminated.
\begin{figure}[t]
  \centering
  \vspace{-0.5cm}
  \parbox{50pt}{\begin{fmffile}{1PISelf}
     \begin{fmfgraph*}(30,80)
        \fmfkeep{1PISelf}
     \fmfset{arrow_len}{2.8mm}
     \fmfset{arrow_ang}{20}
     \fmfbottom{b} \fmftop{t}
     \fmf{phantom, tag=2}{b,v1}
     \fmf{phantom, tag=1}{v1,t}
     \fmfv{d.sh=circle,d.f=empty, d.si=0.8w, b=(1,,1,,1), label=$\Sigma$, label.dist=0}{v1}
    \fmfposition
    \fmfipath{p[]}
    \fmfiset{p1}{vpath1(__v1,__t)}
    \fmfiset{p2}{vpath2(__b,__v1)}
    \fmfi{dbl_plain}{subpath (0,length(p1)/5) of p1}
    \fmfi{dbl_plain}{subpath (4length(p2)/5,length(p2)) of p2}
    \end{fmfgraph*}
  \end{fmffile}}
  \hspace{-0.25cm}
  =
  \hspace{-0.5cm}
\parbox{60pt}{\begin{fmffile}{1SelfEnergy}
  \begin{fmfgraph*}(40,60)
    \fmfkeep{1SelfEnergy}
    \fmfbottom{i}
    \fmftop{o}
    \fmf{phantom, tension=1, tag=1}{v1,o}
    \fmf{dbl_plain, tension=0.7, tag=2}{v1,v1}
    \fmf{phantom, tension=1, tag=3}{i,v1}
    \fmfv{d.shape=circle,d.filled=full,d.size=3thick}{v1}
    \fmfposition
    \fmfipath{p[]}
    \fmfiset{p1}{vpath1(__v1,__o)}
    \fmfiset{p3}{vpath3(__i,__v1)}
    \fmfi{dbl_plain}{subpath (0,length(p1)/3) of p1}
    \fmfi{dbl_plain}{subpath (2length(p3)/3,length(p3)) of p3}
\end{fmfgraph*}
\end{fmffile}}
\hspace{0.2cm}
+
\parbox{60pt}{\begin{fmffile}{Exct2BVertexSelfEnergy}
  \begin{fmfgraph*}(40,60)
    \fmfkeep{Exct2BVertexSelfEnergy}
  \fmfbottom{i}
  \fmftop{o}
  \fmf{phantom, tension=2, tag=1}{i,v1}
  \fmf{phantom, tension=2, tag=2}{v2,o}
  \fmf{dbl_plain}{v1,v2}
  \fmffreeze
  \fmf{dbl_plain, right}{v1,v2}
  \fmffreeze
  \fmf{dbl_plain, right}{v2,v1}
  \fmfv{d.sh=circle,d.f=empty, d.si=0.6w, b=(1,,1,,1), label=$\Gamma^{(2)}$, label.dist=0}{v1}
  \fmfv{d.shape=circle,d.filled=full,d.size=3thick}{v2}
  \fmfposition
  \fmfipath{p[]}
  \fmfiset{p1}{vpath1(__i,__v1)}
  \fmfiset{p2}{vpath2(__v2,__o)}
  \fmfi{dbl_plain}{subpath (0,length(p2)/3) of p2}
  \fmfi{dbl_plain}{subpath (0,length(p1)) of p1}
\end{fmfgraph*}
\end{fmffile}}
\vspace{-0.5cm}
\caption{Diagrammatic representation of Eq.~\eqref{Sigma_2BExactVertex}.}
\label{Fig:Sigma_2BExactVertex}
\end{figure}

For simplicity, we consider from now on the case
where the perturbative part of the Hamiltonian
contains only a two-body interaction.
In other words, we assume that $k_{\text{max}} = 2$
and $v^{(1)}_{\mu_1\mu_2} = v^{(0)} = 0$.
With this, the partitioning of Eqs.~\eqref{GeneralPartitionPT}
remains unchanged but the perturbative part
of Eq.~\eqref{perturbedHamiltonian} becomes
  \begin{align}
    H_1 &\equiv
      \frac{1}{4!}
      \sum_{\mu_1 \mu_2 \mu_3 \mu_4}
            v^{(2)}_{\mu_1 \mu_2 \mu_3 \mu_4} \
            \mathrm{A}^{\mu_1} \mathrm{A}^{\mu_2}
            \mathrm{A}^{\mu_3} \mathrm{A}^{\mu_4} \ .
  \end{align}
In this case, the one-body Green's function couples only to the two-body
Green's function through the equation of motion, Eq.~\eqref{GeneralEoM}.
Therefore, the relation between the self-energy and the two-body Green's function
reads simply
\begin{multline}\label{Sigma_2BGF}
 \sum_{\mu_1}\int^{\beta}_{0} \mathrm{d}s \
  \Sigma_{\mu\mu_1}(\tau,s)
  \mathcal{G}^{\mu_1\nu}(s, \tau') \\
  =
  -\frac{1}{3!} \sum_{\mu_2\mu_3\mu_4}
  v^{(2)}_{[\mu \dot{\mu}_2 \dot{\mu}_3 \dot{\mu}_4]} \
  \mathcal{G}^{\mu_2\mu_3\mu_4\nu}
  (\tau^{+++}, \tau^{++}, \tau^+, \tau') \ ,
\end{multline}
where all the time variables in the two-body Green's function are written explicitly
for clarity.
Using the implicit definition of $\Gamma^{(2)}$, Eq.~\eqref{Def2BExactVertex}, 
the equation of motion
is expressed as a relation between the self-energy, $\Sigma$, and the
exact two-body vertex, $\Gamma^{(2)}$, 
\begin{multline}\label{Sigma_2BExactVertex}
  -\Sigma_{\mu\nu}(\tau, \tau')
  = \frac{1}{2} \sum_{\lambda_1\lambda_2}
      v^{(2)}_{[\mu \dot{\lambda}_1 \dot{\lambda}_2 \nu]}
      \delta(\tau - \tau')
      \mathcal{G}^{\lambda_1\lambda_2}(\tau^+, \tau)
    \\
    - \frac{1}{3!}
    \sum_{\substack{\lambda_1 \lambda_2 \lambda_3 \\\lambda'_1 \lambda'_2 \lambda'_3}}
    v^{(2)}_{[\mu\lambda_1\lambda_2\lambda_3]}
    \int^{\beta}_{0} \mathrm{d}\tau'_1 \mathrm{d}\tau'_2 \mathrm{d}\tau'_3 \ 
        \Gamma^{(2)}_{\lambda'_3\lambda'_2\lambda'_1\nu}
            (\tau'_3,\tau'_2,\tau'_1,\tau') \\
        \times
        \mathcal{G}^{\lambda_1\lambda'_1}(\tau, \tau'_1)
        \mathcal{G}^{\lambda_2\lambda'_2}(\tau, \tau'_2)
        \mathcal{G}^{\lambda_3\lambda'_3}(\tau, \tau'_3) \ .
\end{multline}
We note that, remarkably, the totally and partially antisymmetric vertices
defined in the NCPT of Part~I
appear here but, this time, on the pure basis of non-perturbative arguments.
The same remark applies to the symmetry factors.
Hence, $\Gamma^{(2)}$ can be understood as the kernel for any
self-energy contributions that go beyond the contribution 
associated to the HFB diagram. 
Eq.~\eqref{Sigma_2BExactVertex} is represented diagrammatically  in Fig.~\ref{Fig:Sigma_2BExactVertex}.

\begin{figure}[t]
  \centering
  \parbox{40pt}{\begin{fmffile}{Gamma2Vertex_bis}
     \begin{fmfgraph*}(40,80)
     \fmfset{arrow_len}{2.8mm}
     \fmfset{arrow_ang}{20}
     \fmfbottom{i1,i2} \fmftop{o1,o2}
     \fmf{phantom, tag=1}{i1,v1}
     \fmf{phantom, tag=2}{i2,v1}
     \fmf{phantom, tag=3}{v1,o1}
     \fmf{phantom, tag=4}{v1,o2}
     \fmfv{d.sh=circle,d.f=empty, d.si=0.65w, b=(1,,1,,1), label=$\Gamma^{(2)}$, label.dist=0}{v1}
     \fmfposition
     \fmfipath{p[]}
     \fmfiset{p1}{vpath1(__i1,__v1)}
     \fmfiset{p2}{vpath2(__i2,__v1)}
     \fmfiset{p3}{vpath3(__v1,__o1)}
     \fmfiset{p4}{vpath4(__v1,__o2)}
     \fmfi{dbl_plain}{subpath (2length(p1)/3,length(p1)) of p1}
     \fmfi{dbl_plain}{subpath (2length(p2)/3,length(p2)) of p2}
     \fmfi{dbl_plain}{subpath (0,length(p3)/3) of p3}
     \fmfi{dbl_plain}{subpath (0,length(p4)/3) of p4}
    \end{fmfgraph*}
  \end{fmffile}}
=
\parbox{25pt}{\begin{fmffile}{FirstOrder2B}
  \begin{fmfgraph*}(25,60)
    \fmfkeep{FirstOrder2B}
  \fmfbottom{i1,i2}
  \fmftop{o1,o2}
  \fmf{phantom, tag=1}{i1,v1}
  \fmf{phantom, tag=2}{i2,v1}
  \fmf{phantom, tag=3}{v1,o1}
  \fmf{phantom, tag=4}{v1,o2}
  \fmfv{d.shape=circle,d.filled=full,d.size=3thick}{v1}
  \fmfposition
  \fmfipath{p[]}
  \fmfiset{p1}{vpath1(__i1,__v1)}
  \fmfiset{p2}{vpath2(__i2,__v1)}
  \fmfiset{p3}{vpath3(__v1,__o1)}
  \fmfiset{p4}{vpath4(__v1,__o2)}
  \fmfi{dbl_plain}{subpath (2length(p1)/3,length(p1)) of p1}
  \fmfi{dbl_plain}{subpath (2length(p2)/3,length(p2)) of p2}
  \fmfi{dbl_plain}{subpath (0,length(p3)/3) of p3}
  \fmfi{dbl_plain}{subpath (0,length(p4)/3) of p4}
\end{fmfgraph*}
\end{fmffile}}
+
\parbox{35pt}{\begin{fmffile}{SecondOrder2B}
  \begin{fmfgraph*}(35,60)
    \fmfkeep{SecondOrder2B}
  \fmfbottom{i1,i2}
  \fmftop{o1,o2}
  \fmf{phantom, tension=2.5, tag=1}{i1,v1}
  \fmf{phantom, tension=2.5, tag=2}{i2,v1}
  \fmf{dbl_plain, right=0.65}{v1,v2}
  \fmf{dbl_plain, left=0.65}{v1,v2}
  \fmf{phantom, tension=2.5, tag=3}{v2,o1}
  \fmf{phantom, tension=2.5, tag=4}{v2,o2}
  \fmfv{d.shape=circle,d.filled=full,d.size=3thick}{v1}
  \fmfv{d.shape=circle,d.filled=full,d.size=3thick}{v2}
    \fmfposition
  \fmfipath{p[]}
  \fmfiset{p1}{vpath1(__i1,__v1)}
  \fmfiset{p2}{vpath2(__i2,__v1)}
  \fmfiset{p3}{vpath3(__v2,__o1)}
  \fmfiset{p4}{vpath4(__v2,__o2)}
  \fmfi{dbl_plain}{subpath (2length(p1)/3,length(p1)) of p1}
  \fmfi{dbl_plain}{subpath (2length(p2)/3,length(p2)) of p2}
  \fmfi{dbl_plain}{subpath (0,length(p3)/3) of p3}
  \fmfi{dbl_plain}{subpath (0,length(p4)/3) of p4}
\end{fmfgraph*}
\end{fmffile}}
+
\parbox{35pt}{\begin{fmffile}{ThirdOrder2B_1}
  \begin{fmfgraph*}(35,60)
    \fmfkeep{ThirdOrder2B_1}
  \fmfbottom{i1,i2}
  \fmftop{o1,o2}
  \fmf{phantom, tension=2.5, tag=1}{i1,v1}
  \fmf{phantom, tension=2.5, tag=2}{i2,v1}
  \fmf{dbl_plain, right=0.75}{v1,v2}
  \fmf{dbl_plain, left=0.75}{v1,v2}
  \fmf{dbl_plain, right=0.75}{v2,v3}
  \fmf{dbl_plain, left=0.75}{v2,v3}
  \fmf{phantom, tension=2.5, tag=3}{v3,o1}
  \fmf{phantom, tension=2.5, tag=4}{v3,o2}
  \fmfv{d.shape=circle,d.filled=full,d.size=3thick}{v1}
  \fmfv{d.shape=circle,d.filled=full,d.size=3thick}{v2}
  \fmfv{d.shape=circle,d.filled=full,d.size=3thick}{v3}
    \fmfposition
  \fmfipath{p[]}
  \fmfiset{p1}{vpath1(__i1,__v1)}
  \fmfiset{p2}{vpath2(__i2,__v1)}
  \fmfiset{p3}{vpath3(__v3,__o1)}
  \fmfiset{p4}{vpath4(__v3,__o2)}
  \fmfi{dbl_plain}{subpath (2length(p1)/3,length(p1)) of p1}
  \fmfi{dbl_plain}{subpath (2length(p2)/3,length(p2)) of p2}
  \fmfi{dbl_plain}{subpath (0,length(p3)/3) of p3}
  \fmfi{dbl_plain}{subpath (0,length(p4)/3) of p4}
\end{fmfgraph*}
\end{fmffile}}
+
\parbox{35pt}{\begin{fmffile}{ThirdOrder2B_2}
  \begin{fmfgraph*}(35,60)
    \fmfkeep{ThirdOrder2B_2}
  \fmfbottom{i1,i2}
  \fmftop{o1,o2}
  \fmf{phantom, tension=2.5, tag=1}{i1,v1}
  \fmf{phantom, tension=2.5, tag=2}{i2,v1}
  \fmf{dbl_plain, right=0.25}{v1,v3}
  \fmf{dbl_plain, left=0.25}{v1,v2}
  \fmf{phantom, tension=2.5, tag=3}{v2,o1}
  \fmf{phantom, tension=2.5, tag=4}{v3,o2}
  \fmffreeze
  \fmf{dbl_plain, right=0.65}{v2,v3}
  \fmf{dbl_plain, left=0.65}{v2,v3}
  \fmfv{d.shape=circle,d.filled=full,d.size=3thick}{v1}
  \fmfv{d.shape=circle,d.filled=full,d.size=3thick}{v2}
  \fmfv{d.shape=circle,d.filled=full,d.size=3thick}{v3}
    \fmfposition
  \fmfipath{p[]}
  \fmfiset{p1}{vpath1(__i1,__v1)}
  \fmfiset{p2}{vpath2(__i2,__v1)}
  \fmfiset{p3}{vpath3(__v2,__o1)}
  \fmfiset{p4}{vpath4(__v3,__o2)}
  \fmfi{dbl_plain}{subpath (2length(p1)/3,length(p1)) of p1}
  \fmfi{dbl_plain}{subpath (2length(p2)/3,length(p2)) of p2}
  \fmfi{dbl_plain}{subpath (0,length(p3)/3) of p3}
  \fmfi{dbl_plain}{subpath (0,length(p4)/3) of p4}
\end{fmfgraph*}
\end{fmffile}}
+
\dots

\caption{
Diagrammatic content of the exact two-body vertex up to third order. 
We omit amputated diagrams which are equivalent up
to a permutation of the external legs for the sake of conciseness.}
\label{Fig:Diag2BExactVertex}
\end{figure}
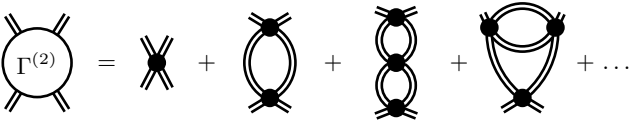

Now that we have a relation between the exact two-body vertex,
$\Gamma^{(2)}$, and the self-energy, $\Sigma$,
we turn to discuss the dependence of $\Gamma^{(2)}$
on the dressed propagator, $\mathcal{G}$.
We express the exact two-body vertex in terms of Feynman
diagrams with dressed propagators,
\begin{multline}\label{ExpansionExact2BVertex}
    \Gamma^{(2)}_{\mu_1\mu_2\mu_3\mu_4}(\tau_1,\tau_2,\tau_3,\tau_4)
        \equiv \\
    \sum_{\mathscr{G}  \in \mathcal{T}'_{SK}}
        \mathcal{A}_{\mu_1\mu_2\mu_3\mu_4}[\mathcal{G}]
        (\tau_1,\tau_2,\tau_3,\tau_4) \ .
\end{multline}
Here, $\mathcal{T}'_\text{SK}$ represents the set of skeleton
Feynman diagrams with four amputated external lines.
$\mathcal{A}_{\mu_1\mu_2\mu_3\mu_4}[\mathcal{G}](\tau_1,\tau_2,\tau_3,\tau_4)$
are the associated amplitudes for a given dressed propagator, $\mathcal{G}$.
These amplitudes can be obtained using the Feynman rules described in Part~I. 
We show Feynman diagrams up to third order in the interaction vertex in 
Fig.~\ref{Fig:Diag2BExactVertex}. 
We have omitted diagrams that are equivalent up to a permutation of
the amputated lines. Those must be included explicitly in
Eq.~\eqref{ExpansionExact2BVertex}\footnote{Alternatively, we could
consider only one diagram per class of equivalence and perform an \emph{ad hoc}
antisymmetrisation procedure.}.

\begin{figure}[t]
  \centering
  \includegraphics[scale=0.3227, trim= 170 50 170 50]{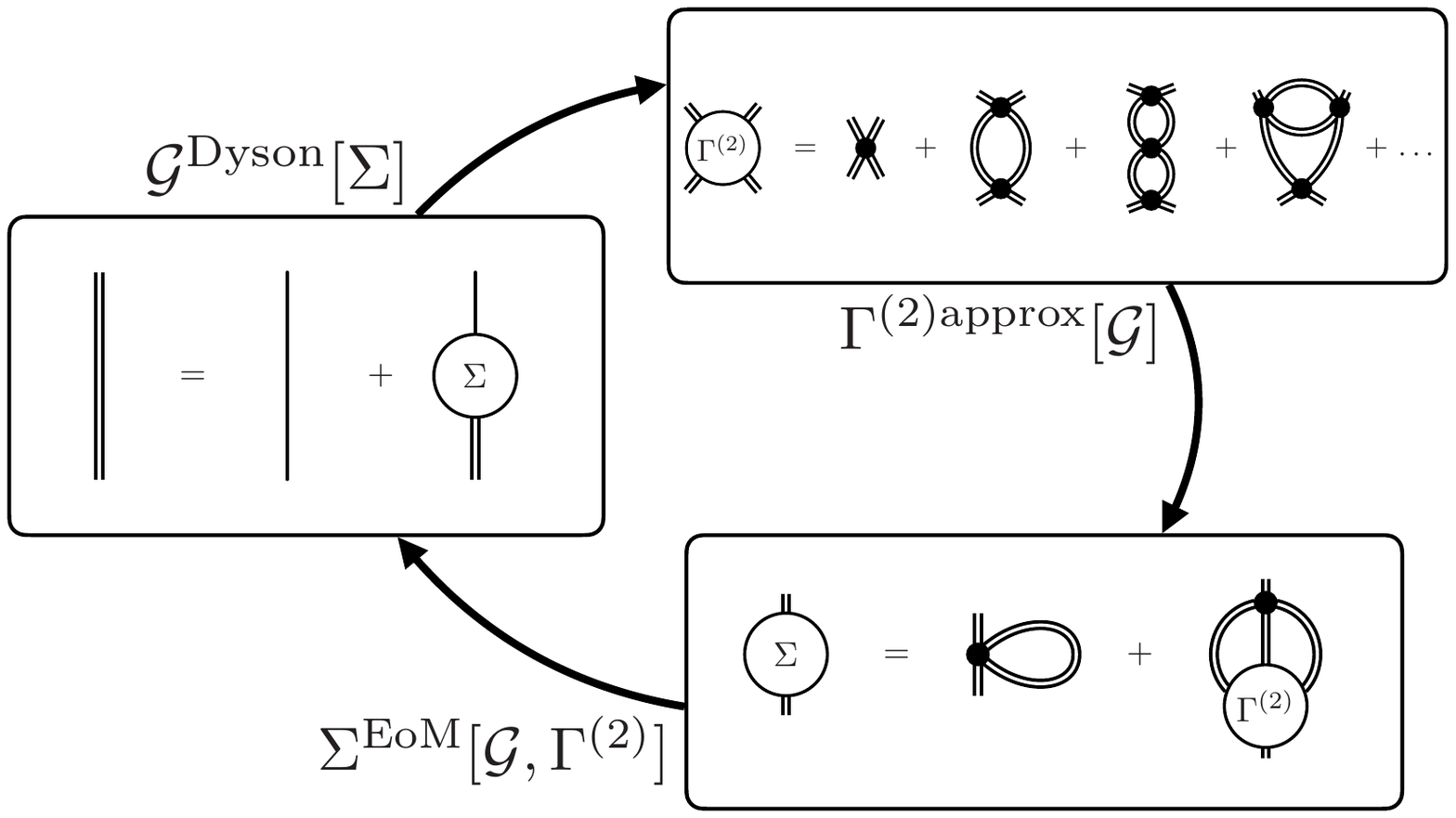}
\caption{Flowchart representing the self-consistent cycle solved by iterating
$\mathcal{G}^{\text{Dyson}}[\Sigma]$,
$\Sigma^{\text{EoM}}[\mathcal{G},\Gamma^{(2)}]$ and
$\Gamma^{(2)\text{approx}}[\mathcal{G}]$.}
\label{Fig:SCGF_cycle_2Bvertex}
\end{figure}

In the two-body interaction case, the self-energy depends only on $\mathcal{G}$ and 
$\Gamma^{(2)}$ via Eq.~\eqref{Sigma_2BExactVertex}.
We denote this self-energy as
$\Sigma^{\text{EoM}}[\mathcal{G},\Gamma^{(2)}]$.
The self-consistent scheme summarised in Eq.~\eqref{SelfConsistentMasterEq} 
can be reformulated in terms of approximated many-body vertices. 
In this formulation, we look for the solutions
$(\mathcal{G}^{\text{SC}},\Sigma^{\text{SC}}, \Gamma^{(2)\text{SC}})$
of
\begin{equation}\label{SelfConsistentPropVertexMasterEq}
    \begin{pmatrix}
        \mathcal{G}^{\text{Dyson}}[\Sigma^{\text{SC}}] \\
        \Sigma^{\text{EoM}}[\mathcal{G}^{SC},\Gamma^{(2)\text{SC}}] \\
        \Gamma^{(2)\text{approx}}[\mathcal{G}^{\text{SC}}]
    \end{pmatrix}
    =
    \begin{pmatrix}
        \mathcal{G}^{\text{SC}} \\
        \Sigma^{\text{SC}} \\
        \Gamma^{(2)\text{SC}}
    \end{pmatrix} \, ,
\end{equation}
where $\Gamma^{(2)\text{approx}}[\mathcal{G}]$ denotes an approximated 
functional of the exact two-body vertex.
This formulation of the self-consistent cycle, given by
Eq.~\eqref{SelfConsistentPropVertexMasterEq}, is shown in terms of diagrams in
the flowchart of Fig.~\ref{Fig:SCGF_cycle_2Bvertex}.
For example, $\Gamma^{(2)\text{approx}}[\mathcal{G}]$ can be any truncation
on the set of Feynman diagrams contributing in Eq.~\eqref{ExpansionExact2BVertex}.

Additionally, we stress that this formulation allows us to introduce
approximated functionals which are self-consistent not only in the propagator,
but also in the interaction vertices. This is done by choosing a functional,
$\Gamma^{(2)\text{approx}}[\mathcal{G}]$, defined as a self-consistent
solution in $\Gamma^{(2)}$ of an auxiliary functional,
$\Gamma^{(2)\text{implicit}}[\mathcal{G},\Gamma^{(2)}]$.
We shall now work out explicitly one of these approximations
and discuss in more detail the relevance of the auxiliary functional. 

\subsection{Bethe-Salpeter equation}\label{subsec:BSeq}
To go beyond any finite order NC-SCGF approximation, we have introduced
in Sec.~\ref{subsec:EoMs} approximations on the exact two-body vertex 
defined by an auxiliary functional,
$\Gamma^{(2)\text{implicit}}[\mathcal{G},\Gamma^{(2)}]$.
A specific approximated functional is typically chosen
depending on the many-body system under study
and on the available computational resources.
In this section, we introduce a class of functionals
based on a Bethe-Salpeter equation satisfied
by the exact two-body vertex.

\subsubsection{Motivations}\label{subsubsec:motivationsBS}

The general rationale underlying the Bethe-Salpeter equation~\cite{Salpeter1951}
is similar to the one motivating the Dyson-Schwinger equation.
In the case of the Dyson-Schwinger equation, we require that the poles in energy
of the unperturbed one-body Green's function \emph{must} be shifted in order
to achieve a precise enough approximation. However, for any correction 
made of a finite number of Feynman diagrams, the poles from the unperturbed
part of the propagator remain un-modified\footnote{See for instance the second- and
third-order corrections to the propagator in Part~I. While new poles are generated
as linear combinations of single-particle energies, the poles from the original
unperturbed contribution are un-affected.}.
To design approximations that can
shift unperturbed poles, the self-energy
and, correspondingly, the Dyson-Schwinger equation are introduced. With this, any 
finite number of 1PI Feynman diagrams
contributing to the self-energy will generate, through the Dyson-Schwinger equation,
an infinite number of Feynman diagrams contributing to the one-body Green's function, 
shifting its poles accordingly.
The HFB approximation of the self-energy is a classical example of this approach.

Similarly, let us assume that the poles in energy of the two-body Green's function
\emph{must} be shifted compared to those obtained by a simple product
of two one-body Green's functions.
An approximation based on a finite number of Feynman diagrams
in $\Gamma^{(2)}$
cannot achieve this goal.
To design approximations
where these shifts are possible,  we distinguish between \emph{irreducible} and \emph{reducible} 
contributions\footnote{This is similar in spirit to the Dyson-Schwinger equation,
where only 1PI diagrams are kept and then iterated to generate
an infinite number of reducible terms.}. For any finite number of
irreducible Feynman diagrams, an infinite number of reducible diagrams are generated
by iteration, so that poles of the two-body Green's function are shifted.
Since the two-body Green's function depends on three independent energies,
poles can be shifted in three directions. These three possibilities are reflected
in three different ways of building reducible diagrams from an irreducible set.
A classical example of a set of equations that relate a two-body irreducible part to 
reducible components is the parquet equations. For a detailed account of parquet
equations in the symmetry-conserving case, we refer the reader to Chap.~15 of Ref.~\cite{Blaizot1986}.

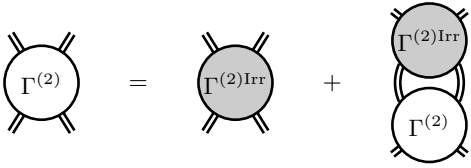
\begin{figure}[t]
  \centering
  \parbox{60pt}{\begin{fmffile}{Gamma2Vertex}
     \begin{fmfgraph*}(40,80)
     \fmfset{arrow_len}{2.8mm}
     \fmfset{arrow_ang}{20}
     \fmfbottom{i1,i2} \fmftop{o1,o2}
     \fmf{phantom, tag=1}{i1,v1}
     \fmf{phantom, tag=2}{i2,v1}
     \fmf{phantom, tag=3}{v1,o1}
     \fmf{phantom, tag=4}{v1,o2}
     \fmfv{d.sh=circle,d.f=empty, d.si=0.7w, b=(1,,1,,1), label=$\Gamma^{(2)}$, label.dist=0}{v1}
     \fmfposition
     \fmfipath{p[]}
     \fmfiset{p1}{vpath1(__i1,__v1)}
     \fmfiset{p2}{vpath2(__i2,__v1)}
     \fmfiset{p3}{vpath3(__v1,__o1)}
     \fmfiset{p4}{vpath4(__v1,__o2)}
     \fmfi{dbl_plain}{subpath (2length(p1)/3,length(p1)) of p1}
     \fmfi{dbl_plain}{subpath (2length(p2)/3,length(p2)) of p2}
     \fmfi{dbl_plain}{subpath (0,length(p3)/3) of p3}
     \fmfi{dbl_plain}{subpath (0,length(p4)/3) of p4}
    \end{fmfgraph*}
  \end{fmffile}}
  =
\parbox{60pt}{\begin{fmffile}{Gamma2_2PI}
  \begin{fmfgraph*}(40,80)
    \fmfkeep{Gamma2_2PI}
    \fmfbottom{i1,i2}
    \fmftop{o1,o2}
     \fmf{phantom, tag=1}{i1,v1}
     \fmf{phantom, tag=2}{i2,v1}
     \fmf{phantom, tag=3}{v1,o1}
     \fmf{phantom, tag=4}{v1,o2}
     \fmfv{d.sh=circle,d.f=empty, d.si=0.7w, b=(0.8,,0.8,,0.8), label=$\Gamma^{(2)\text{Irr}}$, label.dist=0}{v1}
     \fmfposition
     \fmfipath{p[]}
     \fmfiset{p1}{vpath1(__i1,__v1)}
     \fmfiset{p2}{vpath2(__i2,__v1)}
     \fmfiset{p3}{vpath3(__v1,__o1)}
     \fmfiset{p4}{vpath4(__v1,__o2)}
     \fmfi{dbl_plain}{subpath (2length(p1)/3,length(p1)) of p1}
     \fmfi{dbl_plain}{subpath (2length(p2)/3,length(p2)) of p2}
     \fmfi{dbl_plain}{subpath (0,length(p3)/3) of p3}
     \fmfi{dbl_plain}{subpath (0,length(p4)/3) of p4}
\end{fmfgraph*}
\end{fmffile}}
+
\parbox{60pt}{\begin{fmffile}{Gamma2VertexImplicit}
  \begin{fmfgraph*}(40,80)
    \fmfkeep{Gamma2VertexImplicit}
  \fmfbottom{i1,i2}
  \fmftop{o1,o2}
  \fmf{phantom, tension=2, tag=1}{i1,v1}
  \fmf{phantom, tension=2, tag=2}{i2,v1}
  \fmf{phantom, tension=2, tag=3}{v2,o1}
  \fmf{phantom, tension=2, tag=4}{v2,o2}
  \fmf{dbl_plain, right=0.75}{v1,v2}
  \fmf{dbl_plain, right=0.75}{v2,v1}
  \fmfv{d.sh=circle,d.f=empty, d.si=0.7w, b=(1,,1,,1), label=$\Gamma^{(2)}$, label.dist=0}{v1}
  \fmfv{d.sh=circle,d.f=empty, d.si=0.7w, b=(0.8,,0.8,,0.8), label=$\Gamma^{(2)\text{Irr}}$, label.dist=0}{v2}
     \fmfposition
     \fmfipath{p[]}
     \fmfiset{p1}{vpath1(__i1,__v1)}
     \fmfiset{p2}{vpath2(__i2,__v1)}
     \fmfiset{p3}{vpath3(__v2,__o1)}
     \fmfiset{p4}{vpath4(__v2,__o2)}
     \fmfi{dbl_plain}{subpath (2length(p1)/3,length(p1)) of p1}
     \fmfi{dbl_plain}{subpath (2length(p2)/3,length(p2)) of p2}
     \fmfi{dbl_plain}{subpath (0,length(p3)/3) of p3}
     \fmfi{dbl_plain}{subpath (0,length(p4)/3) of p4}
\end{fmfgraph*}
\end{fmffile}}
\caption{Diagrammatic representation of the Bethe-Salpeter
equation, Eq.~\eqref{BSequ}. To easily distinguish
the irreducible part, $\Gamma^{(2)\text{Irr}}$, is represented
with a shaded grey background.}
\label{Fig:BSequ}
\end{figure}

The parquet equations have high potentialities but, unfortunately, their application
to nuclear physics has been hampered by numerical complexity.
Ref.~\cite{Bergli2011} describes an attempt in the symmetry-conserving case.
The main difference with respect to the Dyson-Schwinger equation is that the parquet equations
cannot be expressed as an explicit functional of irreducible diagrams. As a consequence,
computing the infinite set of reducible diagrams contributing to $\Gamma^{(2)}$
from the parquet equations is an iterative process,
starting from a set (its kernel) of chosen irreducible diagrams. 
To the best of our knowledge, even in the simplest case
where just one interaction vertex is used for the irreducible kernel, the numerical complexity to evaluate the whole parquet series has not yet been overcome
for nuclear systems. In particular, the attempts of Ref.~\cite{Bergli2011} were hindered by the uncontrolled energy behaviour near isolated poles of the propagators.
One approach to resolve this problem was suggested in 
Ref.~\cite{Barbieri2003} where the problem was reformulated as an energy-independent eigenvalue problem. While the latter work focused on improving approximations for the Bethe-Salpeter equation, it could also resum one particular channel of the parquet equations. Further computational developments remain however necessary in that direction.

Alternatively,
to bypass the numerical complexity of the parquet equations,
the class of irreducible diagrams can be enlarged so that the reducible contributions
become simpler to compute.
One faces different possibilities when it comes to preselecting these contributions. 
In our case, we want to describe the microscopic properties and the thermodynamics
of superfluid nuclear matter. We choose to keep both
particle-hole ring diagrams as well as particle-particle ladder diagrams.
Iterated particle-hole excitations
impact non-trivially
the collective behaviour of the many-body system in the low-energy and
long-range regime.
These correlations are known to bring important corrections
in the description of, for instance, giant resonances~\cite{Brown1959,Baranger1960}. 
Complementarily, the sum of particle-particle ladders
impacts the short-range and high-energy behaviour of the many-body system,
which is relevant for the macroscopic properties of the system.
The ladders corrections are necessary to properly account for
the strong repulsion (or even a hard core) part of a two-body 
interaction~\cite{Brueckner1954,Brueckner1955,Bethe1956,Day1985,Rios2014}.
In each one of these two cases, the generation of reducible diagrams from an irreducible
set takes the form of a particular Bethe-Salpeter equation.
For more details on those aspects we refer the reader to Chap.~6 of Ref.~\cite{Nozieres1964}.
When particle-number symmetry is broken, the particle-hole and 
particle-particle excitations are coupled to each other. In the following, we specify
the type of irreducible diagrams that we consider to address these correlations. 
We also discuss how a unique Bethe-Salpeter equation
generates a set of reducible diagrams contributing to $\Gamma^{(2)}$,
including particle-particle ladders, particle-hole rings and their coupling.

\begin{figure}[t]
  \centering
\hspace{-2cm}  
\parbox{45pt}{\begin{fmffile}{Irr_Gamma2_2PI}
  \begin{fmfgraph*}(42,60)
    \fmfkeep{Irr_Gamma2_2PI}
    \fmfbottom{i1,i2}
    \fmftop{o1,o2}
     \fmf{phantom, tag=1}{i1,v1}
     \fmf{phantom, tag=2}{i2,v1}
     \fmf{phantom, tag=3}{v1,o1}
     \fmf{phantom, tag=4}{v1,o2}
     \fmfv{d.sh=circle,d.f=empty, d.si=0.65w, b=(0.8,,0.8,,0.8), label=$\Gamma^{(2)\text{Irr}}$, label.dist=0}{v1}
     \fmfposition
     \fmfipath{p[]}
     \fmfiset{p1}{vpath1(__i1,__v1)}
     \fmfiset{p2}{vpath2(__i2,__v1)}
     \fmfiset{p3}{vpath3(__v1,__o1)}
     \fmfiset{p4}{vpath4(__v1,__o2)}
     \fmfi{dbl_plain}{subpath (2length(p1)/3,length(p1)) of p1}
     \fmfi{dbl_plain}{subpath (2length(p2)/3,length(p2)) of p2}
     \fmfi{dbl_plain}{subpath (0,length(p3)/3) of p3}
     \fmfi{dbl_plain}{subpath (0,length(p4)/3) of p4}
     \fmfiv{label=$4$,l.dist=0,l.angle=0}{point 1length(p1)/15 of p1}
     \fmfiv{label=$3$,l.dist=0,l.angle=0}{point 1length(p2)/15 of p2}
     \fmfiv{label=$1$,l.dist=0,l.angle=0}{point 14length(p3)/15 of p3}
     \fmfiv{label=$2$,l.dist=0,l.angle=0}{point 14length(p4)/15 of p4}
\end{fmfgraph*}
\end{fmffile}}
  =
\parbox{50pt}{\begin{fmffile}{Irr_FirstOrder2B}
  \begin{fmfgraph*}(40,60)
    \fmfkeep{Irr_FirstOrder2B}
  \fmfbottom{i1,i2}
  \fmftop{o1,o2}
  \fmf{phantom, tag=1}{i1,v1}
  \fmf{phantom, tag=2}{i2,v1}
  \fmf{phantom, tag=3}{v1,o1}
  \fmf{phantom, tag=4}{v1,o2}
  \fmfv{d.shape=circle,d.filled=full,d.size=3thick}{v1}
  \fmfposition
  \fmfipath{p[]}
  \fmfiset{p1}{vpath1(__i1,__v1)}
  \fmfiset{p2}{vpath2(__i2,__v1)}
  \fmfiset{p3}{vpath3(__v1,__o1)}
  \fmfiset{p4}{vpath4(__v1,__o2)}
  \fmfi{dbl_plain}{subpath (2length(p1)/3,length(p1)) of p1}
  \fmfi{dbl_plain}{subpath (2length(p2)/3,length(p2)) of p2}
  \fmfi{dbl_plain}{subpath (0,length(p3)/3) of p3}
  \fmfi{dbl_plain}{subpath (0,length(p4)/3) of p4}
  \fmfiv{label=$4$,l.dist=0,l.angle=0}{point 1length(p1)/15 of p1}
  \fmfiv{label=$3$,l.dist=0,l.angle=0}{point 1length(p2)/15 of p2}
  \fmfiv{label=$1$,l.dist=0,l.angle=0}{point 14length(p3)/15 of p3}
  \fmfiv{label=$2$,l.dist=0,l.angle=0}{point 14length(p4)/15 of p4}
\end{fmfgraph*}
\end{fmffile}}
+
\parbox{60pt}{\begin{fmffile}{Irr_SecondOrder2B}
  \begin{fmfgraph*}(50,50)
    \fmfkeep{Irr_SecondOrder2B}
  \fmfleft{l1,l2}
  \fmfright{r1,r2}
  \fmf{phantom, tension=2.5, tag=1}{l1,v1}
  \fmf{phantom, tension=2.5, tag=2}{l2,v1}
  \fmf{dbl_plain, right=0.65}{v1,v2}
  \fmf{dbl_plain, left=0.65}{v1,v2}
  \fmf{phantom, tension=2.5, tag=3}{v2,r1}
  \fmf{phantom, tension=2.5, tag=4}{v2,r2}
  \fmfv{d.shape=circle,d.filled=full,d.size=3thick}{v1}
  \fmfv{d.shape=circle,d.filled=full,d.size=3thick}{v2}
    \fmfposition
  \fmfipath{p[]}
  \fmfiset{p1}{vpath1(__l1,__v1)}
  \fmfiset{p2}{vpath2(__l2,__v1)}
  \fmfiset{p3}{vpath3(__v2,__r1)}
  \fmfiset{p4}{vpath4(__v2,__r2)}
  \fmfi{dbl_plain}{subpath (2length(p1)/3,length(p1)) of p1}
  \fmfi{dbl_plain}{subpath (2length(p2)/3,length(p2)) of p2}
  \fmfi{dbl_plain}{subpath (0,length(p3)/3) of p3}
  \fmfi{dbl_plain}{subpath (0,length(p4)/3) of p4}
  \fmfiv{label=$4$,l.dist=0,l.angle=0}{point 1length(p1)/15 of p1}
  \fmfiv{label=$1$,l.dist=0,l.angle=0}{point 1length(p2)/15 of p2}
  \fmfiv{label=$3$,l.dist=0,l.angle=0}{point 14length(p3)/15 of p3}
  \fmfiv{label=$2$,l.dist=0,l.angle=0}{point 14length(p4)/15 of p4}
\end{fmfgraph*}
\end{fmffile}}
\\
\vspace{1cm}
+
\parbox{70pt}{\begin{fmffile}{Irr_SecondOrder3B_0}
  \begin{fmfgraph*}(70,50)
    \fmfkeep{Irr_SecondOrder3B_0}
  \fmfleft{l1,l2}
  \fmfright{r1,r2}
  \fmf{phantom, tension=2.5, tag=1}{l1,v1}
  \fmf{phantom, tension=2.5, tag=2}{l2,v1}
  \fmf{dbl_plain, right=0.65}{v1,v2}
  \fmf{dbl_plain, left=0.65}{v1,v2}
  \fmf{dbl_plain, right=0.65}{v2,v3}
  \fmf{dbl_plain, left=0.65}{v2,v3}
  \fmf{phantom, tension=2.5, tag=3}{v3,r1}
  \fmf{phantom, tension=2.5, tag=4}{v3,r2}
  \fmfv{d.shape=circle,d.filled=full,d.size=3thick}{v1}
  \fmfv{d.shape=circle,d.filled=full,d.size=3thick}{v2}
  \fmfv{d.shape=circle,d.filled=full,d.size=3thick}{v3}
    \fmfposition
  \fmfipath{p[]}
  \fmfiset{p1}{vpath1(__l1,__v1)}
  \fmfiset{p2}{vpath2(__l2,__v1)}
  \fmfiset{p3}{vpath3(__v3,__r1)}
  \fmfiset{p4}{vpath4(__v3,__r2)}
  \fmfi{dbl_plain}{subpath (2length(p1)/3,length(p1)) of p1}
  \fmfi{dbl_plain}{subpath (2length(p2)/3,length(p2)) of p2}
  \fmfi{dbl_plain}{subpath (0,length(p3)/3) of p3}
  \fmfi{dbl_plain}{subpath (0,length(p4)/3) of p4}
  \fmfiv{label=$4$,l.dist=0,l.angle=0}{point 1length(p1)/15 of p1}
  \fmfiv{label=$1$,l.dist=0,l.angle=0}{point 1length(p2)/15 of p2}
  \fmfiv{label=$3$,l.dist=0,l.angle=0}{point 14length(p3)/15 of p3}
  \fmfiv{label=$2$,l.dist=0,l.angle=0}{point 14length(p4)/15 of p4}
\end{fmfgraph*}
\end{fmffile}}
+
\parbox{50pt}{\begin{fmffile}{Irr_ThirdOrder2B_1}
  \begin{fmfgraph*}(50,50)
    \fmfkeep{Irr_ThirdOrder2B_1}
  \fmfleft{l1,l2}
  \fmfright{r1,r2}
  \fmf{phantom, tension=2.5, tag=1}{l1,v1}
  \fmf{phantom, tension=2.5, tag=2}{l2,v1}
  \fmf{dbl_plain, left=0.25}{v1,v3}
  \fmf{dbl_plain, right=0.25}{v1,v2}
  \fmf{phantom, tension=1.5, tag=3}{v2,r1}
  \fmf{phantom, tension=1.5, tag=4}{v3,r2}
  \fmffreeze
  \fmf{dbl_plain, right=0.4}{v2,v3}
  \fmf{dbl_plain, left=0.4}{v2,v3}
  \fmfv{d.shape=circle,d.filled=full,d.size=3thick}{v1}
  \fmfv{d.shape=circle,d.filled=full,d.size=3thick}{v2}
  \fmfv{d.shape=circle,d.filled=full,d.size=3thick}{v3}
    \fmfposition
  \fmfipath{p[]}
  \fmfiset{p1}{vpath1(__l1,__v1)}
  \fmfiset{p2}{vpath2(__l2,__v1)}
  \fmfiset{p3}{vpath3(__v2,__r1)}
  \fmfiset{p4}{vpath4(__v3,__r2)}
  \fmfi{dbl_plain}{subpath (2length(p1)/3,length(p1)) of p1}
  \fmfi{dbl_plain}{subpath (2length(p2)/3,length(p2)) of p2}
  \fmfi{dbl_plain}{subpath (0,length(p3)/3) of p3}
  \fmfi{dbl_plain}{subpath (0,length(p4)/3) of p4}
  \fmfiv{label=$4$,l.dist=0,l.angle=0}{point 1length(p1)/15 of p1}
  \fmfiv{label=$1$,l.dist=0,l.angle=0}{point 1length(p2)/15 of p2}
  \fmfiv{label=$3$,l.dist=0,l.angle=0}{point 16length(p3)/15 of p3}
  \fmfiv{label=$2$,l.dist=0,l.angle=0}{point 16length(p4)/15 of p4}
\end{fmfgraph*}
\end{fmffile}}
+
\parbox{50pt}{\begin{fmffile}{Irr_ThirdOrder2B_2}
  \begin{fmfgraph*}(50,50)
    \fmfkeep{Irr_ThirdOrder2B_2}
  \fmfright{r1,r2}
  \fmfleft{l1,l2}
  \fmf{phantom, tension=2.5, tag=1}{r1,v1}
  \fmf{phantom, tension=2.5, tag=2}{r2,v1}
  \fmf{dbl_plain, right=0.25}{v1,v3}
  \fmf{dbl_plain, left=0.25}{v1,v2}
  \fmf{phantom, tension=1.5, tag=3}{v2,l1}
  \fmf{phantom, tension=1.5, tag=4}{v3,l2}
  \fmffreeze
  \fmf{dbl_plain, right=0.4}{v2,v3}
  \fmf{dbl_plain, left=0.4}{v2,v3}
  \fmfv{d.shape=circle,d.filled=full,d.size=3thick}{v1}
  \fmfv{d.shape=circle,d.filled=full,d.size=3thick}{v2}
  \fmfv{d.shape=circle,d.filled=full,d.size=3thick}{v3}
    \fmfposition
  \fmfipath{p[]}
  \fmfiset{p1}{vpath1(__r1,__v1)}
  \fmfiset{p2}{vpath2(__r2,__v1)}
  \fmfiset{p3}{vpath3(__v2,__l1)}
  \fmfiset{p4}{vpath4(__v3,__l2)}
  \fmfi{dbl_plain}{subpath (2length(p1)/3,length(p1)) of p1}
  \fmfi{dbl_plain}{subpath (2length(p2)/3,length(p2)) of p2}
  \fmfi{dbl_plain}{subpath (0,length(p3)/3) of p3}
  \fmfi{dbl_plain}{subpath (0,length(p4)/3) of p4}
  \fmfiv{label=$3$,l.dist=0,l.angle=0}{point 1length(p1)/15 of p1}
  \fmfiv{label=$2$,l.dist=0,l.angle=0}{point 1length(p2)/15 of p2}
  \fmfiv{label=$4$,l.dist=0,l.angle=0}{point 16length(p3)/15 of p3}
  \fmfiv{label=$1$,l.dist=0,l.angle=0}{point 16length(p4)/15 of p4}
\end{fmfgraph*}
\end{fmffile}}
+
\dots
\caption{Examples of amputated Feynman diagrams contributing to the irreducible part
of the two-body vertex. For clarity, we write explicitly the numbers
associated to each amputated external line.
Two amputated diagrams differing only by a permutation
of the amputated lines can also differ in their irreducible character.
We do not exhaust the contributions up to third order for the sake of conciseness.}
\label{Fig:2BIrrVertex}
\end{figure}
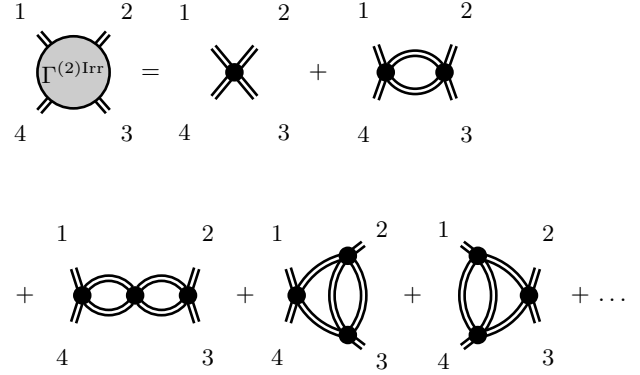

\subsubsection{Bethe-Salpeter equation}
We want to describe
$\Gamma^{(2)}$ in terms of a Bethe-Salpeter equation, to limit the computational complexity
while keeping our ability to describe the relevant phenomenology of
the many-body system. 
We aim to generate particle-hole rings and particle-particle ladders from the Bethe-Salpeter equation
starting from an irreducible set of diagrams.
In the case of an S-wave contact interaction,
the relevant equation was studied by Haussmann~\cite{Haussmann1993,Haussmann1999}.
Similarly, we consider the following Bethe-Salpeter equation
\begin{multline}\label{BSequ}
  \Gamma^{(2)}_{\mu_1\mu_2\mu_3\mu_4}
    (\tau_{1}, \tau_{2}, \tau_{3}, \tau_{4})
  =
  \Gamma^{(2)\text{Irr}}_{\mu_1\mu_2\mu_3\mu_4}
    (\tau_{1}, \tau_{2}, \tau_{3}, \tau_{4})
  \\
  +
  \frac{1}{2}
  \sum_{\substack{\lambda_1 \lambda_2 \\ \lambda'_1 \lambda'_2}}
  \int^{\beta}_{0}
  \mathrm{d}\tau_{\lambda_1} \mathrm{d}\tau_{\lambda_2}
  \mathrm{d}\tau_{\lambda'_1} \mathrm{d}\tau_{\lambda'_2} \ 
  \Gamma^{(2)\text{Irr}}_{\mu_1\mu_2\lambda_2\lambda_1}
    (\tau_{1}, \tau_{2}, \tau_{\lambda_2}, \tau_{\lambda_1})
  \\
  \times
  \mathcal{G}^{\lambda_1\lambda'_1}(\tau_{\lambda_1}, \tau_{\lambda'_1})
  \mathcal{G}^{\lambda_2\lambda'_2}(\tau_{\lambda_2}, \tau_{\lambda'_2}) \\ 
  \times \Gamma^{(2)}_{\lambda'_1\lambda'_2\mu_3\mu_4}
    (\tau_{\lambda'_1}, \tau_{\lambda'_2}, \tau_{3}, \tau_{4}) \ ,
\end{multline}
where $\Gamma^{(2)\text{Irr}}_{\mu_1\mu_2\mu_3\mu_4}
(\tau_{1}, \tau_{2}, \tau_{3}, \tau_{4})$ denotes the sum of irreducible Feynman diagrams
such that, by definition, the two-body vertex obtained by Eq.~\eqref{BSequ} is exact.
Eq.~\eqref{BSequ} is represented diagrammatically in Fig.~\ref{Fig:BSequ}.

To determine the subset of diagrams that contribute to the irreducible part,
let us recall that
$\Gamma^{(2)}$ is the sum of dressed skeleton diagrams with four amputated external lines.
Consequently, the irreducible part, $\Gamma^{(2)\text{Irr}}$, must include
all necessary diagrams such that the whole set of skeleton ones are generated
by iteration of the Bethe-Salpeter equation, Eq.~\eqref{BSequ}.
In this case, it is straightforward to show that the class of irreducible diagrams
is the class of amputated dressed diagram that are both skeleton and $[12]$-simple.
A diagram is said to be $[12]$-simple if, by cutting two internal lines,
it cannot be separated into two disconnected parts, such that one part contains
external lines $1$ and $2$ and the other, external lines $3$ and $4$\footnote{By
line $i$ we mean here the line associated to the external global index $\mu_i$
and imaginary-time $\tau_i$. We follow here the definition given
in Chap.~15 of Ref.~\cite{Blaizot1986}.}.
We show in Fig.~\ref{Fig:2BIrrVertex} some examples contributing to the irreducible
part of $\Gamma^{(2)}$.

\begin{figure}[t]
  \centering
  \includegraphics[scale=0.31, trim= 170 50 170 50]{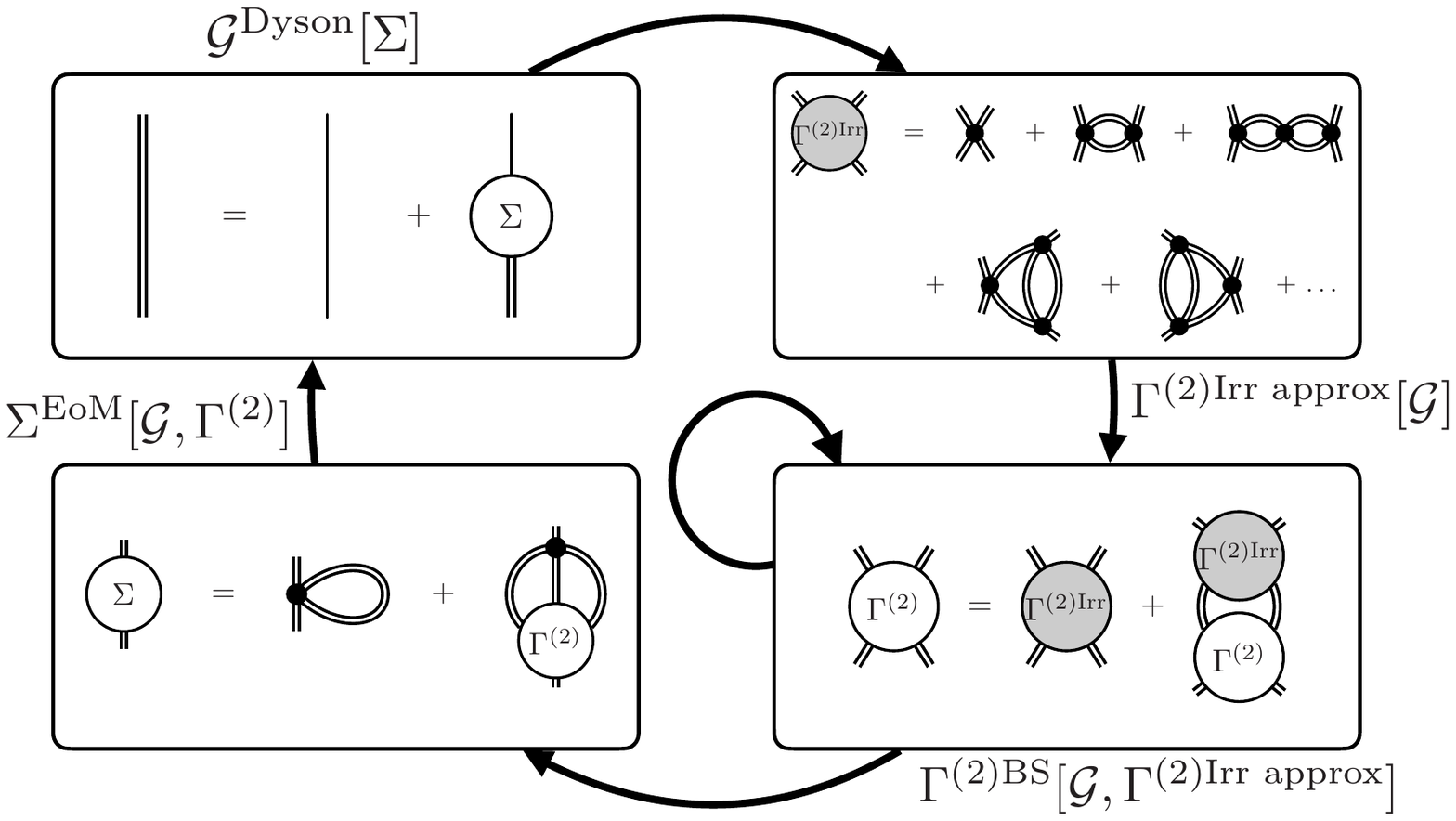}
\caption{Flowchart representing the self-consistent cycle obtained by iterating
$\mathcal{G}^{\text{Dyson}}[\Sigma]$,
$\Sigma^{\text{EoM}}[\mathcal{G}, \Gamma^{(2)}]$,
$\Gamma^{(2)\text{Irr}\text{ approx}}[\mathcal{G}]$
and $\Gamma^{(2)\text{BS}}[\mathcal{G}, \Gamma^{(2)\text{Irr}\text{ approx}}]$.
The inner cycle represents the iterations on $\Gamma^{(2)}$, which are necessary to evaluate
$\Gamma^{(2)\text{BS}}[\mathcal{G}, \Gamma^{(2)\text{Irr}\text{ approx}}]$
from the auxiliary functional $\Gamma^{(2)\text{Implicit-BS}}[\mathcal{G},\Gamma^{(2)}, \Gamma^{(2)\text{Irr}\text{ approx}}]$.}
\label{Fig:SCGF_cycle_BetheSalpeter}
\end{figure}

The Bethe-Salpeter equation~\eqref{BSequ} defines an auxiliary functional
$\Gamma^{(2)\text{Implicit-BS}}[\mathcal{G},\Gamma^{(2)}, \Gamma^{(2)\text{Irr}\text{ approx}}]$.
The solution of Eq.~\eqref{BSequ}, self-consistent in $\Gamma^{(2)}$,
is then denoted by the functional
$\Gamma^{(2)\text{BS}}[\mathcal{G}, \Gamma^{(2)\text{Irr}\text{ approx}}]$.
For a given approximated functional of the irreducible part
$\Gamma^{(2)\text{Irr}\text{ approx}}[\mathcal{G}]$, we obtain
a new self-consistent cycle as shown in Fig.~\ref{Fig:SCGF_cycle_BetheSalpeter}.
Note that the approximated two-body vertex must be computed iteratively
from $\Gamma^{(2)\text{Implicit-BS}}[\mathcal{G},\Gamma^{(2)}, \Gamma^{(2)\text{Irr}\text{ approx}}]$
which increases the numerical cost of the self-consistent scheme
by a non-negligible amount.
As it will be shown in Sec.~\ref{subsec:ladderApprox},
this extra cost can be avoided in the particular case of the
\emph{self-consistent ladder approximation} by working out
an explicit functional of the propagator.

\subsection{Ladder approximation}\label{subsec:ladderApprox}
The ladder approximation is introduced as the first order truncation
on the irreducible part of the two-body vertex in
the Bethe-Salpeter equation, Eq.~\eqref{BSequ}. 
The approximated irreducible two-body vertex 
is given by the first term in Fig.~\ref{Fig:2BIrrVertex} and consequently 
reads
\begin{multline}\label{DefIrr2BVertexLadder}
  \Gamma^{(2)\text{Irr Ladder}}_{\mu_1\mu_2\mu_3\mu_4}
  (\tau_1, \tau_2, \tau_3, \tau_4)
  \equiv \\
    v^{(2)}_{[\mu_1\mu_2\mu_3\mu_4]} \ 
    \delta(\tau_1 - \tau_2) \ 
    \delta(\tau_3 - \tau_4) \ 
    \delta(\tau_1 - \tau_3) \ .
\end{multline}
The resulting approximated two-body vertex is called
(in-medium) $T$-matrix and denoted
$T_{\mu_1\mu_2\mu_3\mu_4}(\tau_1, \tau_2, \tau_3, \tau_4)$.

In this section, we derive an implicit equation on the $T$-matrix.
We then find 
an explicit functional $T[\mathcal{G}]$,
so that the iterative cycle on the $T$-matrix is avoided.
Exact properties satisfied by the $T$-matrix are detailed
inasmuch as they allow us to express the equations of the
self-consistent ladder approximation in terms of spectral functions only.
In addition, we discuss the convergence of the series of ladders
for a HFB propagator. We give a sufficient condition to guarantee the convergence
for any Hermitian two-body interaction.

\subsubsection{Implicit equation}
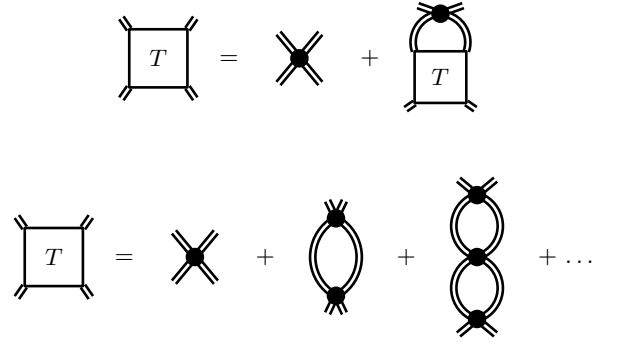
\begin{figure}[t]
  \centering
  \parbox{40pt}{\begin{fmffile}{T_Vertex}
     \begin{fmfgraph*}(40,60)
     \fmfset{arrow_len}{2.8mm}
     \fmfset{arrow_ang}{20}
     \fmfbottom{i1,i2} \fmftop{o1,o2}
     \fmfpolyn{empty, label=$T$}{T}{4}
     \fmf{phantom, tension=2, tag=1}{i1,T1}
     \fmf{phantom, tension=2, tag=2}{i2,T2}
     \fmf{phantom, tension=2, tag=3}{T4,o1}
     \fmf{phantom, tension=2, tag=4}{T3,o2}
         \fmfposition
    \fmfipath{p[]}
    \fmfiset{p1}{vpath1(__i1,__T1)}
    \fmfiset{p2}{vpath2(__i2,__T2)}
    \fmfiset{p3}{vpath3(__T4,__o1)}
    \fmfiset{p4}{vpath4(__T3,__o2)}
    \fmfi{dbl_plain}{subpath (2length(p1)/3,length(p1)) of p1}
    \fmfi{dbl_plain}{subpath (2length(p2)/3,length(p2)) of p2}
    \fmfi{dbl_plain}{subpath (0,length(p3)/3) of p3}
    \fmfi{dbl_plain}{subpath (0,length(p4)/3) of p4}
    \end{fmfgraph*}
  \end{fmffile}}
  =
\parbox{40pt}{\begin{fmffile}{T_Vertex1stOrder}
  \begin{fmfgraph*}(40,60)
    \fmfkeep{T_Vertex1stOrder}
    \fmfbottom{i1,i2}
    \fmftop{o1,o2}
  \fmf{phantom, tag=1}{i1,v1}
  \fmf{phantom, tag=2}{i2,v1}
  \fmf{phantom, tag=3}{v1,o1}
  \fmf{phantom, tag=4}{v1,o2}
  \fmfv{d.shape=circle,d.filled=full,d.size=3thick}{v1}
           \fmfposition
    \fmfipath{p[]}
    \fmfiset{p1}{vpath1(__i1,__v1)}
    \fmfiset{p2}{vpath2(__i2,__v1)}
    \fmfiset{p3}{vpath3(__v1,__o1)}
    \fmfiset{p4}{vpath4(__v1,__o2)}
    \fmfi{dbl_plain}{subpath (2length(p1)/3,length(p1)) of p1}
    \fmfi{dbl_plain}{subpath (2length(p2)/3,length(p2)) of p2}
    \fmfi{dbl_plain}{subpath (0,length(p3)/3) of p3}
    \fmfi{dbl_plain}{subpath (0,length(p4)/3) of p4}
\end{fmfgraph*}
\end{fmffile}}
+
\parbox{40pt}{\begin{fmffile}{T_VertexImplicit}
  \begin{fmfgraph*}(40,60)
    \fmfkeep{T_VertexImplicit}
  \fmfbottom{i1,i2}
  \fmftop{o1,o2}
  \fmfpolyn{empty, label=$T$}{T}{4}
  \fmf{phantom, tension=4, tag=1}{i1,T1}
  \fmf{phantom, tension=4, tag=2}{i2,T2}
  \fmf{phantom, tension=4, tag=3}{v2,o1}
  \fmf{phantom, tension=4, tag=4}{v2,o2}
  \fmf{dbl_plain, right=0.5, tension=2}{T3,v2}
  \fmf{dbl_plain, right=0.5, tension=2}{v2,T4}
  \fmfv{d.shape=circle,d.filled=full,d.size=3thick}{v2}
        \fmfposition
  \fmfipath{p[]}
  \fmfiset{p1}{vpath1(__i1,__T1)}
  \fmfiset{p2}{vpath2(__i2,__T2)}
  \fmfiset{p3}{vpath3(__v2,__o1)}
  \fmfiset{p4}{vpath4(__v2,__o2)}
  \fmfi{dbl_plain}{subpath (2length(p1)/3,length(p1)) of p1}
  \fmfi{dbl_plain}{subpath (2length(p2)/3,length(p2)) of p2}
  \fmfi{dbl_plain}{subpath (0,length(p3)/3) of p3}
  \fmfi{dbl_plain}{subpath (0,length(p4)/3) of p4}
\end{fmfgraph*}
\end{fmffile}}

\vspace{0.5cm}
\parbox{40pt}{\begin{fmffile}{T_Vertex}
   \begin{fmfgraph*}(40,60)
   \fmfset{arrow_len}{2.8mm}
   \fmfset{arrow_ang}{20}
   \fmfbottom{i1,i2} \fmftop{o1,o2}
   \fmfpolyn{empty, label=$T$}{T}{4}
     \fmf{phantom, tension=2, tag=1}{i1,T1}
     \fmf{phantom, tension=2, tag=2}{i2,T2}
     \fmf{phantom, tension=2, tag=3}{T4,o1}
     \fmf{phantom, tension=2, tag=4}{T3,o2}
         \fmfposition
    \fmfipath{p[]}
    \fmfiset{p1}{vpath1(__i1,__T1)}
    \fmfiset{p2}{vpath2(__i2,__T2)}
    \fmfiset{p3}{vpath3(__T4,__o1)}
    \fmfiset{p4}{vpath4(__T3,__o2)}
    \fmfi{dbl_plain}{subpath (2length(p1)/3,length(p1)) of p1}
    \fmfi{dbl_plain}{subpath (2length(p2)/3,length(p2)) of p2}
    \fmfi{dbl_plain}{subpath (0,length(p3)/3) of p3}
    \fmfi{dbl_plain}{subpath (0,length(p4)/3) of p4}
  \end{fmfgraph*}
\end{fmffile}}
=
\parbox{40pt}{\begin{fmffile}{T_Vertex1stOrder}
\begin{fmfgraph*}(40,60)
  \fmfkeep{T_Vertex1stOrder}
  \fmfbottom{i1,i2}
  \fmftop{o1,o2}
  \fmf{phantom, tag=1}{i1,v1}
  \fmf{phantom, tag=2}{i2,v1}
  \fmf{phantom, tag=3}{v1,o1}
  \fmf{phantom, tag=4}{v1,o2}
  \fmfv{d.shape=circle,d.filled=full,d.size=3thick}{v1}
           \fmfposition
    \fmfipath{p[]}
    \fmfiset{p1}{vpath1(__i1,__v1)}
    \fmfiset{p2}{vpath2(__i2,__v1)}
    \fmfiset{p3}{vpath3(__v1,__o1)}
    \fmfiset{p4}{vpath4(__v1,__o2)}
    \fmfi{dbl_plain}{subpath (2length(p1)/3,length(p1)) of p1}
    \fmfi{dbl_plain}{subpath (2length(p2)/3,length(p2)) of p2}
    \fmfi{dbl_plain}{subpath (0,length(p3)/3) of p3}
    \fmfi{dbl_plain}{subpath (0,length(p4)/3) of p4}
\end{fmfgraph*}
\end{fmffile}}
+
\parbox{40pt}{\begin{fmffile}{T_Vertex2ndOrder}
\begin{fmfgraph*}(40,60)
  \fmfkeep{T_Vertex2ndOrder}
  \fmfbottom{i1,i2,i3,i4}
  \fmftop{o1,o2,o3,o4}
    \fmf{phantom, tension=2, tag=1}{i2,v1}
    \fmf{phantom, tension=2, tag=2}{i3,v1}
    \fmf{phantom, tension=2, tag=3}{v2,o2}
    \fmf{phantom, tension=2, tag=4}{v2,o3}
    \fmf{dbl_plain, right=0.6}{v1,v2}
    \fmf{dbl_plain, right=0.6}{v2,v1}
    \fmfv{d.shape=circle,d.filled=full,d.size=3thick}{v1}
    \fmfv{d.shape=circle,d.filled=full,d.size=3thick}{v2}
           \fmfposition
    \fmfipath{p[]}
    \fmfiset{p1}{vpath1(__i2,__v1)}
    \fmfiset{p2}{vpath2(__i3,__v1)}
    \fmfiset{p3}{vpath3(__v2,__o2)}
    \fmfiset{p4}{vpath4(__v2,__o3)}
    \fmfi{dbl_plain}{subpath (2length(p1)/3,length(p1)) of p1}
    \fmfi{dbl_plain}{subpath (2length(p2)/3,length(p2)) of p2}
    \fmfi{dbl_plain}{subpath (0,length(p3)/3) of p3}
    \fmfi{dbl_plain}{subpath (0,length(p4)/3) of p4}
\end{fmfgraph*}
\end{fmffile}}
+
\parbox{40pt}{\begin{fmffile}{T_Vertex3rdOrder}
\begin{fmfgraph*}(40,60)
  \fmfkeep{T_Vertex3rdOrder}
\fmfbottom{i1,i2,i3,i4}
\fmftop{o1,o2,o3,o4}
\fmf{dbl_plain, tension=4}{i2,v1}
\fmf{dbl_plain, tension=4}{i3,v1}
\fmf{dbl_plain, tension=4}{v3,o2}
\fmf{dbl_plain, tension=4}{v3,o3}

\fmf{dbl_plain, right=0.75}{v1,v2}
\fmf{dbl_plain, right=0.75}{v2,v3}
\fmf{dbl_plain, right=0.75}{v3,v2}
\fmf{dbl_plain, right=0.75}{v2,v1}
\fmfv{d.shape=circle,d.filled=full,d.size=3thick}{v1}
\fmfv{d.shape=circle,d.filled=full,d.size=3thick}{v2}
\fmfv{d.shape=circle,d.filled=full,d.size=3thick}{v3}
\end{fmfgraph*}
\end{fmffile}}
+
\dots

\caption{Top: diagrammatic representation of the implicit
equation~\eqref{InoTmatrixEq} on the $T$-matrix.
Bottom: explicit diagrammatic expression of the $T$-matrix.
Since $T$ has equal incoming or outgoing times, it is represented by
a rectangular box.}
\label{Fig:InoTmatrixEq}
\end{figure}

Plugging the first-order approximation of the irreducible
two-body vertex, Eq.~\eqref{DefIrr2BVertexLadder},
into the Bethe-Salpeter equation, Eq.~\eqref{BSequ}, we see that
the $T$-matrix
only depends on two times, $\tau_1$ and $\tau_4$,
\begin{multline}
    T_{\mu_1\mu_2\mu_3\mu_4}(\tau_1, \tau_2, \tau_3, \tau_4)
    = \\
    T_{\mu_1\mu_2\mu_3\mu_4}(\tau_1, \tau_4) \ 
    \delta(\tau_1 - \tau_2) \ 
    \delta(\tau_3 - \tau_4) \ ,
\end{multline}
provided we factorise the two Dirac distribution on $\tau_2$ and $\tau_3$.
The simplified Bethe-Salpeter equation reads
\begin{multline}\label{InoTmatrixEq}
  T_{\mu_1\mu_2\mu_3\mu_4}(\tau_1, \tau_4)
  =
  v^{(2)}_{[\mu_1\mu_2\mu_3\mu_4]} \ \delta(\tau_1 - \tau_4)
  \\
  +
  \frac{1}{2}
  \sum_{\substack{\lambda_1 \lambda_2 \\ \lambda'_1 \lambda'_2}}
  \int^{\beta}_{0}
  \mathrm{d}\tau' \
  v^{(2)}_{[\mu_1\mu_2\lambda_2\lambda_1]} \ 
  \mathcal{G}^{\lambda_1\lambda'_1}(\tau_1, \tau')
  \mathcal{G}^{\lambda_2\lambda'_2}(\tau_1, \tau') \\
  \times T_{\lambda'_1\lambda'_2\mu_3\mu_4}(\tau',\tau_4) \ .
\end{multline}
Eq.~\eqref{InoTmatrixEq} is shown in Fig.~\ref{Fig:InoTmatrixEq}.
From Eq.~\eqref{InoTmatrixEq}, we can show that
$\left( \partial_{\tau_1} + \partial_{\tau_4} \right)
T_{\mu_1\mu_2\mu_3\mu_4}(\tau_1, \tau_4)$ satisfies an homogeneous Fredholm
integral equation. Assuming the kernel to be non-singular implies that
\begin{equation}
\left( \partial_{\tau_1} + \partial_{\tau_4} \right)
  T_{\mu_1\mu_2\mu_3\mu_4}(\tau_1, \tau_4)
  = 0
\end{equation}
so that $T$ only depends on the relative time $\tau \equiv \tau_1 - \tau_4$
and we define
\begin{equation}
    T_{\mu_1\mu_2\mu_3\mu_4}(\tau)
    \equiv T_{\mu_1\mu_2\mu_3\mu_4}(\tau_1 - \tau_4, 0)
    = T_{\mu_1\mu_2\mu_3\mu_4}(\tau_1, \tau_4) \ .
\end{equation}

To further simplify the notation, we introduce the following multi-indices
\begin{subequations}
\begin{align}
  M  &\equiv (\mu_1,\mu_2) \ , \\
  N  &\equiv (\mu_3,\mu_4) \ , \\
  L  &\equiv (\lambda_1,\lambda_2) \ , \\
  L' &\equiv (\lambda'_1,\lambda'_2) \ ,
\end{align}
\end{subequations}
and define the objects
\begin{subequations}
\begin{align}
  T_{MN}(\tau) &\equiv T_{\mu_1\mu_2\mu_3\mu_4}(\tau) \ , \\
  V^{(2)}_{MN} &\equiv v^{(2)}_{[\mu_1\mu_2\mu_3\mu_4]} \ , \\
  \Pi^{LL'}(\tau) &\equiv
      -\mathcal{G}^{\lambda_1\lambda'_1}(\tau)
      \mathcal{G}^{\lambda_2\lambda'_2}(\tau) \ . \label{DefBubble}
\end{align}
\end{subequations}
In keeping with the standard symmetry-conserving nomenclature, 
we refer to  $\Pi^{LL'}(\tau)$  as the ``bubble" propagator or, simply, the bubble. 
As with global indices, we use an intrinsic notation for the multi-indices
$M,N,\dots$ i.e.\ multi-indices are dropped whenever there is no ambiguity.
For example, since $H_1$ is assumed to be Hermitian,
the potential satisfies the symmetries
\begin{subequations}
\begin{align}
    V^{(2)} &= {V^{(2)}}^\dagger \ , \label{HermitianPotTensor} \\
    V^{(2)} &= {V^{(2)}}\transpose \ , \label{AntisymPotTensor}
\end{align}
\end{subequations}
where the Hermitian conjugation of a $(p,q)$-tensor is defined
in~\ref{App:HermitianConjugation} and
where the transposition is defined for $(0,4)$-, $(4,0)$-
and $(2,2)$-tensors respectively by
\begin{subequations}
\begin{align}
    (t\transpose)_{MN} &\equiv t_{NM} \ , \\
    (t\transpose)^{MN} &\equiv t^{NM} \ , \\
    {(t\transpose)^{M}}_{N} &\equiv {t_{N}}^{M} \ .
\end{align}
\end{subequations}
With these additional notations, the Bethe-Salpeter equation reads simply
\begin{multline}\label{TmatrixEqTime_Intrinsic}
  T(\tau) =
    V^{(2)} \ \delta(\tau)
    + \frac{1}{2}
      V^{(2)} 
      \int^{\beta}_{0} \mathrm{d}\tau' \
      \Pi(\tau - \tau') \ 
      T(\tau') \ .
\end{multline}

Finally, using the $\beta$-quasiperiodicity of
the propagator stated in Eq.~\eqref{QuasiPeriodicProp},
we find that the bubble, $\Pi$, and the $T$-matrix, $T$, are $\beta$-periodic functions. 
Consequently, we introduce their energy representations,
$\Pi(\Omega_p)$ and $T(\Omega_p)$, using the Fourier transforms
\begin{align}
    \Pi(\Omega_p)
    &\equiv
    \int^{\beta}_{0}\mathrm{d}\tau \
        e^{i \Omega_p \tau} \Pi(\tau) \ , \\
    T(\Omega_p)
    &\equiv
    \int^{\beta}_{0}\mathrm{d}\tau \
        e^{i \Omega_p \tau} T(\tau) \ ,
\end{align}
where $\Omega_p \equiv 2p \frac{\pi}{\beta}$ are bosonic Matsubara frequencies.
The appearance of these frequencies is a consequence of $\beta$-periodicity, but also
highlights the fact that the bubble and the $T$-matrix are two-fermion functions. 
In other words, they describe (bosonic) two-fermion pair propagation and scattering 
in the medium. 
In the energy representation, the Bethe-Salpeter equation becomes
\begin{equation}\label{TmatrixEqEnergy}
  T(\Omega_p) =
    V^{(2)}
    + \frac{1}{2}
      V^{(2)} \ 
      \Pi(\Omega_p) \ 
      T(\Omega_p) \ .
\end{equation}
Comparing Eqs.~\eqref{EnergyDysonEqs} to Eq.~\eqref{TmatrixEqEnergy},
the similarity between the Bethe-Salpeter equation and the Dyson-Schwinger
appears clearly.

\subsubsection{Exact properties}
The $T$-matrix and the bubble, $\Pi$, verify
exact properties which are similar to those of the propagator, $\mathcal{G}$,
and the self-energy, $\Sigma$.
We briefly recall them here in the energy representation.
First, we introduce the analytical continuations, $\Pi(Z)$ and $T(Z)$,
into the upper and lower complex energy half-planes, together with
the associated spectral functions, $P(\Omega)$ and $\mathscr{T}(\Omega)$,
\begin{subequations}
\begin{align}
    \Pi(Z)
    &\equiv \int^{+\infty}_{-\infty} \frac{\mathrm{d}\Omega}{2\pi} \
      \frac{P(\Omega)}{Z - \Omega} \ , \\
    T(Z) 
    &\equiv V^{(2)}
    + \int^{+\infty}_{-\infty} \frac{\mathrm{d}\Omega}{2\pi}
        \frac{\mathscr{T}(\Omega)}{Z - \Omega} \ ,
\end{align}
\end{subequations}
such that we recover
\begin{subequations}
\begin{align}
    \Pi(Z = i\Omega_p) &= \Pi(\Omega_p) \ , \\
    T(Z = i\Omega_p) &= T(\Omega_p) \ .
\end{align}
\end{subequations}
The $T$-matrix and the bubble verify the analytically extended equation
\begin{equation}\label{TmatrixEqAnalytic}
  T(Z) =
    V^{(2)}
    + \frac{1}{2}
      V^{(2)} \ 
      \Pi(Z) \ 
      T(Z) \ .
\end{equation}
The corresponding retarded and advanced components are defined, as usual,
by functions with arguments infinitesimally close to the real energy axis,
\begin{subequations}
\begin{align}
  \Pi^{R/A}(\Omega) &\equiv \Pi(\Omega \pm i\eta) \ , \\
  T^{R/A}(\Omega) &\equiv T(\Omega \pm i \eta) \ .
\end{align}
\end{subequations}
The corresponding spectral functions can be obtained from the discontinuities across the
real axis, expressed here as differences between advanced and retarded components,
\begin{subequations}
\begin{align}
    P(\Omega) 
        &= i \left[\Pi^R(\Omega) -  \Pi^{A}(\Omega)\right]\ , \\
    \mathscr{T}(\Omega)
        &= i \left[T^R(\Omega) -  T^{A}(\Omega)\right] \ . \label{SpeTmatrix}
\end{align}
\end{subequations}

Second,
we enumerate the symmetry properties on the spectral functions,
which translate directly on the corresponding analytical continuations.
For the bubble and the $T$-matrix, the antisymmetry and Hermitian properties read
\begin{subequations}
\begin{align}
    P(\Omega) &= -P\transpose(-\Omega) \ ,
        \quad P(\Omega) = P^\dagger(\Omega) \ , \\
    \mathscr{T}(\Omega) &= -\mathscr{T}\transpose(-\Omega) \ ,
        \;\, \mathscr{T}(\Omega) = \mathscr{T}^\dagger(\Omega) \ .
\end{align}
\end{subequations}
Regarding the positive definiteness of the spectral functions,
we have\footnote{The positive definiteness of $\mathscr{T}(\Omega)$ is readily obtained
from that of $P(\Omega)$ and the generalised optical theorem satisfied by the
$T$-matrix~\cite{Kadanoff1962,Kraeft1986}, which reads
\begin{equation}
    \mathscr{T}(\Omega) 
        = T^R(\Omega) \ P(\Omega) \ {T^R}^\dagger(\Omega) \ .
\end{equation}}
\begin{subequations}
  \begin{align}
    \forall \ \Omega > 0 \quad , \quad P(\Omega) &\succ 0 \ , 
        \quad \mathscr{T}(\Omega) \succ 0 \ , \\
    P( 0 ) &= 0 \ , 
        \quad \mathscr{T}( 0 ) = 0 \ , \\
    \forall \ \Omega < 0 \quad , \quad P(\Omega) &\prec 0 \ ,
        \quad \mathscr{T}(\Omega) \prec 0 \ .
  \end{align}
\end{subequations}

Last,
the previous expressions allow us to find the following dispersion relations for the bubble
\begin{subequations}\label{DispersionBubble}
\begin{align}
 \xoverline{\Re} \ \Pi^{R/A}(\Omega)
          &= \mathcal{P}\int_{-\infty}^{+\infty}
              \frac{\mathrm{d}\Omega'}{2\pi}
              \frac{P(\Omega')}{\Omega-\Omega'} \label{RePiR/A_from_SpBubble} \ , \\
 \xoverline{\Im} \ \Pi^{R/A}(\Omega)
          &= \mp \frac{1}{2} P(\Omega) \ ,
\end{align}
\end{subequations}
and for the $T$-matrix,
\begin{subequations}\label{DispersionTmatrix}
\begin{align}
 \xoverline{\Re} \ T^{R/A}(\Omega) &=
    V^{(2)}
    +
    \mathcal{P}\int_{-\infty}^{+\infty}
              \frac{\mathrm{d}\Omega'}{2\pi}
              \frac{\mathscr{T}(\Omega')}{\Omega-\Omega'} \ , \\
 \xoverline{\Im} \ T^{R/A}(\Omega) &=
    \mp \frac{1}{2} \mathscr{T}(\Omega) \ .
\end{align}
\end{subequations}

\subsubsection{Explicit solution}
As mentioned earlier, a key advantage of the ladder approximation
is the fact that an explicit solution for the in-medium $T$-matrix can be found. 
The solution of the Bethe-Salpeter equation in $T(Z)$ can be expressed in terms
of the bare two-body interaction, $V^{(2)}$, and the bubble, $\Pi(Z)$, namely
\begin{equation}\label{ExplicitTmatrixAnalytic}
  T(Z) =
      V^{(2)}
      \left(1 - \frac{1}{2}\Pi(Z)V^{(2)}\right)^{-1} \ .
\end{equation}
To have a direct connection with the propagator, we need to relate
explicitly the (two-body) bubble to the (one-body) propagator. We do this in the energy representation.
Fourier transforming Eq.~\eqref{DefBubble}, we find
\begin{equation}\label{PropToBubble}
  \Pi_{MN}(\Omega_p)
  = - \frac{1}{\beta}
    \sum_{q} \mathcal{G}_{\mu_1\nu_1}(\omega_{q}) \ 
      \mathcal{G}_{\mu_2\nu_2}(\Omega_p - \omega_{q}) \ ,
\end{equation}
where the multi-indices are $M = (\mu_1,\mu_2)$ and $N= (\nu_1,\nu_2)$.
Therefore, plugging Eq.~\eqref{PropToBubble} into 
Eq.~\eqref{ExplicitTmatrixAnalytic} with $Z = i\Omega_p$,
we find an explicit functional of the $T$-matrix as a function of the propagator,
$T[\mathcal{G}]$.  
Together with $\mathcal{G}^{\text{Dyson}}[\Sigma]$
and $\Sigma^{\text{EoM}}[\mathcal{G}, T]$, we obtain the complete set of equations
for the self-consistent cycle in the ladder approximation.

For future numerical applications, we represent the expressions in terms of
spectral functions only. 
We start by relating $\mathscr{T}(\Omega)$ to $S(\omega)$.
From Eq.~\eqref{ExplicitTmatrixAnalytic} and Eq.~\eqref{SpeTmatrix},
the spectral function of the $T$-matrix reads
\begin{align}\label{SpTmatrixLadder_from_PiR}
  \mathscr{T}(\Omega)
    &=
    i V^{(2)}
    \left[
        \left(1 - \frac{1}{2}\Pi^{R}(\Omega)V^{(2)}\right)^{-1}
    \right. \nonumber \\
    &\phantom{= i V^{(2)} [ \ ( \ }
    \left.
        - \left(1 - \frac{1}{2}\Pi^{A}(\Omega)V^{(2)}\right)^{-1}
    \right] \ .
\end{align}
Combined with the dispersion relations of the bubble given in 
Eqs.~\eqref{DispersionBubble}, we obtain an explicit functional $\mathscr{T}[P]$.
To complete the relation between $\mathscr{T}(\Omega)$ and $S(\omega)$
we must express the spectral function of the bubble $P(\Omega)$
in terms of $S(\omega)$. The functional $P[S]$ is straightforwardly
obtained from Eq.~\eqref{PropToBubble} which reads, in terms of spectral
functions,
\begin{multline}\label{SpBubble_from_SpProp}
    P_{MN}(\Omega)
    =
    \frac{1}{b(\Omega)}
    \int^{+\infty}_{-\infty} \frac{\mathrm{d}\omega}{2\pi} \
      S_{\mu_{1}\nu_{1}}(\omega) f(\omega) \\ 
      \times S_{\mu_{2}\nu_{2}}(\Omega - \omega) f(\Omega - \omega) \ ,
\end{multline}
where $b(\Omega) \equiv \frac{1}{e^{\beta \Omega} - 1}$
is the Bose-Einstein distribution function.

Having found the relation between $\mathscr{T}(\Omega)$ and $S(\omega)$,
we now need to relate
$\mathscr{T}(\Omega)$ to the width $\Gamma(\omega)$
(and $\Sigma^{\infty}$) by using $\Sigma^{\text{EoM}}[\mathcal{G}, T]$.
In the particular case of the ladder approximation,
the two-loop Feynman diagram of Eq.~\eqref{Sigma_2BExactVertex},
shown in Fig.~\ref{Fig:Sigma_2BExactVertex}, can be simplified.
Eventually, we find that
\begin{multline}\label{WidthLadder_from_SPTmatrix}
  \Gamma_{\mu\nu}(\omega)
  = - \frac{1}{3}
  \sum_{\lambda_1 \lambda_2}
  \int^{+\infty}_{-\infty} \frac{\mathrm{d}\omega'}{2\pi}\
    \left[f(\omega') + b(\omega'-\omega)\right] \\
    \times 
    \mathscr{T}_{\mu\lambda_1\lambda_2\nu}(\omega-\omega')
    S^{\lambda_1\lambda_2}(\omega') \ .
\end{multline}
Regarding the instantaneous part of the self-energy, $\Sigma^{\infty}$,
we obtain it by computing the tadpole given in Eq.~\eqref{Sigma_2BExactVertex}
(also shown in Fig.~\ref{Fig:Sigma_2BExactVertex}).
To express it in terms of spectral functions,
we use the HFB self-energy expression of Eq.~\eqref{DressedSelfHFBkBody_from_SpProp}.
Since we assume only two-body interactions, the instantaneous
self-energy simply reads
\begin{equation}\label{DressedSelfHFB_from_SpProp}
  \Sigma^{\infty}_{\mu_1\mu_2}
  =
  \frac{1}{2}\sum_{\mu_3 \mu_4}
    v^{(2)}_{[\mu_{1} \mu_{2} \dot{\mu}_{3} \dot{\mu}_{4}]}
        \int^{+\infty}_{-\infty} \frac{\mathrm{d}\omega}{2\pi}
        f(-\omega) \ S^{\mu_3\mu_4}(\omega) \ .
\end{equation}
To close the self-consistent cycle in terms of spectral functions only,
all that is left to do is to relate $S(\omega)$ with $\Gamma(\omega)$
and $\Sigma^{\infty}$. This has already been done in Sec.~\ref{subsec:SpectralFromSelf},
with the help of Eq.~\eqref{DirectSpFromSelfEnergy}.
A step-by-step summary on the self-consistent cycle in the ladder approximation
is provided later on, in Sec.~\ref{subsubsec:Synthesis_SCLadder}.
Before this synthesis, however, we discuss the
validity of the explicit solution given in Eq.~\eqref{ExplicitTmatrixAnalytic} 
in terms of its convergence as a series.

\subsubsection{Convergence of the series of ladders}
Although the $T$-matrix equation can be put formally
in the explicit form of Eq.~\eqref{ExplicitTmatrixAnalytic},
we must ensure that the solution
is mathematically well-defined and physically meaningful.
Thouless, in his pioneering work of Ref.~\cite{Thouless1960}, argued that the 
infinite series of ladder diagrams
must be convergent.
Put differently, the kernel of the equation must satisfy
\begin{equation}\label{NecessarySufficientSolTmatrix}
  \forall \Omega_p, \ 
    r\left(\frac{1}{2}\Pi(\Omega_p)V^{(2)}\right) < 1 \ ,
\end{equation}
where $r\left(M\right)$ denotes the spectral radius of the operator $M$, i.e.\ 
\begin{equation}\label{DefSpectralRad}
  r(M) \equiv \sup\left\{ \abs{\lambda}, \ \lambda \in \sigma(M)\right\} \ ,
\end{equation}
and $\sigma(M)$ denotes the spectrum of $M$.
In Thouless' work, the main argument for not allowing any eigenvalue
to be strictly greater than $1$ was that, in the case of homogeneous matter,
the infinite volume limit would be ill-defined.
Anticipating whether the series of ladders will be convergent or not for any general
propagator is a difficult problem. Instead, we focus here on the simpler case
of assuming a HFB unperturbed propagator. This case is relevant
to ensure the convergence of (at least) the first iteration in the self-consistent cycle,
when an HFB propagator is taken as a starting point.

In Ref.~\cite{Thouless1960}, Thouless studied when
the condition~\eqref{NecessarySufficientSolTmatrix}
was satisfied in the context of homogeneous matter.
His derivations assumed an Hermitian two-body
interaction that is separable in terms of the relative incoming
and outgoing momenta.
Under some additional simplifying assumptions on the potential,
Thouless showed that the condition~\eqref{NecessarySufficientSolTmatrix}
is equivalent to using an unperturbed propagator
associated to a local minimum of the Bardeen-Cooper-Schrieffer (BCS) energy.
Eventually, this leads to the well-known statement
that the critical temperature of a superfluid transition, $T_c$, corresponds
to the temperature at which the series of ladders with a normal propagator
(i.e.\ conserving particle-number symmetry)
diverges at zero energy and total momentum. 

In the case of a general Hermitian two-body interaction,
we show in~\ref{App:CvLadders_Nec}
that the stability of the complex general HFB self-energy
is a \emph{necessary condition} for the series of ladders to converge
at any energy.
More precisely, the stability of the HFB self-energy is equivalent to the following
statement about the spectral radius of the $T$-matrix kernel:
\begin{equation}\label{StabilitySelfEnergyHFB}
    r\left(\frac{1}{2}\Pi(0)V^{(2)}\right) < 1 \ .
\end{equation}
We note that, unlike Eq.~\eqref{NecessarySufficientSolTmatrix}, 
the kernel here is evaluated at $\Omega_p=0$. 
When changing the temperature, the series of ladders at
zero energy converges if and only if the HFB self-energy is stable.
With this demonstration, we extend Thouless' criterion to the case 
of a general Hermitian two-body interaction
- without any assumption about its separability.
The critical temperature $T_c$ of a
phase transition corresponds to the temperature at which the series
of ladders at zero energy starts to diverge.

Although this criterion is 
useful to determine the critical temperature, the stability
of the HFB self-energy does not appear to be sufficient to ensure
the convergence of the series of ladders at \emph{any} energy.
This is crucial
in our case
to have a well-defined many-body approximation.
This convergence problem was studied at zero temperature by
Balian and Mehta~\cite{Balian1962} for a generic
pairing interaction. Eventually, their proof relating the stability of the BCS
energy and the convergence of the series of ladders at any energy
turned out to be incomplete, due to the presence of what
they refer as essential singular points~\cite{Balian1963a}.
This suggests that the original argument of Thouless 
to ensure convergence of the series of ladders at any energy relies
too strongly on the simplifying assumptions made in Ref.~\cite{Thouless1960}.

Having said that,
in practical applications such as in the study of
the BCS-BEC crossover~\cite{Strinati2012},
the series of ladders in the normal phase
are typically observed to diverge 
at non-zero energies \emph{before} reaching the superfluid phase.
When this happens, the divergence is interpreted as the occurrence of
a pseudogap in the weak-coupling regime, or a bosonic mode in the strong-coupling
regime~\cite{Perali2002,Yanase1999,Maly1999}.
Those pre-pairing effects occur between the critical temperature $T_c$
and a temperature $T_d$, which we choose to define as
\begin{equation}\label{DynamicalPairing_DefTemp}
    T_d \equiv 
        \sup \left\{ 
                T, \ 
                \exists \Omega_p, 
                     r\left(\frac{1}{2}\Pi(\Omega_p)V^{(2)}\right) > 1
            \right\} \ .
\end{equation}
We refer to $T_d$ as the \emph{dynamical pairing temperature}, since it characterises
a regime where the $T$-matrix is divergent at an energy which is not necessarily zero.
In contrast, $T_c$ is purely related to static, zero-energy effects.
We note that, in the literature,
many different ways of evaluating the temperature regime
where pre-pairing effects occur have been proposed.
For example, Ref.~\cite{Tsuchiya2009} introduces two different pseudogap temperatures 
and Ref.~\cite{Palestini2014} defines a crossover temperature 
which is compared to a pair dissociation temperature. To avoid confusions,
we chose a notation which departs from the usual $T^*$,
which is typically used to denote all these analogous
(yet different) temperatures.

Being able to estimate $T_d$ in a general case, to have an idea of the
regime where pre-pairing effects can play an important role,
is crucial. These effects matter in various areas of physics, such as in the study of
high-$T_c$ superconductors, of ultra-cold Fermi gases and of nuclear systems.
For reviews on those systems and how their superfluid properties are connected,
we refer the reader to Refs.~\cite{Strinati2018,Chen2005}.
Regarding the existence of a pseudogap and its impact on observables,
the question still remains open for
both the unitary Fermi
gas~\cite{Jensen2019,Randeria2014}
and infinite nuclear matter~\cite{Durel2020,Schnell1999}. 
In this context, we have derived a \emph{sufficient condition}
that ensures the convergence of
the series of ladders at \emph{any energy}, thus allowing to exclude
dynamical pairing effects whenever such condition is satisfied.
Similarly to Thouless' criterion, the condition is formulated as a strong
stability condition on the HFB self-energy.
To be precise, we show in~\ref{App:CvLadders_Suf} that
the series of ladders converges at any energy if
\begin{equation}\label{StrongStabilitySelfEnergyHFB}
    \norm{\frac{1}{2} \Pi(0) V^{(2)}}_{\mathcal{S}_{\infty}} < 1 \ ,
\end{equation}
where $\norm{M}_{\mathcal{S}_{\infty}}$
denotes the supremum of the singular values of an operator $M$, i.e.\ 
\begin{equation}\label{DefInfSchattenNorm}
  \norm{M}_{\mathcal{S}_{\infty}}
    \equiv 
    \sup \left\{\sqrt{s}, \ s \in \sigma\left(M^\dagger M\right)\right\} \ .
\end{equation}
The condition~\eqref{StrongStabilitySelfEnergyHFB} is said to
be strong because it
implies the standard stability
condition~\eqref{StabilitySelfEnergyHFB}, thanks to the following property
\begin{equation}\label{spRad_vs_normInf}
    \forall M , \; r(M) \leq \norm{M}_{\mathcal{S}_{\infty}} \ .
\end{equation}
More details on that condition and its derivation are given in~\ref{App:CvLadders_Suf}.

Let us consider a standard scenario, where the temperature of the physical system
decreases steadily, starting from the normal phase down to the superfluid phase.
Starting from the normal phase, we first hit the temperature $T_s$, defined as the 
temperature where the equality
\begin{equation}\label{UpperBound_DynPairing_DefTemp}
    \norm{\frac{1}{2}  \Pi(0) V^{(2)}}_{\mathcal{S}_{\infty}} = 1 \ 
\end{equation}
is satisfied. 
The physical system will be in a normal phase for $T > T_s$ and in a superfluid phase for $T < T_c$.
Dynamical pairing effects can only occur in
the regime $T_c < T < T_s$, where the conditions
\begin{equation}
    r\left(\frac{1}{2}  \Pi(0) V^{(2)}\right)
    < 1 < \norm{\frac{1}{2}  \Pi(0) V^{(2)}}_{\mathcal{S}_{\infty}} \ 
\end{equation}
are satisfied. 
In this scenario, Eq.~\eqref{UpperBound_DynPairing_DefTemp}
defines an upper-bound on $T_d$, i.e.\ 
\begin{equation}
    T_c \leq T_d \leq T_s \ .
\end{equation}
It is remarkable that, under some quite general assumptions,
the dynamical pairing instability that generates a divergence of the 
series of ladders (at an energy which is not necessarily zero), can be estimated on
the basis of purely static considerations at the mean-field level.

There are two interesting cases where
the conditions~\eqref{StabilitySelfEnergyHFB}
and~\eqref{StrongStabilitySelfEnergyHFB} become equivalent
to the convergence of the series of ladders at any energy.
The first one is the limit of a weakly interacting many-body system.
In this trivial case, Thouless' criterion
becomes asymptotically valid, as observed in the context of ultra-cold
Fermi gases~\cite{Strinati2012}.
A second, more interesting case, detailed in~\ref{App:CvLadders_Sep},
concerns two-body interactions that are separable in the sense
that there exist two tensors $v$ and $v'$ such that
\begin{equation}\label{SeparableInteraction_Def}
    V^{(2)}_{MN} = v_M v'_N \ .
\end{equation}
This case is in close connection with the one originally studied by
Thouless~\cite{Thouless1960}.
However, the two-body interactions considered in Ref.~\cite{Thouless1960}
were assumed to be separable only for relative incoming and outgoing momenta.
Because of the total momentum conservation and the spin dependence,
the two-body interactions were not separable in the same sense as
Eq.~\eqref{SeparableInteraction_Def}, thus leaving open the possibility
that $T_c < T_s$.
In the sub-case where the two-body interaction is separable for relative
incoming/outgoing momenta \emph{and} non-zero only
between spin singlet states of zero total momentum, 
the separability of Eq.~\eqref{SeparableInteraction_Def}
is recovered and Thouless' criterion holds as it was shown by an exact calculation
in Ref.~\cite{Gaudin1960b}.
These considerations shed some new light on the shortcomings of
Thouless' criterion regarding dynamical pairing instabilities and on the existence of
pre-pairing effects in the case of strongly interacting fermions like in the
BCS-BEC crossover or in nuclear systems.

In this section, considerations on the convergence of the series of ladders
have been quite general. We
do not discuss whether (or how) better bounds on $T_d$ can be found in the general case.
Further investigations on physical systems of interest, where the interactions
are known, are also left out of the scope of this paper.
In particular, if the criterion of Eq.~\eqref{UpperBound_DynPairing_DefTemp}
is to be useful in practice,
one should check whether or not $T_s$
remains close to $T_d$ in typical cases of interest.

%
%
  \begin{table}
    \centering
    \begin{tabular}{|c||c|c|}

      \hline
      Step &
      Eq. &
      Instruction \\

      \hline
      1 &
      \eqref{SpBubble_from_SpProp} &
      \begin{minipage}{0.677\linewidth}
        \begin{multline*}
          P_{MN}(\Omega)
            =
            \frac{1}{b(\Omega)}
            \int^{+\infty}_{-\infty} \frac{\mathrm{d}\omega}{2\pi} \
              S_{\mu_1\nu_1}(\omega)f(\omega) \\
              \times S_{\mu_2\nu_2}(\Omega - \omega)f(\Omega - \omega)
        \end{multline*}
       \vspace{0cm}
     \end{minipage}
      \\

      \hline
      2 &
      \eqref{RePiR/A_from_SpBubble} &
      \begin{minipage}{0.677\linewidth}
       \vspace{0.2cm}
       \begin{equation*}
         \xoverline{\Re} \ \Pi^{R/A}(\Omega)
                 = \mathcal{P}\int_{-\infty}^{+\infty}
                     \frac{\mathrm{d}\Omega'}{2\pi}
                     \frac{P(\Omega')}{\Omega-\Omega'}
      \end{equation*}
      \vspace{0cm}
    \end{minipage}
      \\

      \hline
      3 &
      \eqref{SpTmatrixLadder_from_PiR} &
      \begin{minipage}{0.677\linewidth}
       \vspace{0.2cm}
       \begin{multline*}
         \mathscr{T}(\Omega) =
           i V^{(2)}
           \left[
              \left(1 - \frac{1}{2}\Pi^{R}(\Omega)V^{(2)}\right)^{-1}
           \right. \\
           \left.
              -
              \left(1 - \frac{1}{2}\Pi^{A}(\Omega)V^{(2)}\right)^{-1}
           \right]
        \end{multline*}
        \vspace{0cm}
      \end{minipage}
      \\

      \hline
      4 &
      \eqref{WidthLadder_from_SPTmatrix} &
      \begin{minipage}{0.677\linewidth}
       \vspace{0.2cm}
       \begin{multline*}
         \Gamma_{\mu\nu}(\omega)
         = \\
         - \frac{1}{3}
          \sum_{\lambda_1 \lambda_2}
          \int^{+\infty}_{-\infty} \frac{\mathrm{d}\omega'}{2\pi}\
            \left[f(\omega') + b(\omega'-\omega)\right] \\
            \times
            \mathscr{T}_{\mu\lambda_1\lambda_2\nu}(\omega-\omega') \ 
            S^{\lambda_1\lambda_2}(\omega')
       \end{multline*}
       \vspace{0cm}
     \end{minipage}
     \\
   
      \hline
       5&
       \eqref{DressedSelfHFB_from_SpProp}&
       \begin{minipage}{0.677\linewidth}
        \vspace{0cm}
        \begin{multline*}
          \Sigma^{\infty}_{\mu_1\mu_2}
          =
          \frac{1}{2}\sum_{\mu_3 \mu_4}
            v^{(2)}_{[\mu_{1} \mu_{2} \dot{\mu}_{3} \dot{\mu}_{4}]} \\
          \times
            \int^{+\infty}_{-\infty} \frac{\mathrm{d}\omega}{2\pi} \
            f(-\omega) \ S^{\mu_3\mu_4}(\omega)
        \end{multline*}
        \vspace{0cm}
      \end{minipage}
      \\

      \hline
      6&
      \eqref{DispSelfEnergy}&
      \begin{minipage}{0.677\linewidth}
       \vspace{0.2cm}
       \begin{equation*}
         \xoverline{\Re} \ \Sigma^{R/A}(\omega) =
           \Sigma^{\infty}
           + \mathcal{P}\int^{+\infty}_{-\infty}
               \frac{\mathrm{d}\omega'}{2\pi} \
               \frac{\Gamma(\omega')}{\omega-\omega'}
       \end{equation*}
       \vspace{0cm}
     \end{minipage}
     \\

     \hline
     7&
     \eqref{SPFunctionFromSelfEnergy}&
     \begin{minipage}{0.677\linewidth}
      \vspace{0.2cm}
      \begin{multline*}
        S(\omega) =
          i\left(
            \omega - U - \Sigma^{R}(\omega)
          \right)^{-1}
          \\
          -
          i\left(
            \omega - U - \Sigma^{A}(\omega)
          \right)^{-1}
      \end{multline*}
      \vspace{0cm}
    \end{minipage}
    \\


   \hline
    \end{tabular}
    \caption{Equations to be solved numerically for
    the self-consistent ladder approximation.}
    \label{Table:SetEqsSCGFladder}
  \end{table}

\subsubsection{Summary}\label{subsubsec:Synthesis_SCLadder}
We recapitulate here the set of equations which are to be solved
in the self-consistent ladder approximation. These are summarised succinctly for ease of access, 
and showcase the potential of NC-SCGF to generate formally simple expressions to
describe physical systems in symmetry-broken phases.

Before doing so, we want to emphasise that the equations for the self-consistent cycle
can be written down in several ways. 
One possible formulation of the self-consistent cycle could focus on the
different objects $(\mathcal{G},\Sigma,T,\Pi)$.
Working directly on these objects is problematic numerically.
In particular, at zero-temperature, the self-energy undergoes drastic variations around its poles~\cite{Bergli2011}.
This issue stems from the presence of isolated poles for finite systems and it is worsened when working with discrete single-particle bases.
In practice, this problem is circumvented
by reformulating the Dyson-Schwinger and the Bethe-Salpeter
equations as energy-independent eigenvalue
problems~\cite{Schirmer1982,Danovich2011,Barbieri2003,Degroote2011}.
Their solutions give the spectroscopic amplitudes and pole positions
of the Green's functions~\cite{Barbieri2017lnp,Barbieri2001}.
Combining this approach to Krylov projection techniques~\cite{Schirmer1989,Soma2014},
SCGF calculations are routinely carried out for medium-mass nuclei~\cite{Soma2020} and reached masses of A=138~\cite{Arthuis2020}.

Alternatively, in the context of infinite nuclear matter,
SCGF calculations at non-zero temperature were carried out
in the ladder approximation
without symmetry-breaking~\cite{Ramos1989,Bozek1999,Bozek2002,Frick2003,Soma2006,Rios2006a,Rios2008,Soma2008}
in a continuous plane-wave single-particle basis.
In this case, the spectral functions are continuous 
and the self-consistent problem is more easily handled numerically
when expressed in terms of those.
Note that the two numerical approaches above are complementary in the sense
that one takes care of isolated poles on the real axis (associated to a
discrete set of states) and the other takes care of branch cuts
(associated to a continuous set of states). 

\begin{figure*}[t]
  \centering
  \includegraphics[scale=0.481, trim= 170 50 170 50]{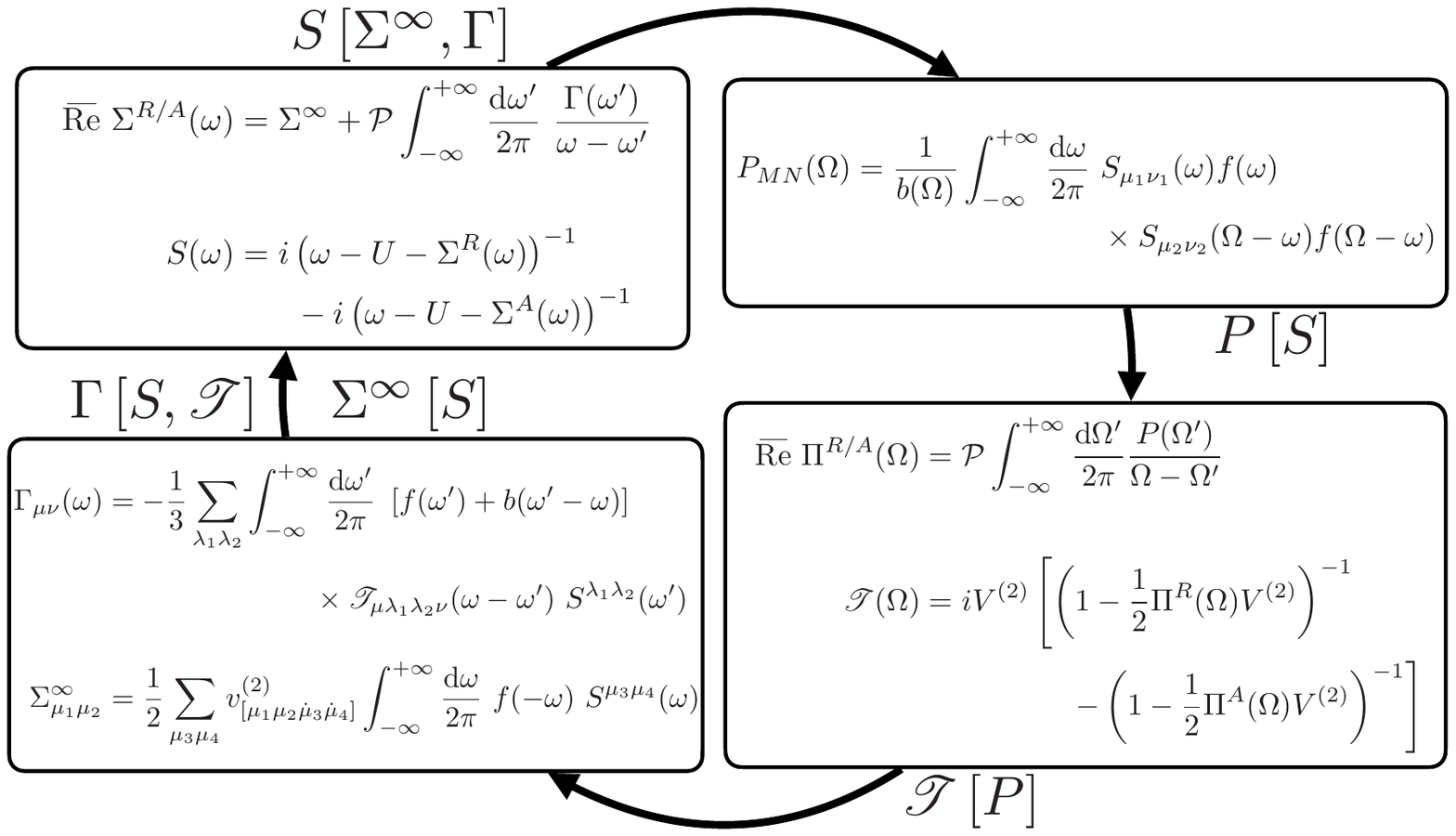}
\caption{Flowchart representing the self-consistent cycle obtained by iterating
the self-consistent ladder approximation expressed in terms of
$(S,\Sigma^{\infty},\Gamma,\mathscr{T},P)$. Equations displayed in the flowchart
corresponds to those gathered in Table.~\ref{Table:SetEqsSCGFladder}.}
\label{Fig:SCGF_cycle_LadderEqsSpectral}
\end{figure*}

For simplicity, we concentrate here on the second approach.
In this case, we choose to focus on the spectral functions 
$(S,\Sigma^{\infty},\Gamma,\mathscr{T},P)$.
The set of equations to be solved for the self-consistent ladder approximation
are gathered in Table~\ref{Table:SetEqsSCGFladder}. The corresponding iterative
cycle is also pictured as a flowchart in
Fig.~\ref{Fig:SCGF_cycle_LadderEqsSpectral} for clarity.
Suppose we start the iterative cycle
with a given spectral function $S^{(n)}(\omega)$. 
The equations displayed in Table~\ref{Table:SetEqsSCGFladder}
give back, after a full cycle, an updated spectral function $S^{(n+1)}(\omega)$.
The calculation is considered to be converged when changes in the spectral function
from one cycle to the next remain below a certain tolerance. At that point,
we obtain the spectral functions $(S,\Sigma^{\infty},\Gamma,\mathscr{T},P)$
in the self-consistent ladder approximation.
We stress that after the equations of Table~\ref{Table:SetEqsSCGFladder} are solved 
self-consistently, the knowledge of the spectral functions allows us
to derive the one- and two-body Green's functions in the
self-consistent ladder approximation.
From these, we can obtain any one- and two-body observables, including macroscopic
properties, like the total energy and the thermodynamics, or microscopic data, 
like pairing gaps and single-particle spectra.  

Finally, let us stress that all the equations of Table~\ref{Table:SetEqsSCGFladder}
have been derived in the Nambu-covariant formalism.
Since all equations are expressed in terms of Nambu tensors,
the equations for the self-consistent ladder
approximation remain valid after any Bogoliubov transformation.

\section{Conclusions}\label{Sec:Conclusions}

The theory of Self-Consistent Green's Function (SCGF) has been reformulated
in a Nambu-covariant fashion, a substantial formal advance to treat symmetry-broken systems. 
We have dubbed this new formalism Nambu-covariant Self-Consistent Green's function (NC-SCGF).
This step forward is achieved by expressing Green's functions and
other many-body objects in terms of Nambu tensors, as introduced in Part~I of our work~\cite{part1}.
This formalism can be applied to study many-body systems at non-zero temperatures,
and can incorporate the effect of two-, three- and higher many-body interactions.
While most of the exact properties have been shown to remain valid
under any Bogoliubov transformation, we have also exposed examples
which remain valid only up to the action of a restricted group.
For example, this is the case of the GMK sum rule which, in its
standard formulation, remains valid only up to a change of single-particle bases.

In addition, taking advantage of the synthetic Nambu-covariant formalism,
several exact properties have been revisited.
We have shown that the positivity bound on the diagonal elements
of the spectral function is a consequence of a more general
definite positiveness property. 
From it, we have deduced additional positivity
bounds on diagonal and off-diagonal elements of the spectral function.

We have also revisited the standard
interpretation of the spectral function
of the propagator, $S(\omega)$,
as a combination of quasiparticle Lorentzian-like resonances embedded in a
smooth background.
In the case where the tensor $U+\Sigma^{R/A}(\omega)$ is not normal,
the tensor $U+R(\omega)$ does not commute with $\Gamma(\omega)$,
precluding the previous interpretation. 
This led us to introduce the line-shape tensor, $\Theta(\omega)$, which 
can be interpreted physically
as a characterisation of interferences between the damped propagation of
a quasiparticle state in the residual medium and the excitation of a
continuum of modes displayed by the residual medium.
The quasiparticle resonances in the spectrum of $S(\omega)$
become Fano-like resonances, and their line shapes are related to $\Theta(\omega)$.
Eventually, we have argued that $\Theta(\omega)$ provides an interesting indicator of the
combined importance of correlations and symmetry-breaking within a many-body system.

Building on the NC-SCGF formalism, we can formulate
Nambu-covariant approximations that are self-consistent
not only in the propagator, but also in the two-body vertex. 
We have paid specific attention to the self-consistent ladder approximation by
giving it an explicit formulation, valid for symmetry-broken phases
and for a general two-body interaction. The self-consistent cycle boils
down to seven equations, shown in Table~\ref{Table:SetEqsSCGFladder},
for the spectral functions of the propagator, the self-energy,
the in-medium $T$-matrix and the bubble propagator. 
Thanks to the Nambu-covariant formalism introduced in Part~I,
these equations display a formal complexity which is similar to those in the
symmetry-conserving case.
This is a crucial step towards an efficient numerical implementation of the
self-consistent ladder approximation in symmetry-breaking phases.
Applications to
superfluid nuclear matter
will be reported in a future work.

Along these lines, we have also revisited the question of the convergence 
of the series of ladders for a
complex general HFB propagator.
We have shown that 
Thouless' criterion, commonly used to determine
the critical temperature $T_c$,
remains valid in the case of
a complex general HFB propagator and a general Hermitian two-body interaction.
We have also proposed a new criterion to determine a pre-pairing temperature, $T_s$,
such that dynamical pairing instabilities generating singularities in
the (non self-consistent) $T$-matrix can only occur
at temperatures $T_c < T < T_s$.

Finally, let us mention two immediate developments that could stem
from this work.
First, to go beyond the self-consistent ladder approximation,
one can consider corrections to the irreducible part of the Bethe-Salpeter
equation. For example, this has been done in the symmetry-conserving case
at zero temperature for application to atomic nuclei~\cite{Barbieri2003}.
Alternatively, the $T$-matrix can be used as
an effective interaction vertex in the spirit of the Brueckner-Bethe-Goldstone
method (see Ref.~\cite{Day1967} for an introductory review).
Second, beyond purely diagrammatic considerations,
the formalism of NC-SCGF opens up new avenues towards the restoration of symmetries.
While many-body approximations such as CC and MBPT have been
extended to include breaking and
restoration of symmetries~\cite{Duguet2014,Duguet2016},
no restoration procedure has been designed and implemented for SCGF approximations,
despite its critical role in applications to finite
systems such as atomic nuclei~\cite{Qiu2017,Qiu2019,Bally2021}.
In Refs.~\cite{Duguet2014,Duguet2016}, the restoration of symmetries
was designed at zero temperature by mixing several single-reference calculations
with vacua related by non-unitary Bogoliubov transformations.
Studying the problem of symmetry-restoration in terms of Nambu tensors,
which, by design, are covariant with respect to non-unitary Bogoliubov
transformations, could shed new light on the restoration of symmetries
within SCGF schemes.

\section*{Acknowledgments}
The authors thank J.~W.~T.~Keeble for proofreading the manuscript.
M.~D.\ would like to thank T.~Duguet for pointing him to
the work of R.~Balian and M.~L.~Mehta on the convergence of the series
of ladders at zero temperature~\cite{Balian1962,Balian1963a}.

This work is supported by STFC, through Grants Nos 
ST/L005743/1 and ST/P005314/1; 
by the Spanish MICINN
through the ``Ram\'on y Cajal" program with grant RYC2018-026072 and
the ``Unit of Excellence Mar\'ia de Maeztu 2020-2023" award to the Institute of Cosmos Sciences (CEX2019-000918-M).
TRIUMF receives federal funding via a contribution agreement with the National Research Council of Canada. 


\appendix

\section{Hermitian conjugate tensor}\label{App:HermitianConjugation}
In this appendix, we give a precise definition
of the Hermitian conjugate of a tensor.
First we focus on $(1,1)$-tensors where
the definition is identical to the adjoint of an operator
with respect to a given Hermitian product.
Then, we extend the concept of Hermitian conjugation to
$(p,q)$-tensors where $p+q=2k$ and $k$ is a natural number.

\subsection{Conjugate of (1,1)-tensors}
Let $h\left( \ . \ , \ . \ \right)$ be the Hermitian product 
defined on $\mathscr{H}^f$ canonically associated to $\mathcal{B}^f$,
as defined in~\eqref{WorkingFieldBasis}.
We recall that in this case $\mathcal{B}^f = \set{a^\dagger_b} \cup \set{a_c}$
where the creation and annihilation operators are associated to an orthonormal
single-particle basis $\mathcal{B}$, and the covariant Nambu fields reads simply
\begin{subequations}
\begin{align}
    \mathrm{A}_{(b, 1)} &\equiv a^\dagger_b \ , \\
    \mathrm{A}_{(b, 2)} &\equiv a_{b} \ .
\end{align}
\end{subequations}
In this case, we can decompose any two vector $u$ and $v$ in $\mathscr{H}^f$ as
\begin{subequations}
\begin{align}
    u &= \sum_{\mu} u^\mu \ \mathrm{A}_{\mu} \ , \\
    v &= \sum_{\mu} v^\mu \ \mathrm{A}_{\mu} \ ,
\end{align}
\end{subequations}
and their Hermitian product reads by definition as
\begin{equation}\label{FieldHermitianPdt}
    h(u,v) \equiv \sum_{\mu} (u_{\mu})^* v_{\mu} \ .
\end{equation}
Let us stress that $h\left( \ . \ , \ . \ \right)$ is distinct from
the metric tensor introduced in Part~I. In general,
\begin{equation}
    h(\mathrm{A}_\mu, \mathrm{A}_\nu) \neq \Set{\mathrm{A}_\mu, \mathrm{A}_\nu} = g_{\mu\nu} \ .
\end{equation}
We use this Hermitian product to define the notion of orthogonality of a field
basis, i.e.\ a basis ${\mathcal{B}^f}' = \Set{\mathrm{A}'_{\mu}}$ is orthogonal
if and only if
\begin{equation}
    \forall \mu, \nu, \quad h\left(\mathrm{A}'_{\mu},\mathrm{A}'_{\nu}\right) = \delta_{\mu'\nu'} \ .
\end{equation}
We also use this Hermitian product
to define the Hermitian conjugate of a linear operator $t$,
acting on $\mathscr{H}^f$, as the unique linear operator $t^\dagger$ verifying
\begin{equation}
    \forall u, v \in \mathscr{H}^f, \quad
    h(u, t(v)) \equiv h(t^\dagger(u), v) \ .
\end{equation}
Eventually, the Hermitian conjugation is transported to $(1,1)$-tensors
by using the canonical identification of $(1,1)$-tensors to linear operators.
In practice, $t^\dagger$ is the unique $(1,1)$-tensor whose coordinates verify
\begin{equation}\label{HermitianConjTensor_orthoBase}
    (t^\dagger{)^{\mu}}_{\nu} = \left({t^{\nu}}_{\mu}\right)^*
\end{equation}
in any \emph{orthogonal} basis ${\mathcal{B}^f}'$.

If ${\mathcal{B}^f}'$ is not orthogonal, the
relation in Eq.~\eqref{HermitianConjTensor_orthoBase} no longer holds.
Conversely, if, in an orthogonal basis, two $(1,1)$-tensors $r$ and $s$ verify 
\begin{equation}\label{typicHermitianRelation_Unit}
    {r^{\mu}}_{\nu} = \left({s^{\nu}}_{\mu}\right)^* \, ,
\end{equation}
then they verify
\begin{equation}\label{typicHermitianRelation_NonUnit}
    {r^{\mu}}_{\nu} = (s^\dagger{)^{\mu}}_{\nu} 
\end{equation}
in \emph{any} basis.
We say that $r$ is the Hermitian conjugate tensor of $s$.

This notion of Hermitian conjugation allows us
to introduce the unitary group $\U(\mathscr{H}^f)$
defined as the sub-group of $\GL(\mathscr{H}^f)$ characterised by 
a transformation
${\mathcal{W}^{\mu}}_{\nu}$ verifying
\begin{equation}
  \sum_{\lambda} 
    {\mathcal{W}^{\mu}}_{\lambda} \ 
    (\mathcal{W}^\dagger{)^{\lambda}}_{\nu}
    = {g^{\mu}}_{\nu} = \delta_{\mu\nu} \ .
\end{equation}
Note that $\GL(\mathscr{H}^f)$ contains the sub-group
$\mathrm{O}(\mathscr{H}^f, g)$ which is a faithful representation of Bogoliubov
transformations,
while $\U(\mathscr{H}^f)$ only contains the sub-group
$\mathrm{O}(\mathscr{H}^f, g) \cap \U(\mathscr{H}^f)$ which is
a faithful representation of \emph{unitary} Bogoliubov transformations.

In this paper, we  make use extensively of the Hermitian conjugation of tensors.
This allows us to write equations which
are invariant under Bogoliubov
transformations, rather than invariant under
unitary Bogoliubov transformations only.
To see this, compare for example Eq.~\eqref{typicHermitianRelation_Unit}
and Eq.~\eqref{typicHermitianRelation_NonUnit} whose groups of invariance
are, respectively, $\U(\mathscr{H}^f)$ and $\GL(\mathscr{H}^f)$.
In addition, in this paper our working basis $\mathcal{B}^f$ is orthogonal, so
whenever we derive an equation like Eq.~\eqref{typicHermitianRelation_Unit}
in $\mathcal{B}^f$, we can and will straightforwardly generalise it to an equation of 
the type of Eq.~\eqref{typicHermitianRelation_NonUnit}.

\subsection{Extensions}
So far we have only defined the Hermitian conjugate of a $(1,1)$-tensor.
To generalise this definition to $(p,q)$-tensors with $p+q=2k$, 
we first focus on $(k,k)$-tensors.
We will subsequently extend the definition by compatibility with the raising and lowering
of indices with the metric $g$.
\subsubsection{(k,k)-tensors}
Tensors of type $(k,k)$ are canonically associated with linear operators acting
on ${\mathscr{H}^f}^{\otimes k}$. To define the Hermitian
conjugate of a linear operator $t$ acting on ${\mathscr{H}^f}^{\otimes k}$,
we use the Hermitian product defined by
\begin{equation}
    h^{(k)}
    \left(
        u_1 \otimes \dots \otimes u_{k},
        v_1 \otimes \dots \otimes v_{k}
    \right)
    \equiv
    \prod^{k}_{i=1}
        h\left(u_i,v_i\right) \ ,
\end{equation}
where $u_i$ and $v_i$ are vectors of $\mathscr{H}^f$.
We associate a unique linear operator $t^\dagger$ to a linear operator $t$ acting on ${\mathscr{H}^f}^{\otimes k}$,
verifying 
\begin{multline}
    h^{(k)}
    \left(
        u_1 \otimes \dots \otimes u_{k},
        t
        (v_1 \otimes \dots \otimes v_{k})
    \right) \\
    \equiv
    h^{(k)}
    \left(
        t^\dagger
        (u_1 \otimes \dots \otimes u_{k}),
        v_1 \otimes \dots \otimes v_{k}
    \right) \ .
\end{multline}
The Hermitian conjugation of $(k,k)$-tensors is then defined
by canonical identification of $(k,k)$-tensors with linear operators
acting on ${\mathscr{H}^f}^{\otimes k}$.
In practice, for any $(k,k)$-tensor $t$, its Hermitian conjugate is the unique 
$(k,k)$-tensor $t^\dagger$ whose coordinates verify in any \emph{orthogonal} basis
\begin{equation}\label{HermitianConjVector_p+q}
    (t^\dagger{)^{M}}_{N} 
        = ({t^{N}}_{M})^* \ .
\end{equation}
For convenience we have used the $k$-dimensional multi-indices
\begin{subequations}
\begin{align}
    M &\equiv (\mu_1, \dots, \mu_k) \ , \\
    N &\equiv (\nu_1, \dots, \nu_k) \ ,
\end{align}
\end{subequations}
so that tensor coordinates of a $(k,k)$-tensor are denoted as
\begin{equation}
    {t^{M}}_{N} \equiv {t^{\mu_1 \dots \mu_k}}_{\nu_1 \dots \nu_k}  \ .
\end{equation}
In particular, let us stress that the $k^{\text{th}}$ tensor power of our working field basis
${\mathcal{B}^{f}}^{\otimes k}$ is orthogonal with respect to
$h^{(k)}\left( \ . \ , \ . \ \right)$.
Note also that this definition of the Hermitian conjugation
is compatible with the previous one for $(1,1)$-tensors.

\subsubsection{Metric compatibility}
We define the Hermitian conjugate of a $(p,q)$-tensor with $p+q=2k$ and $k$
a natural number, by compatibility with the raising and lowering
of indices with the metric $g$.
In our working basis $\mathcal{B}^f$, which is orthogonal
with respect to $h\left( \ . \ , \ . \ \right)$, the metric verifies
\begin{equation}\label{ComplexCjgDualMetric}
    g^{\mu\nu}= (g_{\nu\mu})^*  \ .
\end{equation}
Consequently, we can define the Hermitian conjugate of a $(0,2k)$-tensor
as the unique $(2k,0)$-tensor whose coordinates verify
\begin{equation}\label{HermitianConjTensor_orthoBase_02k}
    (t^\dagger)^{MN} 
        = \left(t_{NM}\right)^* \ ,
\end{equation}
in any orthogonal basis. Similar definitions hold for all the associated
type $(p,q)$ of tensors with $p+q=2k$.

A $(p,q)$-tensor $t$ (with $p + q = 2k$) is said 
to be Hermitian or anti-Hermitian, respectively, when\footnote{We
employ a slight abuse of notation here. In principle $t^\dagger$
is a $(q,p)$-tensor which is not of the same type as $t$. Here $t^\dagger$
is to be understood as the $(p,q)$-tensor obtained after the appropriate
raising and lowering operation with the metric, so that it is of the same type as $t$.}
\begin{subequations}
\begin{align}
    t^\dagger &= t \ , \\
    t^\dagger &= -t \ .
\end{align}
\end{subequations}
For example, since $\mathcal{B}^f$ is orthogonal and the metric satisfies
Eq.~\eqref{ComplexCjgDualMetric}, we have 
\begin{equation}\label{ComplexCjgDualMetric_TensorEq}
    g^\dagger = g
\end{equation}
i.e.\ the metric is Hermitian.
Finally, a $(p,q)$-tensor $t$ with $p+q=2k$ will be said to be unitary if
and only if 
\begin{equation}
    t^\dagger \ t = t \ t^\dagger = g \ ,
\end{equation}
where the product of tensors is defined
in~\ref{App:FunctionalCalculus}.

\section{Functional calculus}\label{App:FunctionalCalculus}
In this appendix, we specify the functional calculus that 
is used in this paper and that is necessary to develop the NC-SCGF formalism.
In other words, we provide a definition for functions depending on tensors.
We follow the same approach as the previous Appendix and focus first on 
defining functions on $(1,1)$-tensors.
Then, the definition is extended to $(p,q)$-tensors
where $p+q=2k$ and $k$ is a natural number.

\subsection{Functional calculus for (1,1)-tensors}
Let ${t^{\mu}}_{\nu}$ be a $(1,1)$-tensor and $T^{1,1}(\mathscr{H}^f)$
be the vector space of $(1,1)$-tensors.
We define functions on the space of $(1,1)$ tensors in the same way as it is usually done for operators or matrices.
A formal power series on $(1,1)$-tensors
is defined such that $\forall i \in \mathbbm{N}^*$,
\begin{subequations}
\begin{equation}\label{powerTensorDef}
        {(t^i)^{\mu}}_{\nu} 
        \equiv
        \sum_{\alpha_1 \dots \alpha_{i-1}}
            {t^{\mu}}_{\alpha_1} \ 
            {t^{\alpha_1}}_{\alpha_2} \ 
            \dots \ 
            {t^{\alpha_{i-2}}}_{\alpha_{i-1}} \ 
            {t^{\alpha_{i-1}}}_{\nu} \ .
\end{equation}
For $i=0$, we define
\begin{equation}\label{nullPowerTensorDef}
        {(t^0)^{\mu}}_{\nu} \equiv {g^{\mu}}_{\nu} = \delta_{\mu\nu} \ .
\end{equation}
Moreover, for any formal power series $g_1$ and $g_2$ and for $\lambda \in \mathbbm{C}$, 
we define
\begin{equation}
    (\lambda g_1 + g_2 )(t) \equiv \lambda g_1(t) + g_2(t) \ .
\end{equation}
\end{subequations}
Let $f(X)$ be the formal power series
\begin{equation}
    f(X) \equiv \sum^{+\infty}_{i=0} c_i \ X^i \ ,
\end{equation}
with $c_i \in \mathbbm{C}$. With the above definitions, the
function
$f$ on $(1,1)$-tensors reads
\begin{align}
        f: T^{1,1}(\mathscr{H}^f) &\to T^{1,1}(\mathscr{H}^f) \nonumber \\
        t &\mapsto f(t) = \sum^{+\infty}_{i=0} c_i \ t^i \ .
\end{align}
Writing down the coordinates explicitly, one finds
\begin{equation}
        {f(t)^{\mu}}_{\nu} = \sum^{+\infty}_{i=0} c_i \ {(t^i)^{\mu}}_{\nu} \ .
\end{equation}
Note that, throughout this paper, we use 
the shorthand notation $t^0 = 1$.

We extend the functional calculus
on $(1,1)$-tensors
to holomorphic functions by using
Cauchy's integral formula, in analogy to the case of operators and matrices. 
Briefly, we recall that this amounts to define
\begin{equation}
    f(t) \equiv 
        \frac{1}{2\pi i}
        \int_{C} f(\lambda) (\lambda - t)^{-1} \mathrm{d}\lambda
\end{equation}
for any function $f$ which is holomorphic in a open set, including the spectrum
of $t$, and for $C$ a contour enclosing it.
For more details on holomorphic calculus we refer to classical textbooks
such as Ref.~\cite{Dunford1958}.

\subsection{Extensions}
We extend the functional calculus on $(1,1)$-tensors to $(0,2)$- and $(2,0)$-tensors as follows.
Let $t_{\mu\nu}$ be a $(0,2)$-tensor.
Powers $(t^i)_{\mu\nu}$ are defined such that they are compatible
with raising and lowering indices starting from the definitions given
in Eqs.~\eqref{powerTensorDef} and~\eqref{nullPowerTensorDef}.
To be concrete, we have $\forall i \in \mathbbm{N}^*$
\begin{subequations}
\begin{equation}
        (t^i)_{\mu\nu} 
        \equiv
        \sum_{\alpha_1 \dots \alpha_{i-1}}
            t_{\mu\alpha_1} \ 
            {t^{\alpha_1}}_{\alpha_2} \ 
            \dots \ 
            {t^{\alpha_{i-2}}}_{\alpha_{i-1}} \ 
            {t^{\alpha_{i-1}}}_{\nu} \ ,
\end{equation}
and, for $i=0$,
\begin{equation}
        (t^0)_{\mu\nu} \equiv g_{\mu\nu} \ .
\end{equation}
\end{subequations}
In analogy to $(1,1)$-tensors, these definitions are extended
to any formal power series by linearity. The extension to holomorphic functions
is performed, again, using Cauchy's integral formula.
Functions of $(2,0)$-tensors are defined analogously to functions
of $(0,2)$-tensors.

Finally, we extend the functional calculus to tensors of type $(p,q)$
where $p+q=2k$ for a natural number $k$.
As we just did for $(0,2)$- and $(2,0)$-tensors, we only need to define
functions on $(k,k)$-tensors.
Functions on a $(p,q)$-tensor are then obtained by enforcing 
compatibility with the raising and lowering operation of indices.
Let ${t^{\mu_1 \dots \mu_k}}_{\nu_1 \dots \nu_k}$ be a $(k,k)$-tensor.
For convenience, we use the same $k$-dimensional multi-indices
\begin{subequations}
\begin{align}
    M &\equiv (\mu_1, \dots, \mu_k) \ , \\
    N &\equiv (\nu_1, \dots, \nu_k) \ ,
\end{align}
\end{subequations}
as in~\ref{App:HermitianConjugation}.
Powers of a $(k,k)$-tensor are defined such that, $\forall i \in \mathbbm{N}^*$
\begin{subequations}
\begin{equation}
    {(t^i)^{M}}_{N} 
        \equiv
        \sum_{L_1 \dots L_{i-1}}
            {t^{M}}_{L_1} \ 
            {t^{L_1}}_{L_2} \ 
            \dots \ 
            {t^{L_{i-2}}}_{L_{i-1}} \ 
            {t^{L_{i-1}}}_{N} \ ,
\end{equation}
where $L_p$ are dummy $k$-dimensional multi-indices.
Again, for $i=0$, we define
\begin{equation}
    {(t^0)^{M}}_{N} \equiv  \prod^{k}_{p=1} {g^{\mu_p}}_{\nu_p} = \delta_{MN} \ .
\end{equation}
\end{subequations}
Similarly to $(1,1)$-tensors, those definitions are extended
to formal power series by linearity and to holomorphic functions
using Cauchy's integral formula.

\section{Interpretation of \texorpdfstring{$\Theta(\omega)$}{O(w)}}
\label{App:ThetaInterpretation}
In Sec.~\ref{subsubsec:DirectSpectralFromSelf}, we introduce a tensor,
$\Theta(\omega)$,
which allows us to relate the self-energy to
the spectral function of the propagator.
In this section, we motivate the physical interpretation
of $\Theta(\omega)$ as a tensor characterising
the interferences between the damped propagation of 
quasiparticle states and the excitation
of a continuum of non-resonant modes displayed by the residual medium. 
To justify this interpretation, we look at the impact of $\Theta(\omega) \neq 0$ in
two approximations that are often considered in the symmetry-conserving case. We first
discuss the \emph{peak
approximation}, which is obtained assuming that all the self-energy components are constant 
around a given quasiparticle peak energy. We then turn our attention to
the \emph{quasiparticle approximation}, which incorporates additional dispersive corrections associated to the 
energy dependence of $\Sigma$ around the peak. These two cases turn out to 
provide a concordant physical picture for $\Theta(\omega)$. 

\subsection{Peak approximation}
The peak approximation of the spectral function $S(\omega)$ can be interpreted
as a degraded version of the standard quasiparticle approximation~\cite{Rios2012}.
In this approximation, we focus on describing the spectral function around
one peak, centred at an energy $\omega_{\text{qp}}$, which is associated to
a quasiparticle state.
A crude way of locating these peaks consists in fixing their locus
as the solutions of
\begin{equation}\label{Pk_SpFunction}
  \det\left[ \omega_{\text{qp}} - U - R(\omega_{\text{qp}}) \right] = 0 \ .
\end{equation}
The quasiparticle states are then the eigenvectors of
$U + R(\omega_{\text{qp}})$ associated to the eigenvalues $\omega_{\text{qp}}$.
For energies $\omega \simeq \omega_{\text{qp}}$, the spectral function
can be approximated by assuming that all the energy-dependent components
of the self-energy are roughly constant and independent of the energy, 
\begin{subequations}\label{PkApproxSelfEnergy}
\begin{align}
    R(\omega) &\simeq  R(\omega_{\text{qp}}) \ , \\
    \Gamma(\omega) &\simeq  \Gamma(\omega_{\text{qp}}) \ , \\
    \Theta(\omega) &\simeq  \Theta(\omega_{\text{qp}}) \ .
\end{align}
\end{subequations}
This approximation, albeit crude, is already sufficient
to motivate a physical interpretation of $\Gamma(\omega_{\text{qp}})$
in the symmetry-conserving case.
The spectral function in this peak approximation reads
\begin{multline}
\label{PK_NCSpFunctionFormula}
  S_{\text{pk}}(\omega) \equiv
    \left[\vphantom{\left(\left(\frac{\Gamma(\omega)}{2}\right)^2\right)}
      \Gamma(\omega_{\text{qp}})
      +
      2 \left( \omega - U - R(\omega_{\text{qp}}) \right)
      \Theta(\omega_{\text{qp}})
    \right]
     \\  \times 
    \left[
      \left(
      \left(\omega - U - R(\omega_{\text{qp}})\right)^2
      +
      \left(\frac{\Gamma(\omega_{\text{qp}})}{2}\right)^2
      \right)
      \left(\vphantom{\left(\frac{\Gamma(\omega)}{2}\right)^2}
        1 + \Theta^2(\omega_{\text{qp}})
      \right)
    \right]^{-1} \, .
\end{multline}
%
When $\Theta(\omega_{\text{qp}}) = 0$, we recover the well-known
Lorentzian shape of the spectral function.
We thus interpret $\Gamma(\omega_{\text{qp}})$ as
a tensor generalisation of the width of a quasiparticle resonance,
whose inverse is related to the life-time and mean-free path of
a quasiparticle state propagating at energy $\omega_{\text{qp}}$~\cite{Rios2012}.
In other words, $\Gamma(\omega_{\text{qp}})$ characterises the damping
of a quasiparticle state propagating at energy $\omega_{\text{qp}}$ in the medium.

In contrast, when $\Theta(\omega_{\text{qp}}) \neq 0$,
Eq.~\eqref{PK_NCSpFunctionFormula} is no longer a simple Lorentzian,
but resembles instead a \emph{Fano function} (or a Fano line-shape profile)~\cite{Fano1961}.
We recall that a normalised Fano resonance, $F_{q}(\omega)$,
is expressed in terms of the position of the resonance, $\omega_{\text{qp}}$;
its width, $\Gamma_{\text{qp}}$; and its line-shape parameter, $q$, as
\begin{equation}\label{FanoProfileExpanded}
    F_{q}(\omega) =
        \frac{
                \left(\frac{\Gamma_{\text{qp}}}{2}\right)^2
                + 2 \left(\frac{\Gamma_{\text{qp}}}{2}\right)
                    (\omega - \omega_{\text{qp}}) q^{-1}
                +
                \left(\frac{\omega - \omega_{\text{qp}}}{q}\right)^{2}
            }{\left(
                (\omega - \omega_{\text{qp}})^2 
                + \left(\frac{\Gamma_{\text{qp}}}{2}\right)^2
                \right)
                \left(
                    \vphantom{\left(\frac{\Gamma_{\text{qp}}}{2}\right)^2}
                    1+q^{-2}
                \right)
            } \ .
\end{equation}
For an introduction and an historical perspective on Fano functions,
we refer the reader to Refs.~\cite{Miroshnichenko2010,Rau2004}.
The family of Fano functions, indexed over the line-shape parameter $q$,
can be seen as an extension of a Lorentzian function.
The latter is recovered in the limit $q \to +\infty$.

At finite $q$, the constant term in the numerator is related to the quasiparticle resonance;
the quadratic term, to the non-resonant background; and the linear term,
to quasiparticle-background interferences~\cite{Miroshnichenko2010}.
For $\abs{\omega - \omega_{\text{qp}}} \ll q$ we can drop the quadratic term
in the numerator and we recover in essence Eq.~\eqref{PK_NCSpFunctionFormula},
i.e.\ a Lorentzian whose numerator is shifted by a linear contribution
in $\omega - \omega_{\text{qp}}$. Another way to show the similarity
between Eq.~\eqref{PK_NCSpFunctionFormula} and a Fano function 
consists in rewriting Fano functions as
\begin{equation}\label{FanoProfileFactorised}
   F_{q}(\omega) =
        \frac{1}{4}
        \frac{ 
            \left(\Gamma_{\text{qp}} + 2 (\omega - \omega_{\text{qp}}) q^{-1}\right)^2
        }{
         \left(
                (\omega - \omega_{\text{qp}})^2 
                + \left(\frac{\Gamma_{\text{qp}}}{2}\right)^2
                \right)
                \left(
                    \vphantom{\left(\frac{\Gamma_{\text{qp}}}{2}\right)^2}
                    1+q^{-2}
                \right)   
        }   \ .    
\end{equation}
This formula suggests a direct analogy between the line-shape parameter, $q$, and the inverse 
of the line-shape tensor, $\Theta^{-1}(\omega_\text{qp})$.
Note, however, the difference between the numerator in Eq.~\eqref{PK_NCSpFunctionFormula}, which is
squared, and that of Eq.~\eqref{FanoProfileFactorised}, which is not.
A similar difference already appears in the symmetry-conserving case,
where the numerator in the Lorentzian spectral function typically involves a linear (rather than quadratic) width. 

In physical applications, the Fano line-shape parameter $q$ is interpreted as the
consequence of interferences between a discrete state and a competing continuum
of states around the same energy $\omega_{\text{qp}}$~\cite{Miroshnichenko2010}.
In the time domain, the interferences are related to a phase shift,
$-2\arctan(q^{-1})$, and a relative scaling factor, $q^2+1$, between
the quasiparticle and the background contributions~\cite{Ott2013}.
This motivates
an interpretation of $\Theta(\omega_{\text{qp}})$ as a tensor characterising 
the interferences between the propagation
of a quasiparticle state of energy $\omega_{\text{qp}}$ in the medium,
and the excitation of a continuum of modes that the background displays
around the energy $\omega_{\text{qp}}$.

Despite their usefulness, these interpretations of
$\Gamma(\omega_{\text{qp}})$ and $\Theta(\omega_{\text{qp}})$
are only valid for the crude peak approximation employed here.
If this approximation is refined, the direct connection between
$\Gamma(\omega_{\text{qp}})$ and the width of a quasiparticle resonance becomes
more tenuous. The same is true for the association between $\Theta(\omega_{\text{qp}})$
and the line-shape parameter of the resonance.
Moreover, it may be possible that the relation between $S_{\text{pk}}(\omega)$
and Fano functions is merely symbolic.
To discard the last possibility, we now turn to the more realistic
quasiparticle approximation of the spectral function and show
that the spectrum of the approximated
spectral function, $S_{\text{qp}}(\omega)$,
displays Fano-like resonances.

\subsection{Quasiparticle approximation}

In the symmetry-conserving case, one can prove that the spectral
function remains Lorentzian even when some dispersion corrections, which incorporate
energy-dependent effects, are included in the description
of the self-energy~\cite{Rios2007,Rios2012}.
Similarly, we now proceed to show that the shape of
the resonances displayed by the spectrum of the spectral function
is, in the quasiparticle approximation,
closely related to a Fano resonance.

The quasiparticle approximation we consider consists in assuming that
the analytic propagator, $\mathcal{G}(z)$, contains only simple isolated
poles in non-physical Riemann sheets\footnote{Non-physical
Riemann sheets are obtained by analytically continuing the propagator
through the real axis cut~\cite{Rios2012}. For technical details,
we refer to the review of Ref.~\cite{Farid1999}.}.
From Eq.~\eqref{DirectPropFromSelf}, the locus  $z_{\text{qp}}$ of poles 
in the complex energy plane arise from the solutions of
\begin{equation}\label{QP_Poles}
  \det\left[ z_{\text{qp}} - U - \Sigma(z_{\text{qp}}) \right] = 0 \ .
\end{equation}
Combining the Mittag-Leffler's theorem\footnote{See for example pp.~299-301
of~Ref. \cite{Markushevich1965}.}, the asymptotic property of the
analytic propagator given in Eq.~\eqref{AsymptoticExpProp},
and the assumption of simple poles,
the exact propagator can be decomposed in partial fractions and reads,
for $\Im z > 0$,
\begin{equation}\label{PartialFraction_Prop}
    \mathcal{G}(z) = \sum_{\substack{z_{\text{qp}} \\ \Im z_{\text{qp}} \leq 0}}
        \frac{G_{\text{qp}}}{z - z_{\text{qp}}}
\end{equation}
where $G_{\text{qp}}$ are the residues associated to the poles $z_{\text{qp}}$,
which verify
\begin{equation}
   G_{\text{qp}} = \frac{1}{2\pi i} 
    \int_{C_{\text{qp}}}  \left( z - (U + \Sigma(z)) \right)^{-1} \mathrm{d}z \ ,
\end{equation}
where $C_{\text{qp}}$ is a positively-oriented, arbitrarily small
closed path around $z_{\text{qp}}$.
We stress that in Eq.~\eqref{PartialFraction_Prop}
the sum runs over poles with negative imaginary parts,
$\Im z_{\text{qp}}\leq0$, and the equality is only valid
in the positive half-plane, $\Im z > 0$.
The extension of Eq.~\eqref{PartialFraction_Prop}
to $\Im z \leq 0$ would give the propagator
in a non-physical Riemann sheet.

We can combine Eqs.~\eqref{PartialFraction_Prop},~\eqref{PropToSpectralFunction}
and~\eqref{RetardedAdvancedHermitianCjg}, to find the quasiparticle spectral function
\begin{equation}\label{FanoSpectrumSp}
    S_{\text{qp}}(\omega)
    =
    \sum_{\substack{z_{\text{qp}} \\ \Im z_{\text{qp}} \leq 0}}
        \frac{
            \xoverline{\Re} \left( G_{\text{qp}} \right)
            \Gamma_{\text{qp}}
            -
            2 \ \xoverline{\Im} \left( G_{\text{qp}} \right)
            (\omega-\omega_{\text{qp}})
        }
        {
            (\omega-\omega_{\text{qp}})^2 + \left(\frac{\Gamma_{\text{qp}}}{2}\right)^2
        } \ ,
\end{equation}
where  $\omega_{\text{qp}}$ and $-\frac{\Gamma_{\text{qp}}}{2}$
are the real and imaginary parts of $z_{\text{qp}}$, i.e.\ 
\begin{equation}
    z_{\text{qp}} \equiv \omega_{\text{qp}} - i \frac{\Gamma_{\text{qp}}}{2} \ .
\end{equation}

To have a more direct relation between the residues and the self-energy,
we make the additional approximation that
\begin{equation}\label{TaylorExpSelfEnergy}
    \forall z \in C_{\text{qp}} , \quad
    \Sigma(z) \simeq 
        \Sigma(z_{\text{qp}}) 
        + (z - z_{\text{qp}}) \frac{\partial \Sigma}{\partial z}(z_{\text{qp}}) \ .
\end{equation}
In other words, we assume that the (complex) energy dependence of the 
self-energy around the pole is smooth and can be accurately
captured by a first-order Taylor expansion. 
With a bit of algebra, and using the Laurent expansion of
a resolvent\footnote{See for example Eq.~(6.32) of Ref.~\cite{Kato1980}}
around $z_{\text{qp}}$, we obtain
\begin{equation}\label{QP_ApproxResidue}
    G_{\text{qp}} 
        \simeq \mathcal{Z_{\text{qp}}} P_{\text{qp}} \ ,
\end{equation}
where $P_{\text{qp}}$ is the eigenprojection of
$U + \Sigma(z_{\text{qp}})$ associated to $z_{\text{qp}}$\footnote{Since
the eigenspace is non-degenerate, the eigenprojection $P_{\text{qp}}$ is simply
the outer product of the unique right and left eigenvector of
$U + \Sigma(z_{\text{qp}})$ associated to the eigenvalue $z_{\text{qp}}$.} and
\begin{equation}
    \mathcal{Z_{\text{qp}}} = 
        \frac{1}
        {1 - 
            \Tr_{\mathscr{H}^f}\left[ 
                P_{\text{qp}} \ 
                \frac{\partial \Sigma}{\partial z}(z_{\text{qp}})
                \right]
        } \ ,
\end{equation}
where $\Tr_{\mathscr{H}^f}$ denotes the trace over $\mathscr{H}^f$.
Since $U + \Sigma(z_{\text{qp}})$ is not necessarily
Hermitian, neither is the projector $P_\text{qp}$
and $\mathcal{Z_{\text{qp}}}$ is a complex number.
In the symmetry-conserving case, $\mathcal{Z_{\text{qp}}}$
is usually referred to as the \emph{renormalisation factor}
of the quasiparticle resonance. 
Using Eq.~\eqref{FanoSpectrumSp}, the quasiparticle spectral function reads
\begin{multline}\label{FanoSpectrumSp_1stOrder}
    S_{\text{qp}}(\omega)
    = \\
    \sum_{\substack{z_{\text{qp}} \\ \Im z_{\text{qp}} \leq 0}}
        \frac{
            \xoverline{\Re} \left(\mathcal{Z_{\text{qp}}} P_{\text{qp}} \right)
            \Gamma_{\text{qp}}
            -
            2 \ \xoverline{\Im} \left(\mathcal{Z_{\text{qp}}} P_{\text{qp}} \right)
            (\omega-\omega_{\text{qp}})
        }
        {
            (\omega-\omega_{\text{qp}})^2 + \left(\frac{\Gamma_{\text{qp}}}{2}\right)^2
        } \ .
\end{multline}
To make the Fano resonant structure more clear, we assume the projectors
to be orthogonal, so that
\begin{equation}\label{OrthoBasisSelfQp}
    P^\dagger_{\text{qp}} = P_{\text{qp}}
\end{equation}
and, eventually, we obtain
\begin{multline}\label{OrthogonalFanoSpectrumSp}
    S_{\text{qp}}(\omega)
    = \\
    \sum_{\substack{z_{\text{qp}} \\ \Im z_{\text{qp}} \leq 0}}
        {\Re}\left(\mathcal{Z_{\text{qp}}}\right)
        \frac{
            \Gamma_{\text{qp}}
            -
            2 \ \frac{{\Im}\left(\mathcal{Z_{\text{qp}}}\right)}{{\Re}\left(\mathcal{Z_{\text{qp}}}\right)}
            (\omega-\omega_{\text{qp}})
        }
        {
            (\omega-\omega_{\text{qp}})^2 + \left(\frac{\Gamma_{\text{qp}}}{2}\right)^2
        }
        P_{\text{qp}} \ .
\end{multline}
Comparing Eq.~\eqref{OrthogonalFanoSpectrumSp} to Eq.~\eqref{FanoProfileFactorised}, 
we clearly see that the spectrum of
$S_{\text{qp}}(\omega)$ is made of resonances with line-shapes that are similar to Fano functions. 
These resonances occur
at energies around $\omega_{\text{qp}}$; have a width $\Gamma_{\text{qp}}$;
a fragmentation ${\Re}\left(\mathcal{Z_{\text{qp}}}\right)$; 
and a line-shape factor $q_{\text{qp}}$ defined as
\begin{equation}
    q_{\text{qp}} \equiv
        -\frac{{\Re}\left(\mathcal{Z_{\text{qp}}}\right)}
             {{\Im}\left(\mathcal{Z_{\text{qp}}}\right)} \ .
\end{equation}
The quasiparticle state associated to a resonance is non-degenerate,
and corresponds to the eigenstate of $U+\Sigma(z_{\text{qp}})$
associated to the eigenvalue $z_{\text{qp}}$.
Interestingly, the additional assumptions made in Eqs.~\eqref{TaylorExpSelfEnergy}
and~\eqref{OrthoBasisSelfQp} imply that the analytic continuation
of the line-shape tensor verifies $\Theta(z_\text{qp}) = 0$.
However, we still have, in general, $\Theta(\omega_\text{qp}) \neq 0$,
which is reflected on the non-Lorentzian line-shape of
the resonance at $\omega_{\text{qp}}$.

The above analysis in the quasiparticle approximation further
supports our interpretation of the tensor $\Theta(\omega)$ as a characterisation
of the interferences between the propagation of a state
in the medium and the excitation of a continuum of non-resonant modes
displayed by the medium. In the peak approximation, the inverse of $\Theta(\omega_\text{qp})$
was directly proportional to a tensor generalisation of the line-shape parameter $q$.
Here, in the quasiparticle approximation, we gain further insight
by showing how the spectrum of the spectral function
does display Fano-like resonances when we assume Eqs.~\eqref{TaylorExpSelfEnergy}
and~\eqref{OrthoBasisSelfQp} to hold.
Future numerical implementations of the NC-SCGF approach will be able to discern 
the importance of Fano structures in the spectral function of many-body systems. 

\section{Convergence of the series of ladders}\label{App:CvLadders}
In this section, we study the convergence of the series of ladders.
We assume an Hermitian two-body interaction
and a complex general HFB propagator.
First, we derive
a straightforward extension of Thouless' criterion~\cite{Thouless1960}.
We show that, in general, the stability of the HFB self-energy
is equivalent to the convergence of the series of ladders at zero energy.
Second, we work a sufficient condition for the series to converge
at \emph{any} energy.
Finally, we identify separable interactions as a special case,
where the extension of Thouless' criterion is simultaneously 
necessary and sufficient.

\subsection{Necessary condition}\label{App:CvLadders_Nec}
Let us show that the stability of the HFB self-energy,
$\Sigma^{\text{HFB}}$, is a necessary condition
for the convergence of the series of ladders with an HFB propagator.
First, we recall that the HFB self-energy $\Sigma^{\text{HFB}}$
is an implicit solution of Eq.~\eqref{ImplicitHFBSelfenergy}.
Since we are restricting ourselves to the case of a two-body interaction,
this is equivalent to saying that the HFB self-energy is a fixed point of
the functional $\mathcal{F}$ defined by
\begin{equation}
    \mathcal{F}\left[\Sigma\right]_{\mu\nu}
    =
    -\frac{1}{2}
    \sum_{\lambda_2\lambda_{3}}
        v^{(2)}_{[\mu \dot{\lambda}_2 \dot{\lambda}_3 \nu]}
            \frac{1}{\beta} \sum_{\omega_{l}} \mathcal{G}[\Sigma]^{\lambda_{2}\lambda_{3}}(\omega_{l})
            e^{-i\omega_{l} \eta}
    \ ,
\end{equation}
where we recall that
\begin{equation}
    \mathcal{G}[\Sigma](\omega_{l}) = (i\omega_l - (U + \Sigma(\omega_l))^{-1} \ .
\end{equation}
Physically speaking, the stability of the HFB self-energy, as a fixed point of $\mathcal{F}$,
is important to ensure that the associated HFB state of the many-body system
will not decay after an infinitesimally small external (one-body) perturbation.

To study the linear stability of the fixed point, $\Sigma^{\text{HFB}}$,
of $\mathcal{F}$, we compute the effect of a small deviation
$\delta\Sigma$ from it.
Since $\mathcal{F}$ gives a self-energy which is both antisymmetric
and energy independent, we restrict our linear stability analysis to
perturbations $\delta\Sigma$ with the same properties.
From the differential of the inverse, we find the relation between     
$\delta\mathcal{G}$ and $\delta\Sigma$,
\begin{align}
    \delta\mathcal{G}[\Sigma^{\text{HFB}}](\omega_{l})
    &\equiv
        \mathcal{G}[\Sigma^{\text{HFB}} + \delta\Sigma](\omega_{l})
        -\mathcal{G}[\Sigma^{\text{HFB}}](\omega_{l}) \nonumber \\
    &=
        -
        \mathcal{G}[\Sigma^{\text{HFB}}](\omega_{l}) \ 
        \delta\Sigma \ 
        \mathcal{G}[\Sigma^{\text{HFB}}](\omega_{l})  \ .
\end{align}
The differential of $\mathcal{F}$ at $\Sigma^{\text{HFB}}$ reads explicitly
\begin{align}
    \delta\mathcal{F}[\Sigma^{\text{HFB}}]_{\mu\nu}
    &\equiv
        \mathcal{F}\left[\Sigma^{\text{HFB}} + \delta\Sigma\right]_{\mu\nu}
        - \mathcal{F}\left[\Sigma^{\text{HFB}}\right]_{\mu\nu}
        \nonumber \\
    &=
        \frac{1}{2\beta}
        \sum_{\omega_{l}}
        \sum_{\lambda_2\lambda_{3}}
        e^{-i\omega_{l} \eta}
        v^{(2)}_{[\mu \dot{\lambda}_2 \dot{\lambda}_3 \nu]}
        \mathcal{G}[\Sigma^{\text{HFB}}]^{\lambda_{2}\alpha_2}(\omega_{l})
        \nonumber \\
    &\phantom{=\frac{1}{2\beta}
                \sum_{\omega_{l}}\sum_{\lambda_2\lambda_{3}}e
             }
         \times
         \delta\Sigma_{\alpha_2\alpha_3} \ 
         \mathcal{G}[\Sigma^{\text{HFB}}]^{\alpha_3\lambda_{3}}(\omega_{l})
        \ ,
\end{align}
which eventually simplifies to
\begin{equation}
   \delta\mathcal{F}[\Sigma^{\text{HFB}}]_{\mu\nu}
    =
    \frac{1}{2}
    \sum_{\lambda_2\lambda_{3}}
    v^{(2)}_{[\mu \nu \lambda_2 \lambda_3]}
    \Pi^{(\lambda_2,\alpha_2)(\lambda_3,\alpha_3)}(0)
    \ 
    \delta\Sigma_{\alpha_2\alpha_3} \ ,
\end{equation}
where we have used the antisymmetry property of the linear perturbation
of the self-energy, as well as the energy representation
of the bubble propagator, $\Pi$, in Eq.~\eqref{DefBubble}. 
We rewrite the previous expression using multi-indices, 
\begin{equation}
    \delta\mathcal{F}[\Sigma^{\text{HFB}}]^{M}
    =
    \sum_{N}
    \left(
    \frac{1}{2}
    V^{(2)}
    \Pi(0)\right)^{MN}
   \delta\Sigma_{N} \ .
\end{equation}
The Jacobian $J_{\mathcal{F}}[\Sigma^{\text{HFB}}]$ of $\mathcal{F}$ at
$\Sigma^{\text{HFB}}$ thus reads
\begin{equation}
    J_{\mathcal{F}}[\Sigma^{\text{HFB}}]
    =
    \frac{1}{2}
    V^{(2)}
    \Pi(0) \ .
\end{equation}
Let us recall that a fixed point $x_0$ of a functional $g[x]$ is said to be stable
if and only if
\begin{equation}
    r\left(J_{g}[x_0]\right) < 1 \ .
\end{equation}
In the case of $\mathcal{F}$, this means that
the HFB self-energy is stable if and only if\footnote{We recall that,
for any operator $A$ and $B$, we have
\begin{equation}
    r\left(AB\right) = r\left(BA\right) \ .
\end{equation}}
\begin{equation}\label{StabilitySelfEnergyHFB_App}
    r\left(\frac{1}{2} \Pi(0) V^{(2)}\right) < 1 \ .
\end{equation}

As a direct consequence, the stability of the HFB self-energy is equivalent to
the convergence of the series of ladders at the Matsubara frequency $\Omega_p = 0$.
We have not been able to prove that the stability of the HFB self-energy
is a sufficient condition for the ladders to converge at
any Matsubara frequency $\Omega_p$, unless further assumptions are made.
Thus, to the best of our knowledge, the stability of the HFB self-energy
is only a \emph{necessary condition} for the series of ladders to converge
at any energy.

\subsection{Sufficient condition}\label{App:CvLadders_Suf}
In this section we demonstrate how the assumption of a stronger stability condition
on the HFB self-energy allows us to prove the convergence of the ladders
at \emph{any} Matsubara frequency.

\subsubsection{Rationale}
Let us assume that, in addition of condition~\eqref{StabilitySelfEnergyHFB_App},
$\Sigma^{\text{HFB}}$ is a fixed point of $\mathcal{F}$
verifying the stronger stability constraint
\begin{equation}\label{StrongStabilitySelfEnergyHFB_App}
    \norm{\frac{1}{2}  \Pi(0) V^{(2)}}_{\mathcal{S}_{\infty}} < 1 \ .
\end{equation}
We recall that $\norm{M}_{\mathcal{S}_{\infty}}$
denotes the supremum of the singular values of an operator $M$, 
see Eq.~\eqref{DefInfSchattenNorm}.
In practice, this means that we require the singular values
of the Jacobian $J_{\mathcal{F}}[\Sigma^{\text{HFB}}]$
to be strictly smaller than $1$.
To prove that condition~\eqref{StrongStabilitySelfEnergyHFB_App}
is sufficient for the series of ladders to converge,
we prove in~\ref{App:DemoLemma} the following lemma,
\begin{equation}\label{DecreasingNorm}
    \forall \Omega_p , \; 
    \norm{\frac{1}{2}  \Pi(\Omega_p) V^{(2)}}_{\mathcal{S}_{\infty}}
    \leq \ 
    \norm{\frac{1}{2}  \Pi(0) V^{(2)}}_{\mathcal{S}_{\infty}} \ .
\end{equation}
Then, using the useful property
\begin{equation}\label{spRad_vs_normInf_App}
    \forall M , \; r(M) \leq \norm{M}_{\mathcal{S}_{\infty}} \ ,
\end{equation}
we have
\begin{equation}
    r\left(\frac{1}{2}  \Pi(\Omega_p) V^{(2)}\right)
    \leq \norm{\frac{1}{2}  \Pi(\Omega_p) V^{(2)}}_{\mathcal{S}_{\infty}}
\end{equation}
and the following implication is proven to hold
\begin{equation}
    \norm{ \frac{1}{2}  \Pi(0) V^{(2)}}_{\mathcal{S}_{\infty}} < 1
    \implies
    \forall \Omega_p, \;
    r\left(\frac{1}{2}  \Pi(\Omega_p) V^{(2)}\right) < 1 \ .
\end{equation}
Therefore, the strong stability
condition~\eqref{StrongStabilitySelfEnergyHFB_App}
on the HFB self-energy is a \emph{sufficient} condition
for the series of ladders to converge at any energy.

\subsubsection{Demonstration}\label{App:DemoLemma}
To prove lemma~\eqref{DecreasingNorm}, 
we first decompose $\Pi(\Omega_p)$ in its eigenbasis.
Let us recall that in the HFB approximation, the bubble propagator can be written as
\begin{align}\label{bubbleFromPropHFB}
    \Pi_{MN}(\Omega_p)
    = - \frac{1}{\beta}
    \sum_{q} &(i\omega_q - (U + \Sigma^{\text{HFB}}))^{-1}_{\mu_1\nu_1} 
        \nonumber \\
        &\times
        (i(\Omega_p - \omega_{q})- (U + \Sigma^{\text{HFB}}))^{-1}_{\mu_2\nu_2} \ ,
\end{align}
where the multi-indices are defined by
\begin{subequations}
\begin{align}
    M &\equiv (\mu_1,\mu_2) \ , \\
    N &\equiv (\nu_1,\nu_2) \ .
\end{align}
\end{subequations}
For simplicity, we assume that the spectrum of $U + \Sigma^{\text{HFB}}$
is non-degenerate. Since $U + \Sigma^{\text{HFB}}$ is Hermitian,
the decomposition of the analytic propagator in its eigenbasis reads
\begin{equation}\label{eigenDecompoPropHFB}
    (z - (U + \Sigma^{\text{HFB}}))^{-1} 
        = \sum_{i} \frac{P_i}{z - \epsilon_i} \ ,
\end{equation}
where $\epsilon_i$ are the real eigenvalues of $U + \Sigma^{\text{HFB}}$ and
$P_i$ are the Hermitian projectors on the associated eigenspaces.
The Hermitian projectors verify
\begin{subequations}
\begin{align}
    P_i P_j = \delta_{ij} P_i \ , \label{projOrthogonality} \\ 
    P_i^\dagger = P_i \ . \label{projHermitianity}
\end{align}
\end{subequations}
Plugging Eq.~\eqref{eigenDecompoPropHFB} into Eq.~\eqref{bubbleFromPropHFB}
and performing the Matsubara sum, we obtain the following expression for the 
bubble propagator:
\begin{equation}
    \Pi_{MN}(\Omega_p)
    =
    \sum_{ij} 
    (P_i)_{\mu_1\nu_1} (P_j)_{\mu_2\nu_2}
    \frac{1- (f(\epsilon_i) + f(\epsilon_j))}{\epsilon_i + \epsilon_j - i\Omega_p}
    \ .
\end{equation}
We can use this expression to rewrite the kernel of the $T$-matrix as
\begin{align}\label{eigenDecompoKernelLadder}
   \frac{1}{2} (V^{(2)}\Pi(\Omega_p) )_{MN}
   &= \nonumber\\
   \frac{1}{2}
   \sum_{ij}
   \sum_{\lambda_1\lambda_2}
        &\left( 
            v^{(2)}_{[\mu_1\mu_2 \lambda_1\lambda_2]}
            {(P_i)^{\lambda_1}}_{\nu_1} {(P_j)^{\lambda_2}}_{\nu_2}
        \right) \nonumber \\
        &\times
        \frac{1-(f(\epsilon_i) + f(\epsilon_j))}
        {\epsilon_i + \epsilon_j - i\Omega_p} \ .
\end{align}

To study the singular values of the kernel $\frac{1}{2}\Pi(\Omega_p) V^{(2)}$,
we must, by definition, study the spectrum of
$\frac{1}{4} (\Pi(\Omega_p) V^{(2)})^\dagger (\Pi(\Omega_p) V^{(2)})$.
Using the Hermitian property of $V^{(2)}$ and $\Pi(\Omega_p)$
we have
\begin{equation}
    \frac{1}{4} (\Pi(\Omega_p) V^{(2)})^\dagger (\Pi(\Omega_p) V^{(2)})
    =
    \frac{1}{4} V^{(2)} \Pi(-\Omega_p) \Pi(\Omega_p) V^{(2)} \ .
\end{equation}
Then, from Eq.~\eqref{eigenDecompoKernelLadder}, we have
\begin{align}
   \frac{1}{4} &\left(V^{(2)} \Pi(-\Omega_p) \Pi(\Omega_p) V^{(2)}\right)_{MN}
   = \nonumber \\
   &\frac{1}{4}
   \sum_{\substack{ij \\ kl}}
   \sum_{\substack{\lambda_1\lambda_2 \\ \kappa_1 \kappa_2}}
   \sum_{\alpha_1\alpha_2}
        \left( 
            v^{(2)}_{[\mu_1\mu_2 \lambda_1\lambda_2]}
            {(P_i)^{\lambda_1}}_{\alpha_1} {(P_j)^{\lambda_2}}_{\alpha_2}
        \right. \nonumber \\
    &\phantom{\frac{1}{4}
              \sum_{\substack{ij \\ kl}}
              \sum_{\substack{\lambda_1\lambda_2 \\ \kappa_1 \kappa_2}}
              \sum_{\alpha_1\alpha_2} ( \ 
             }
        \left.
        \times \ 
            (P_k)^{\alpha_1\kappa_1} (P_l)^{\alpha_2\kappa_2}
            v^{(2)}_{[\kappa_1\kappa_2 \nu_1\nu_2]}
        \right) \nonumber \\
    &\phantom{\frac{1}{4}
              \sum_{\substack{ij \\ kl}}
             }    
        \times
        \frac{1-(f(\epsilon_i) + f(\epsilon_j))}
        {\epsilon_i + \epsilon_j + i\Omega_p}
        \ 
        \frac{1-(f(\epsilon_k) + f(\epsilon_l))}
        {\epsilon_k + \epsilon_l - i\Omega_p}
        \ .
\end{align}
Using Eq.~\eqref{projOrthogonality} the expression simplifies to
\begin{align}
   \frac{1}{4} &\left(V^{(2)} \Pi(-\Omega_p) \Pi(\Omega_p) V^{(2)}\right)_{MN}
   = \nonumber \\
   &\frac{1}{4}
   \sum_{ij}
   \sum_{\substack{\lambda_1\lambda_2 \\ \kappa_1\kappa_2}}
        \left( 
            v^{(2)}_{[\mu_1\mu_2 \lambda_1\lambda_2]}
            (P_i)^{\lambda_1\kappa_1} (P_j)^{\lambda_2\kappa_2}
            v^{(2)}_{[\kappa_1\kappa_2 \nu_1\nu_2]}
        \right) \nonumber \\
    &\phantom{\frac{1}{4}
              \sum_{ij}
              \sum_{\substack{\lambda_1\lambda_2 \\ \kappa_1\kappa_2}}
              (
             }
        \times
        \frac{(1-(f(\epsilon_i) + f(\epsilon_j)))^2}
        {(\epsilon_i + \epsilon_j)^2 + \Omega_p^2}
        \ .
\end{align}
Using the idempotence of the projectors $P_i$ we have
\begin{align}
   \frac{1}{4} &\left(V^{(2)} \Pi(-\Omega_p) \Pi(\Omega_p) V^{(2)}\right)_{MN}
   = \nonumber \\
   &\frac{1}{4}
   \sum_{\substack{ij \\ kl}}
   \sum_{\substack{\lambda_1\lambda_2 \\ \kappa_1 \kappa_2}}
   \sum_{\alpha_1\alpha_2}
        \left( 
            v^{(2)}_{[\mu_1\mu_2 \lambda_1\lambda_2]}
            {(P_i)^{\lambda_1}}_{\alpha_1} {(P_j)^{\lambda_2}}_{\alpha_2}
        \right. \nonumber \\
    &\phantom{\frac{1}{4}
              \sum_{\substack{ij \\ kl}}
              \sum_{\substack{\lambda_1\lambda_2 \\ \kappa_1 \kappa_2}}
              \sum_{\alpha_1\alpha_2} ( \ 
             }
        \left.
        \times \ 
            (P_i)^{\alpha_1\kappa_1} (P_j)^{\alpha_2\kappa_2}
            v^{(2)}_{[\kappa_1\kappa_2 \nu_1\nu_2]}
        \right) \nonumber \\
    &\phantom{\frac{1}{4}
              \sum_{\substack{ij \\ kl}}
             }    
        \times
        \frac{(1-(f(\epsilon_i) + f(\epsilon_j)))^2}
        {(\epsilon_i + \epsilon_j)^2 + \Omega_p^2}
        \ .
\end{align}
To make the structure of the previous expression clearer
we introduce
\begin{equation}
    (P^{(2)}_{ij})_{(\mu_1,\mu_2)(\nu_1,\nu_2)}
    \equiv 
    (P_i)_{\mu_1\nu_1} (P_j)_{\mu_2\nu_2}
\end{equation}
so that
\begin{align}
   \frac{1}{4} V^{(2)} \Pi(-\Omega_p) \Pi(\Omega_p) V^{(2)}
   =
   \frac{1}{4}
   &\sum_{ij}
           V^{(2)}
            P^{(2)}_{ij}
            P^{(2)}_{ij}
            V^{(2)} \nonumber \\
    &\phantom{1}
        \times
        \frac{(1-(f(\epsilon_i) + f(\epsilon_j)))^2}
        {(\epsilon_i + \epsilon_j)^2 + \Omega_p^2}
        \ .
\end{align}
Since $P_i$ is Hermitian (see Eq.~\eqref{projHermitianity})
so is $P^{(2)}_{ij}$, i.e.\ 
\begin{equation}
    {P^{(2)}_{ij}}^{\dagger} = P^{(2)}_{ij} \ .
\end{equation}
Therefore, using the fact that $V^{(2)}$ is Hermitian
(see Eq.~\eqref{HermitianPotTensor}) we have
\begin{equation}
    V^{(2)} P^{(2)}_{ij} P^{(2)}_{ij} V^{(2)}
    =
     \left( P^{(2)}_{ij} V^{(2)} \right)^\dagger
     \left( P^{(2)}_{ij} V^{(2)}  \right)
\end{equation}
which implies that $V^{(2)} P^{(2)}_{ij} P^{(2)}_{ij} V^{(2)}$
is Hermitian semi-definite positive, i.e.\ 
\begin{equation}
    V^{(2)} P^{(2)}_{ij} P^{(2)}_{ij} V^{(2)} \succeq 0 \ .
\end{equation}
Concretely, this means that 
\begin{equation}
    \left( V^{(2)} P^{(2)}_{ij} P^{(2)}_{ij} V^{(2)}\right)^\dagger
    =
    V^{(2)} P^{(2)}_{ij} P^{(2)}_{ij} V^{(2)}
\end{equation}
and that for any $(2,0)$-tensor $X$ we have,
in any orthogonal basis,
\begin{equation}\label{SemiDefPositivePropVPPV}
    \sum_{MN} \ 
        (X^M)^* \ 
        {( V^{(2)} P^{(2)}_{ij} P^{(2)}_{ij} V^{(2)})^{M}}_{N} \ 
        X^N
        \geq 0 \ .
\end{equation}

To extract information on the eigenvalues from the Hermitian
semi-definite positiveness property~\eqref{SemiDefPositivePropVPPV},
we study its consequence on the Rayleigh quotient.
The Rayleigh quotient of a $(2,2)$-tensor $t$
and a $(2,0)$-tensor $X$ is defined in any orthogonal basis by
\begin{equation}
    \mathcal{R}(t,X)
        \equiv
        \frac{\sum_{MN} (X^M)^* \ {t^{M}}_{N}\  X^N}
        {\sum_{M} (X^M)^* X^M} \ .
\end{equation}
Using inequality~\eqref{SemiDefPositivePropVPPV} and
\begin{equation}
  0 \leq
        \frac{(1-(f(\epsilon_i) + f(\epsilon_j)))^2}
        {(\epsilon_i + \epsilon_j)^2 + \Omega_p^2}
  \leq
        \frac{(1-(f(\epsilon_i) + f(\epsilon_j)))^2}
        {(\epsilon_i + \epsilon_j)^2}
\end{equation}
we find that, for any $(2,0)$-tensor $X$
\begin{multline}
    \mathcal{R}
    \left(\frac{1}{4} V^{(2)} \Pi(-\Omega_p) \Pi(\Omega_p) V^{(2)}, X\right) \\
    \leq
    \mathcal{R}\left(\frac{1}{4} V^{(2)} \Pi(0) \Pi(0) V^{(2)}, X\right) \ .
\end{multline}
Since the supremum of the Rayleigh quotient is the spectral
radius, i.e.\ 
\begin{multline}
    \sup_{X} 
        \mathcal{R}\left(
            \frac{1}{4} V^{(2)} \Pi(-\Omega_p) \Pi(\Omega_p) V^{(2)},
            X
            \right)
    \\ =
    r\left(\frac{1}{4} V^{(2)} \Pi(-\Omega_p) \Pi(\Omega_p) V^{(2)}\right) \ ,
\end{multline}
we obtain
\begin{multline}
    r\left(\frac{1}{4} V^{(2)} \Pi(-\Omega_p) \Pi(\Omega_p) V^{(2)}\right) \\
    \leq
    r\left(\frac{1}{4} V^{(2)} \Pi(0) \Pi(0) V^{(2)}\right) \ .
\end{multline}
Hence, we finally have proven that
\begin{equation}
    \norm{ \frac{1}{2}  \Pi(\Omega_p) V^{(2)}}_{\mathcal{S}_{\infty}}
    \leq
    \norm{ \frac{1}{2}  \Pi(0) V^{(2)}}_{\mathcal{S}_{\infty}} \ .
\end{equation}
Note that, with a similar analysis, we can make the stronger statement that
$\norm{ \frac{1}{2}  \Pi(\Omega_p) V^{(2)}}_{\mathcal{S}_{\infty}}$
is an even function of $\Omega_p$ which decreases for $\Omega_p > 0$.  

\subsection{Separable interaction}\label{App:CvLadders_Sep}
In this final section, we study the particular case where the interaction
$V^{(2)}$ is separable.
By separable we mean that we assume the existence
of two tensors $v$ and $v'$ such that
\begin{equation}\label{SeparableInteraction}
    V^{(2)}_{MN} = v_M v'_N \ .
\end{equation}
We start from the stability condition~\eqref{StabilitySelfEnergyHFB_App}
on the HFB self-energy, $\Sigma^{\text{HFB}}$.
Since $V^{(2)}$ is separable, so is the product
$\frac{1}{2} \Pi(\Omega_p) V^{(2)}$
and we thus have
\begin{equation}\label{SpRad_SingularVal_Separable}
    r\left(\frac{1}{2} \Pi(\Omega_p) V^{(2)}\right) 
        = \norm{\frac{1}{2} \Pi(\Omega_p) V^{(2)}}_{\mathcal{S}_\infty} \ .
\end{equation}
Then, combining~\eqref{StabilitySelfEnergyHFB_App} and
Eq.~\eqref{SpRad_SingularVal_Separable} with $\Omega_p = 0$, we obtain
\begin{equation}
    \norm{\frac{1}{2} \Pi(0) V^{(2)}}_{\mathcal{S}_\infty} < 1 \ .
\end{equation}
Using lemma~\eqref{DecreasingNorm}, we have
\begin{equation}
    \norm{\frac{1}{2} \Pi(\Omega_p) V^{(2)}}_{\mathcal{S}_\infty}
    \leq 
    \norm{\frac{1}{2} \Pi(0) V^{(2)}}_{\mathcal{S}_\infty}
    < 1 \ .
\end{equation}
Eventually, using again Eq.~\eqref{SpRad_SingularVal_Separable},
we find that
\begin{equation}
    r\left(\frac{1}{2}  \Pi(\Omega_p) V^{(2)}\right) 
        \leq 
            r\left(\frac{1}{2}  \Pi(0) V^{(2)}\right) < 1  \ .
\end{equation}
Therefore, whenever the interaction is separable,
the stability of the HFB self-energy is a \emph{necessary and sufficient}
condition to the convergence of the series of ladders at any energy.

\bibliographystyle{elsarticle-num} 
\bibliography{biblio}





\end{document}